\documentclass[a4paper]{aa}
\usepackage{lscape}
\usepackage{graphicx} 
\usepackage{latexsym,amsmath,amssymb}
\usepackage{natbib}
\bibpunct{(}{)}{;}{a}{}{,}

\def \xmm {XMM-Newton}
\def \chandra {\it Chandra}
\def \suzaku {\it Suzaku}
\def \src {4U\,1630$-$47}

\def \nh {N${\rm _H}$}

\def \hcm {\hbox {\ifmmode $ atom cm$^{-2}\else atom cm$^{-2}$\fi}}
\def \arcmin {\hbox{$^\prime$}}
\def \arcsec {\hbox{$^{\prime\prime}$}}
\def \deg {$^{\circ}$}
\def \chisq {$\chi ^{2}$}
\def \rchisq {$\chi_{\nu} ^{2}$}
\def \approxgt{\mathrel{\hbox{\rlap{\lower.55ex \hbox {$\sim$}}
        \kern-.3em \raise.4ex \hbox{$>$}}}}
\def \approxlt{\mathrel{\hbox{\rlap{\lower.55ex \hbox {$\sim$}}
        \kern-.3em \raise.4ex \hbox{$<$}}}}

\newcommand{\mc}{\multicolumn}
\newcommand {\Msun}{M_\odot}

\newcommand {\fetfour} {\ion{Fe}{xxiv}}
\newcommand {\fetfive} {\ion{Fe}{xxv}}
\newcommand {\fetsix} {\ion{Fe}{xxvi}}
\newcommand {\ka} {K$\alpha$}
\newcommand {\kb} {K$\beta$}

\newcommand {\ssixteen} {\ion{S}{xvi}}

\newcommand {\nitseven} {\ion{Ni}{xxvii}}
\newcommand {\niteight} {\ion{Ni}{xxviii}}

\def\countsec{\hbox{count s$^{-1}$}}

\def \twelve {XB\,1254$-$690}
\def \grs {GRS\,1915+105}
\def \gro {GRO\,J1655$-$40}
\def \gx {GX\,13+1}

\def \seventeen {H\,1743$-$322}

\def \xtefifteen {XTE\,J1550$-$564}
\def \xtee {XTE\,J1807$-$294}

\def \nhwarmabs {$N{\rm _H^{warmabs}}$}

\def \xiunit {\hbox{erg cm s$^{-1}$}}
\def \logxi {$\log(\xi)$}

\def \countsec{\hbox{counts s$^{-1}$}}

\newcommand {\egau} {$E_{\rm gau}$}

\newcommand {\ktdbb} {$kT_{\rm dbb}$}

\newcommand {\ew} {$EW$}

\def \nh {$N{\rm _H}$}
\def \nhabs {$N{\rm _H^{abs}}$}

\def \xiunit {\hbox{erg cm s$^{-1}$}}
\def \logxi {$\log(\xi)$}
\def \xil {$\xi$}

\newcommand {\sigmav} {$\sigma_{\rm v}$}
\newcommand {\kms} {km~s$^{-1}$}

\def \kdbb {$k_{\rm dbb}$}
\def \kgau {$k_{\rm gau}$}

\begin{document}

\title{XMM-Newton observations reveal the disappearance of the wind in \src}

\author{M. D{\'i}az Trigo\inst{1} \and S. Migliari\inst{2} \and J.~C.~A. Miller-Jones\inst{3} \and M. Guainazzi\inst{4}}
\institute{
ESO, Karl-Schwarzschild-Strasse 2, D-85748 Garching bei M\"unchen, Germany
              \and
              Department of Astronomy and Meteorology \& Institute of Cosmic Science, University of Barcelona, Mart\'i i Franqu\`es 1, 08028 Barcelona, Spain
\and
              International Centre for Radio Astronomy Research, Curtin University, GPO Box U1987, Perth,
Western Australia 6845, Australia
\and
	XMM-Newton Science Operations Centre, Science Operations Department, ESAC, P.O. Box 78, E-28691 Villanueva de la Ca\~nada, Madrid, Spain
}

\date{Received ; Accepted:}

\authorrunning{D{\'i}az Trigo et al.}

\titlerunning{\src\ }

\abstract{We report on \xmm\ observations of the black hole X-ray binary \src\ during its 2012--2013 outburst. The first five observations monitor the source as its luminosity increases across the high--soft state of accretion. In the sixth observation the source has made a transition to an ``anomalous'' state, characterised by a significant contribution of electron scattering. A thermally/radiatively driven disc wind is present in the first four observations, which becomes more photoionised as the luminosity increases with time. In the fifth observation, the wind is not observed any more as a consequence of strong photoionisation and the low sensitivity of this observation. This overall trend is then consistent with a fully ionised wind causing the electron scattering characteristic of the anomalous state in the sixth observation. 
A broad iron emission line co-exists with the absorption features from the wind in the first four observations but is not visible in the last two observations. We find that the changes in the state of the wind as measured from modelling the absorption features with a self-consistent warm absorber model are correlated to the changes in the broad iron line. When the latter is modeled with a reflection component we find that the reflection fraction decreases as the illumination increases. We propose that the changes in both the absorption and broad emission lines are caused by the increasing luminosity and temperature of the accretion disc along the soft state. Such changes ultimately enable the transition to a state where the wind is fully ionised and consequently Comptonisation plays a significant role.
} \maketitle

\section{Introduction}
\label{sect:intro}

In the last decade we have witnessed a wealth of discoveries of hot atmospheres and winds in low-mass X-ray binaries (LMXBs)  \citep[e.g.][]{1254:boirin03aa, 1916:boirin04aa, 1323:boirin05aa, ionabs:diaz06aa, igrj17091:king12apj, 1915:kotani00apj, 1630:kubota07pasj, 1915:lee02apj, 1655:miller06nat, 1743:miller06apj,  igrj17480:miller11apj, 1655:ueda98apj, gx13:ueda01apjl, gx13:ueda04apj, cirx1:schulz08apj, 1658:sidoli01aa, gx13:sidoli02aa, 1624:parmar02aa, 1305:shidatsu13apj}. The presence of highly-ionised plasma in a cylindrical geometry around the compact object has been observed in all the high-inclination neutron star (NS) and black hole (BH) LMXBs \citep{ionabs:diaz06aa,ponti12mnras} i.e. when looking close to the disc. The plasma has been detected as an outflowing wind in 85\% of these BH systems and 30\% of the NS LMXBs, and as a static atmosphere in the remaining cases \citep{review:diaz12}.

Thermal, radiation pressure or magnetic mechanisms can produce a wind but the dominant driving mechanism may vary considerably between different sources \citep[e.g.][]{proga02apj}. A different mechanism for launching the winds in NSs and BHs would mean that the amount of mass expelled could be very different in those systems. This has important consequences for the impact of winds on the LMXBs themselves (e.g. in the disc structure) and on their environment.  
For NS LMXBs, the fact that outflowing winds are only detected for high luminosity, $\approxgt$~0.15 L/L$_{Edd}$ (where L$_{Edd}$ is the Eddington luminosity), and/or long orbital period sources,
indicates that a thermal/radiative mechanism is a viable explanation for the presence of the wind \citep{review:diaz12}. In BH LMXBs, highly ionised winds were detected in the high-soft states (HSS) of \src\ \citep{1630:kubota07pasj}, \gro\ \citep{1655:miller06nat,1655:diaz07aa}, \seventeen\ \citep{1743:miller06apj} and \grs\ \citep{1915:ueda09apj}. In all cases but one, the winds were launched at relatively large radii, $\approxgt$\,10$^{10}$--10$^{12}$\,cm, and therefore consistent with being driven by a combination of thermal and radiative pressure. In one case, a $\chandra$ observation of \gro\ in 2005, a smaller radius of $\approxlt$\,2$\times$10$^{9}$\,cm was derived and consequently a magnetic mechanism invoked \citep{1655:miller06nat}, although this claim has been controversial \citep{1655:netzer06apj}.

Interestingly, sources with disc winds in the HSS  \citep[e.g. \gro,][]{1655:miller06nat,1655:diaz07aa} do not show such winds in low-hard states (LHS) of accretion \citep{1655:takahashi08pasj}. The presence of winds in the HSS and its absence in the LHS has been recently confirmed with a systematic study of all BH LMXBs observed with $\chandra$, \xmm\ and $\suzaku$ \citep[][]{ponti12mnras}. Based on the observation that jets are detected precisely in the LHS, when winds are absent, \citet{1915:neilsen09nat} proposed that a possible explanation for the observed anti-correlation between jets and winds is that the wind observed during the soft state carries enough mass away from the disc to halt the flow of matter into the jet. However, there is one detection of a weak wind in a ``hard" (or ``C") state of \grs\ \citep{1915:lee02apj}, for which simultaneous radio emission is observed, indicating that winds and jets may not be strictly anti-correlated but rather linked to a given state of accretion. In the latter case, a possibility for the absence of winds in the LHS is that the spectral hardness during that state fully ionises the wind, which rather than disappearing becomes simply ``transparent'', i.e. undetectable via line absorption. Alternatively, the wind could be thermodynamically unstable and therefore unobservable during the LHS  \citep{chakravorty13mnras}. 
While significant changes in the ionisation and column density of the wind with X-ray luminosity have already been observed during the HSS of
BH outbursts \citep[e.g.][]{1655:diaz07aa,1630:kubota07pasj, 1915:ueda09apj}, a systematic study of the evolution of the wind during the whole outburst taking into account the changes in the spectral energy distribution (SED) and the luminosity of the source has yet to be done to determine the reason for the absence of winds in the LHS. We note that \citet{1305:shidatsu13apj} have recently reported the detection of low-ionisation absorption both during the hard and soft states of the black hole candidate MAXI~J1305--704. However, they associate the absorption with clumpy, compact structures from the dips instead of a homogeneous wind (but see \citet{1630:miller13apj} for an alternative interpretation of the low-ionisation absorption as an infalling, failed, wind). 

\src\ is a black hole candidate with X-ray outbursts that repeat with an interval of 600-700~days \citep{1630:jones76apj,1630:parmar95apj}. The source was classified as a black hole based on the similarity of its spectral and timing properties to those of black hole transients with known black hole masses and the lack of type I X-ray bursts \citep{1630:parmar86apj, 1630:kuulkers97apj,remillard06araa}.  The detection of X-ray dips \citep{1630:kuulkers98apj, 1630:tomsick98apj} in its light curves indicates an inclination of $\sim$\,60--75\deg. Based on infrared observations during its 1998 outburst, \citet{1630:augusteijn01aa} proposed a variable source at K\,=\,16.1 mag as the counterpart to the X-ray source and concluded that the system contained most likely a relatively early type secondary in a long orbital period around the black hole, defining the system as a LMXB. 

Radio emission from \src\ was first detected during its 1998 outburst \citep{1630:hjellming99apj}, when the source was making a state transition from a hard to a soft state. The emission was optically thin, with a spectral index of $\alpha \sim$ -0.8 (where $S_{\nu} \propto \nu^{\alpha}$), and was attributed to a radio jet ejected over a period of time. 

In 2006, \src\ was observed by $\suzaku$ during the decline of its outburst in a HSS of accretion \citep{1630:kubota07pasj}. These observations
revealed for the first time the presence of a highly ionised disc wind in this source, with an outflow velocity of $\sim$1000~km s$^{-1}$.
The ionisation of the wind decreased together with the flux as the source proceeded through the outburst. Column densities 
of $\sim$10$^{23}$~cm$^{-2}$ were measured and the location of the wind was estimated to be $\sim$10$^{10}$~cm. Based on these parameters, \citet{1630:kubota07pasj} estimated that the wind was in the range of composite thermal/radiation pressure-driven winds if L/L$_{Edd} >$\,0.1. However, they noted that magnetic processes such as those invoked by \citet{1655:miller06nat} to explain the wind of \gro, could be also in play for \src.

We applied for a series of \xmm, VLA and ATCA triggered simultaneous observations of a black hole LMXB at high inclination to investigate further the dominant mechanism in driving disc winds in LMXBs, the variability of the disc wind as a function of accretion state, the extent to which the amount of mass lost in a wind relates to the accreting power, and the possible connection between winds and radio jets. In this paper and \citet{1630:diaz13nat} we report on this monitoring program, which we triggered on \src. We focus here on the evolution of the wind and propose a scenario to explain its disappearance as the system transited from the HSS to an ``anomalous'', very high, state. In the last two observations of this program the wind was not visible anymore and instead radio emission and Doppler-shifted narrow emission lines appeared in the last observation, which we associate to the appearance of a jet. Based on the identification of two of the three detected lines as red- and blue-shifted \fetsix\, we constrained the inclination angle of the jet axis relative to the line of sight to be 65\deg, in agreement with the inclination determined from the existence of dips (assuming the jet is perpendicular to the disc). Therefore, we use an inclination of 65\deg\ hereafter and refer the reader to \citet{1630:diaz13nat} for a detailed discussion on those observations. Since the distance to \src\ is unknown, we follow previous authors \citep[e.g.][]{1630:tomsick05apj,1630:abe05pasj} and use a distance of 10~kpc throughout this paper.

\section{XMM-Newton observations}
\label{sec:observations}

The XMM-Newton Observatory \citep{xmm:jansen01aa} includes three
1500~cm$^2$ X-ray telescopes each with a European Photon Imaging 
Camera (EPIC, 0.1--15~keV) at the focus. Two of the EPIC imaging
spectrometers use Metal Oxide Semiconductor (MOS) CCDs \citep{xmm:turner01aa} and one uses pn CCDs
\citep{xmm:struder01aa}. The Reflection Grating Spectrometers \citep[RGS, 0.35--2.5~keV,][]{xmm:denherder01aa} 
are located behind two of the
telescopes. In addition, there is a co-aligned 30~cm diameter 
Optical/UV Monitor telescope \citep[OM,][]{xmm:mason01aa}, 
providing simultaneous coverage with the X-ray instruments.
Data products were reduced using the Science Analysis
Software (SAS) version 13.5.0 and calibration files (CCF) available in May 2014. 
The EPIC MOS cameras were not used during the
observation in order to allocate their telemetry to the EPIC pn camera
and avoid ``full scientific buffer'' in the latter. We present here the analysis 
of EPIC pn data, and RGS data from both gratings. 
 
\begin{table*}
\begin{center}
\caption[]{XMM-Newton observations of \src. $T$ is the
total EPIC pn exposure time and $C$ is the EPIC-pn 2--10~keV average
count rate in the 24(16) inner columns for timing(burst) exposures, respectively. In all cases the EPIC 
pn thin filter was used. }
\begin{tabular}{cclllcc}
\hline \noalign {\smallskip}
Obs  & Observation & Mode &  \mc{2}{c}{Observation Times (UTC)} & $T$  & $C$   \\
Num & ID   & & Start  & End & (ks) & (s$^{-1}$) \\
        &        &  & (year~month~day hr:mn) & (year~month~day hr:mn) & \\
\hline \noalign {\smallskip}
1 & 0670671501 & Timing & 2012 March 04 11:07 & 2012 March 05 09:25 & 78 & 528 \\
2 & 0670671301 & Timing & 2012 March 20 19:37 & 2012 March 21 02:31 & 22 & 572 \\
3 & 0670672901 & Timing & 2012 March 25 03:37 & 2012 March 25 22:12 & 65 & 582 \\
4T & 0670673001 & Timing & 2012 September 09 20:22 & 2012 September 10 07:49 & 38 & 732\\
4B & 0670673001 & Burst & 2012 September 10 08:28 & 2012 September 10 15:59 & 27 & 763\\
5 & 0670673101 & Burst & 2012 September 11 20:14 & 2012 September 12 05:39 & 31 & 979 \\
6 & 0670673201 & Burst & 2012 September 28 06:33 & 2012 September 28 21:50 & 52 & 1237 \\
\noalign {\smallskip} \hline \label{tab:obslog}
\end{tabular}
\end{center}
\end{table*}

Table~\ref{tab:obslog} is a summary of the XMM-Newton observations. We
used the EPIC pn in timing and burst mode. 
In timing mode only one CCD chip is
operated and the data are collapsed into a one-dimensional row
(4\farcm4) and read out at high speed, the second dimension being
replaced by timing information. This allows a time resolution of
30~$\mu$s and photon pile-up occurs only for count rates $\approxgt$\,800 ~s$^{-1}$. 
The burst mode is a special flavour of the timing mode, which offers a very high time resolution 
of 7~$\mu$s, but has a duty cycle of only 3\%. Photon pile-up occurs only for count
rates above 60000~s$^{-1}$ in this mode. We chose to change the pn mode from timing to burst during observation (obs)~4
due to severe telemetry problems during that observation. Hereafter, we call obs~4T and 4B the two
consecutive parts of obs~4 taken in timing and burst mode and use them to evaluate the cross-calibration differences between both modes (see 
Sect.~\ref{model1}). 

A Rate Dependent Charge Transfer Inefficiency (RDCTI) effect has been observed in EPIC 
pn timing and burst modes when high count rates are present\footnote[1]{More 
information about the CTI correction can be 
found in the {\it Spectral calibration accuracy in EPIC-pn fast modes} (Guainazzi et al. 2013)
and in the Current Calibration File (CCF) release notes {\it Rate-dependent CTI correction for EPIC-pn timing modes} 
(Guainazzi et al. 2009) and {\it Post-XRL Rate-Dependent CTI correction for EPIC-pn Timing
Mode} (Guainazzi 2013) at http:$\slash\slash$xmm2.esac.esa.int$\slash$external$\slash$xmm$\_$sw$\_$cal$\slash$calib}. 
As of SAS v13.5.0, the correction for this RDCTI effect is applied through the parameter ``runepfast'' in {\tt epproc/epchain} (set to ``yes'' by default) and does
not have to be applied independently (via the task {\tt epfast}) as in previous SAS versions. We used this analysis method for the burst mode observations.
For the timing mode observations, we chose to use instead the new, energy-dependent, ``RDPHA" method for correcting the energy scale\footnote[2]{See the CCF release notes {\it Calibration of the Rate-Dependent PHA (RDPHA) correction
for EPIC-pn Timing Mode} (Guainazzi 2013) and {\it RDPHA calibration in the Fe line regime for EPIC-pn Timing Mode} (Guainazzi 2014).} (parameter ``withrdpha'' in {\tt epevents} set to ``yes''\footnote[3]{The 
parameter ``runepreject'' must also be set to ``yes'' for the ``RDPHA" correction to apply.}), which has already proven to give better results
than ``RDCTI'' and is expected to replace the latter in future SAS versions (but is not yet available for burst mode observations). 
Ancillary response files were generated using the SAS task
{\tt arfgen} with the PSF model set to {\tt extended} and following the recommendations of the {\it XMM-Newton SAS
User guide} for piled-up observations in timing mode, whenever applicable. 
Response matrices were generated using the SAS task {\tt rmfgen}.

Light curves were generated with the SAS task
{\tt epiclccorr}, which corrects for a number of effects like
vignetting, bad pixels, encircled energy fraction in the extraction region and
accounts for time dependent corrections within an exposure, like dead
time and time lost due to solar flares.

The SAS task {\tt rgsproc} 
was used to produce calibrated RGS event lists, spectra, and response
matrices.
We also chose 
the option {\tt keepcool=no} to discard single
columns that give signals a few percent below the values expected from
their immediate neighbours. Such columns are likely to be important
when studying weak absorption features in spectra with high
statistics. We used the SAS task {\tt rgsbkgmodel} to compute model background spectra from RGS background templates. We generated RGS light curves with the SAS task 
{\tt rgslccorr}. 

\subsection{Pile-up and X-ray loading in the EPIC pn camera}

For observations performed in timing mode, the count rate in the 
EPIC pn was close to, or above, the
800~\countsec\ level, at which X-ray loading and pile-up effects
become significant. 

Pile-up occurs when more than one photon is read in a pixel during a
read-out cycle. This causes photon loss, pattern
migration from lower to higher pattern types and hardening of the spectrum,
because the charge deposited by more than one photon is added up before being read 
out\footnote[4]{See {\it XMM-Newton Users Handbook}
for more information on pile-up.}. 
In addition, when high
count rates are present, the offset map calculated at the beginning of
an exposure may be contaminated by X-ray events from the source, the
so-called ``X-ray loading''. As a consequence pattern migration from
higher to lower pattern types and a shift to lower energy for all the
events associated with the contaminated pixel occur\footnote[5]{More 
information about X-ray loading can be 
found in the CCF release note {\it PN X-ray loading} (Smith 2013)}. 
Both X-ray loading and pile-up cause significant spectral distortion. However, starting with SAS v13.5.0 the events
files are corrected for X-ray loading effects 
(but note that the former effect can be only present in obs~1--3 since, as of the public release of SAS v12, it has been prevented by taking offset maps prior to exposures in EPIC pn timing mode with the CLOSED optical blocking filter). 
Therefore, 
we investigated in detail only the presence of pile-up effects before extracting the spectra. 
As a diagnostic tool in the pn camera timing mode
data, we used the SAS task {\tt
epatplot}, which utilizes the relative ratios of single- and
double-pixel events, which deviate from standard values in case of
significant pile-up. We found that all the timing spectra were affected by pile-up. Next, we
extracted several spectra selecting single and double timing mode
events (patterns 0 to 4) but different spatial regions for the
source. Source events were first extracted from a 58\arcsec\ (14
columns) wide box centred on the source position (Region~1). Then we
excluded 1, 3, 5, 7 and 9 columns from the centre of Region~1
(Regions~2--6) and extracted one spectrum for each of the defined
regions. While the {\tt epatplot} task indicated that the spectra were free of pile-up once the
inner 5 columns were excluded, the values of the neutral column density showed still significant 
changes between Regions~4, 5 and 6. However, since the relative changes between observations
were similar whether Region~4, 5 or 6 was chosen and since the absolute changes of all the other 
parameters of the fit were consistent within the errors for those three regions, we followed the indications
of {\tt epatplot} and use the spectra of Region~4 hereafter. However, we caution that the absolute values
of the neutral column density could be affected by small errors if the spectra still contain residual pile-up 
effects.

\subsection{Background subtraction}
\label{subsec:bkg}
In the EPIC pn timing and burst modes, there are no source-free background
regions, since the PSF of the telescope extends further than the
central CCD boundaries. The central CCD has a field of view of 
13\farcm6 $\times$ 4\farcm4 in the pn. In timing and burst modes, the largest
column is the one in which the data are collapsed into a one-dimensional
row. Therefore, the maximum angle for background extraction is 2\arcmin, 
compared to 5\arcmin\ for imaging modes. Since \src\ is very bright, its
spectrum will not be significantly modified by the
``real'' background which contributes less than 1\% to the total count
rate in most of the bandwidth. Conversely, subtracting the background
extracted from the outer columns of the central CCD will modify the
source spectrum, since the PSF is energy--dependent and the source
photons scattered to the outer columns do not show the same energy
dependence as the photons focused on the inner columns.
Therefore, we chose not to 
subtract the ``background'' extracted from 
the outer regions of the central CCD \citep[see also][]{gx339:done10mnras,ng10aa}. We expect a contribution from the background to the total count rate of 
more than 1\% below 2~keV. Therefore, below this energy  we used only the RGS spectra, for which background templates were available.

As a consistency check we fitted one timing and one burst mode spectra (from obs~4T and 4B, respectively) with the most simple model (see Model~1 in Sect.~\ref{model1} below) and compared the parameters obtained when  subtracting the background extracted from the ``filter wheel closed'' event files\footnote[6]{Available at {\tt http://xmm2.esac.esa.int/external/} {\tt xmm\_sw\_cal/\tt background/filter\_closed/pn/index.shtml}} or the outer columns of the CCD with those obtained without background subtraction. The parameters of the fit were all consistent within the errors for fits performed not subtracting any background or subtracting the background from the ``filter wheel closed'' event files. In contrast, when using the background extracted from the outer columns of the CCD, the value of the column density of the neutral absorber increased from  7.72\,$\pm$\,0.06 $\times 10^{22}$ cm$^{-2}$ to 7.84\,$\pm$\,0.06 $\times 10^{22}$ cm$^{-2}$ for obs~4T and from 8.19\,$\pm$\,0.08 $\times 10^{22}$ cm$^{-2}$ to 8.43\,$\pm$\,0.10 $\times 10^{22}$ cm$^{-2}$ for obs~4B. For the latter observation, the disc temperature changed also marginally from 1.74\,$\pm$\,0.01 to 1.71\,$\pm$\,0.01 keV. All the other parameters were consistent within the errors. 

\section{Light curves}

\begin{figure*}[!ht]\centerline{\hspace{-1cm}\includegraphics[angle=90,width=18cm]{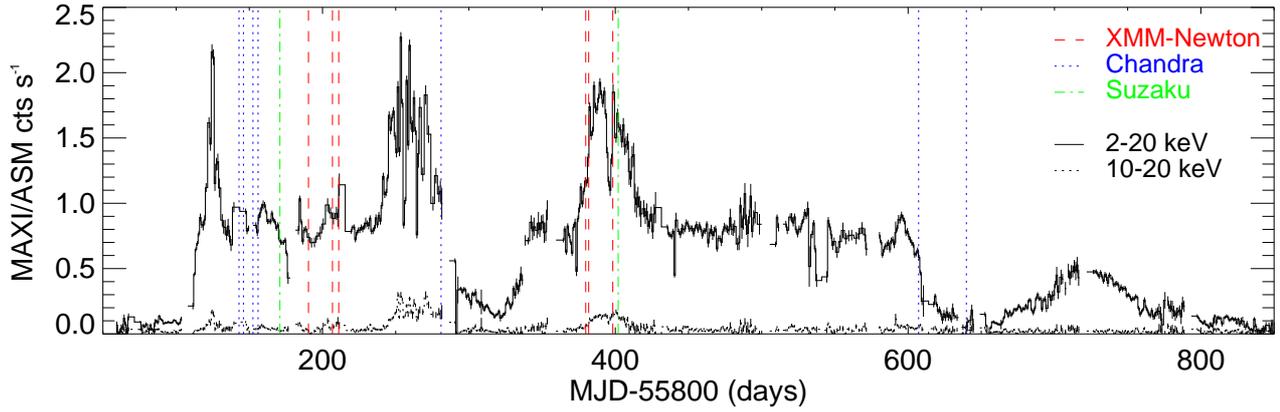}}
\caption{MAXI/ASM lightcurve of \src. The times of the
XMM-Newton observations reported here are indicated
with red dashed lines. The black solid and dotted lines show the 2--20 and 10--20 keV light
curves, respectively.} \label{fig:maxi}
\end{figure*}

Fig.~\ref{fig:maxi} shows the light curve of the 2011-2013 outburst covered by MAXI/ASM. The times of the six \xmm\ observations presented here are marked with red, dashed, lines. The first three observations were performed during the first of three intervals of relatively constant count rate. The second series of \xmm\ observations was performed during the last of the three high luminosity peaks of the outburst. Those peaks occurred at days 55925, 56060 and 56190 (the highest count rates in the 10--20~keV MAXI/ASM and 15--50~keV Swift/BAT energy range were found at days 55925, 56060 and 56202). Interestingly, the spacing between the peaks is similar, $\sim$\,135 days.  

\begin{figure*}[!ht]
\includegraphics[angle=0.0,width=0.25\textheight]{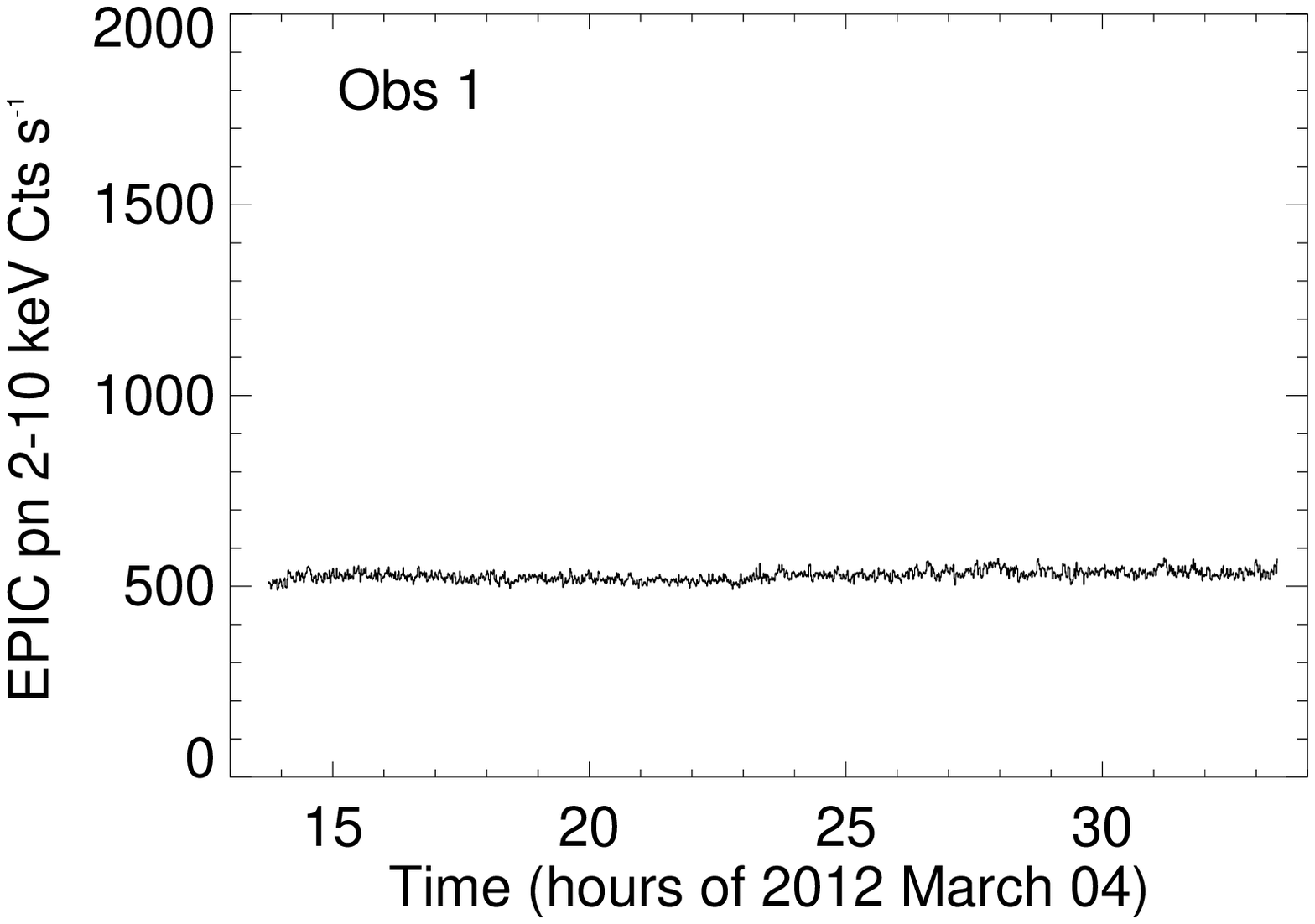}
\includegraphics[angle=0.0,width=0.25\textheight]{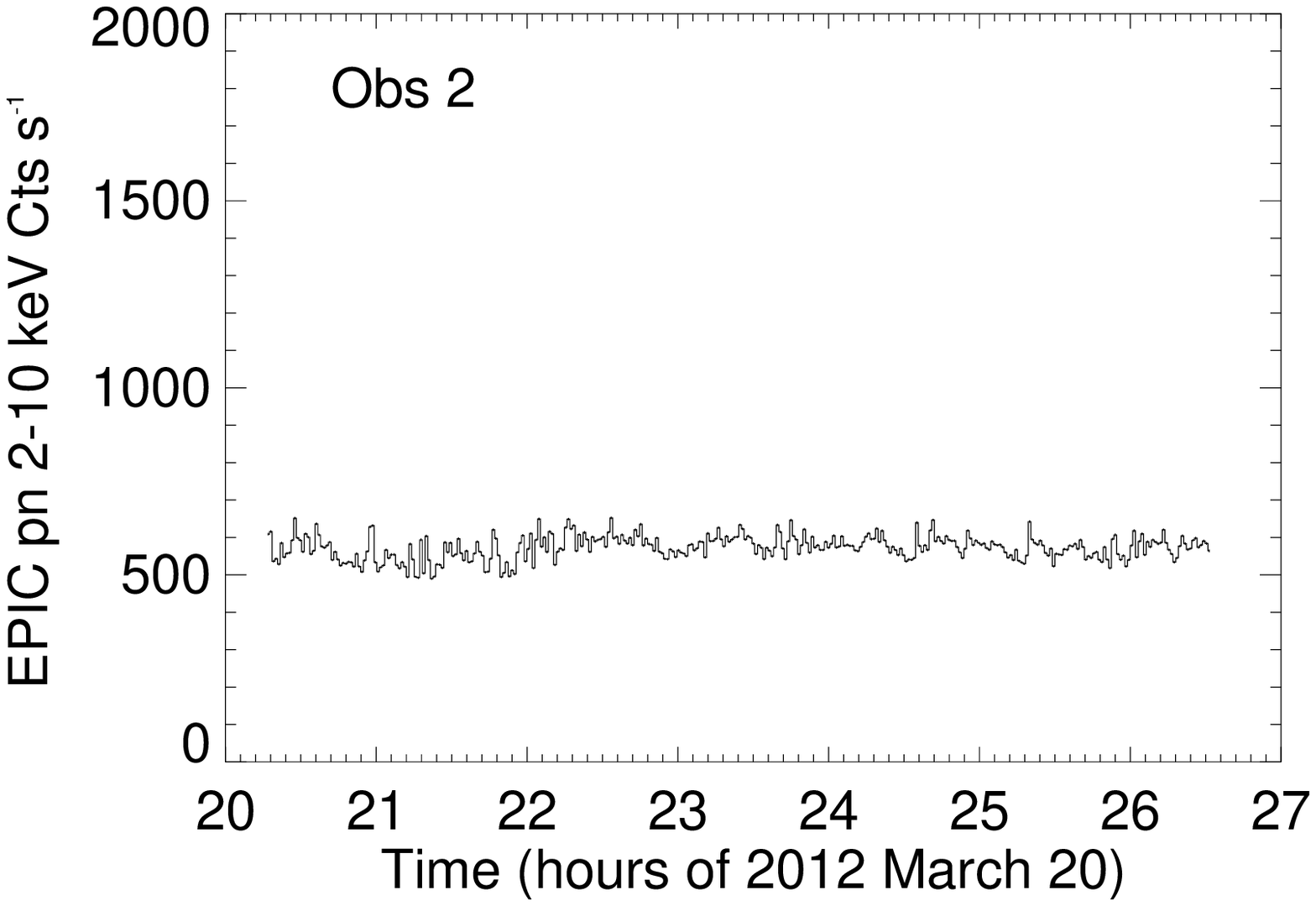}
\includegraphics[angle=0.0,width=0.25\textheight]{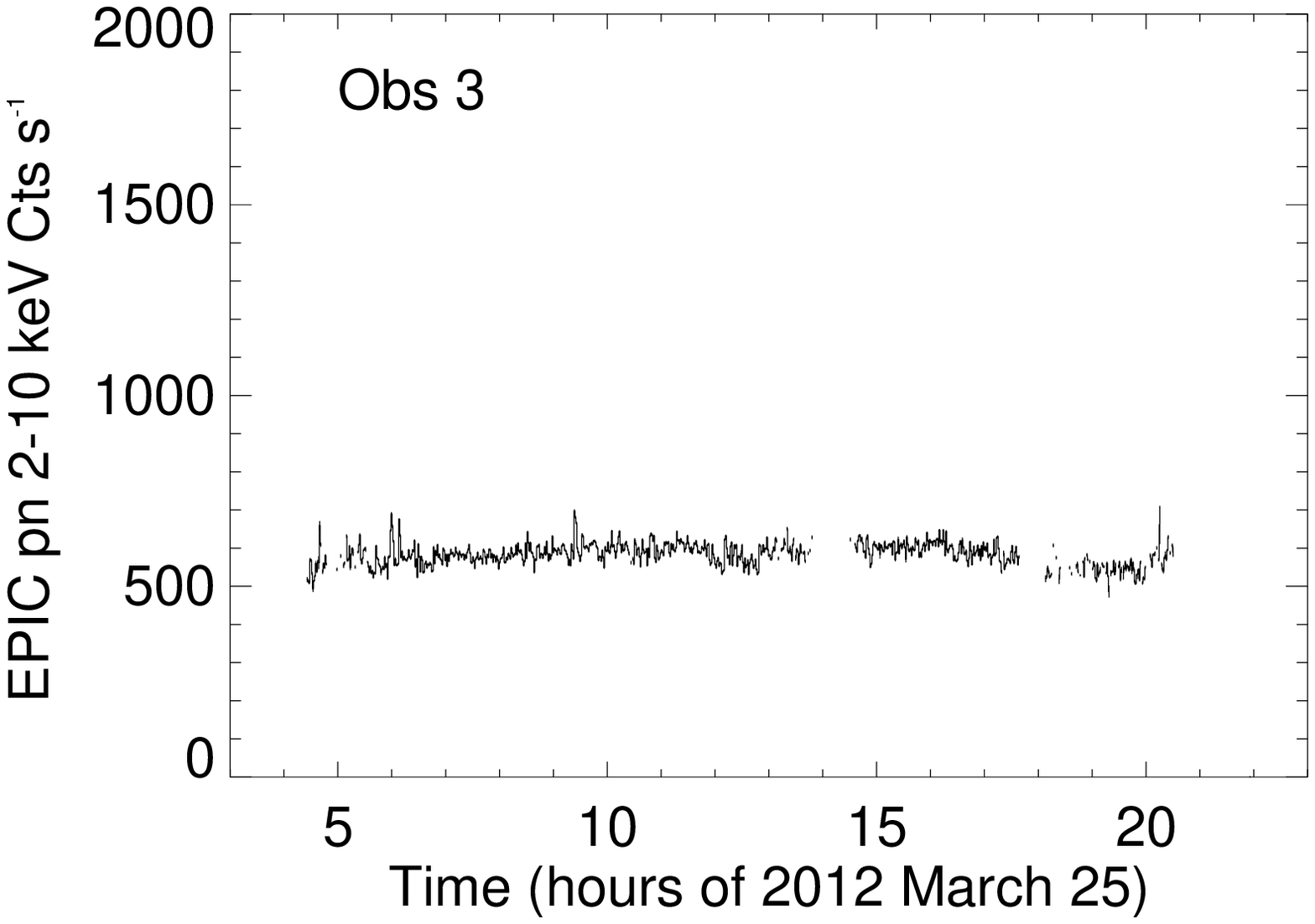}
\includegraphics[angle=0.0,width=0.25\textheight]{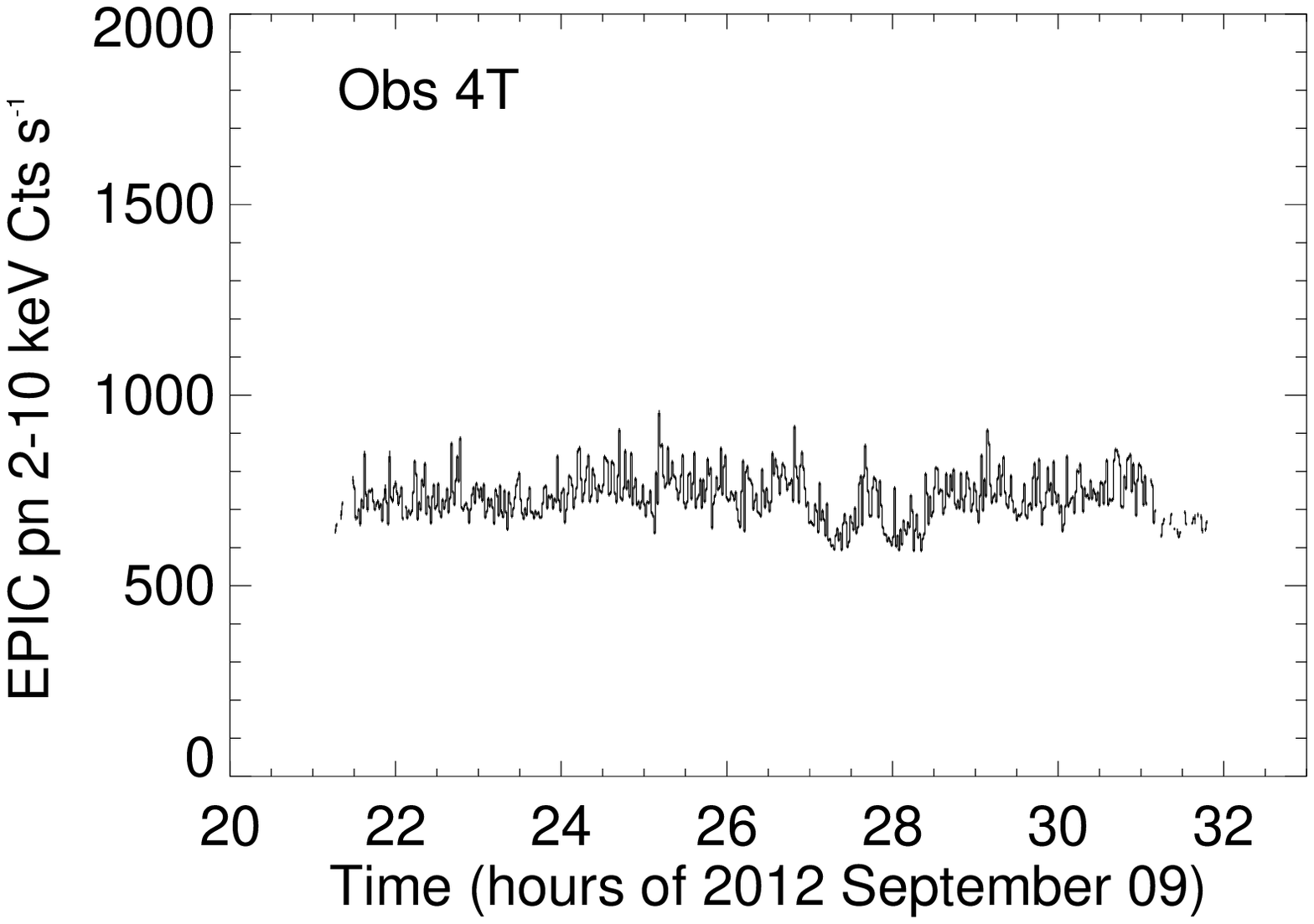}
\includegraphics[angle=0.0,width=0.25\textheight]{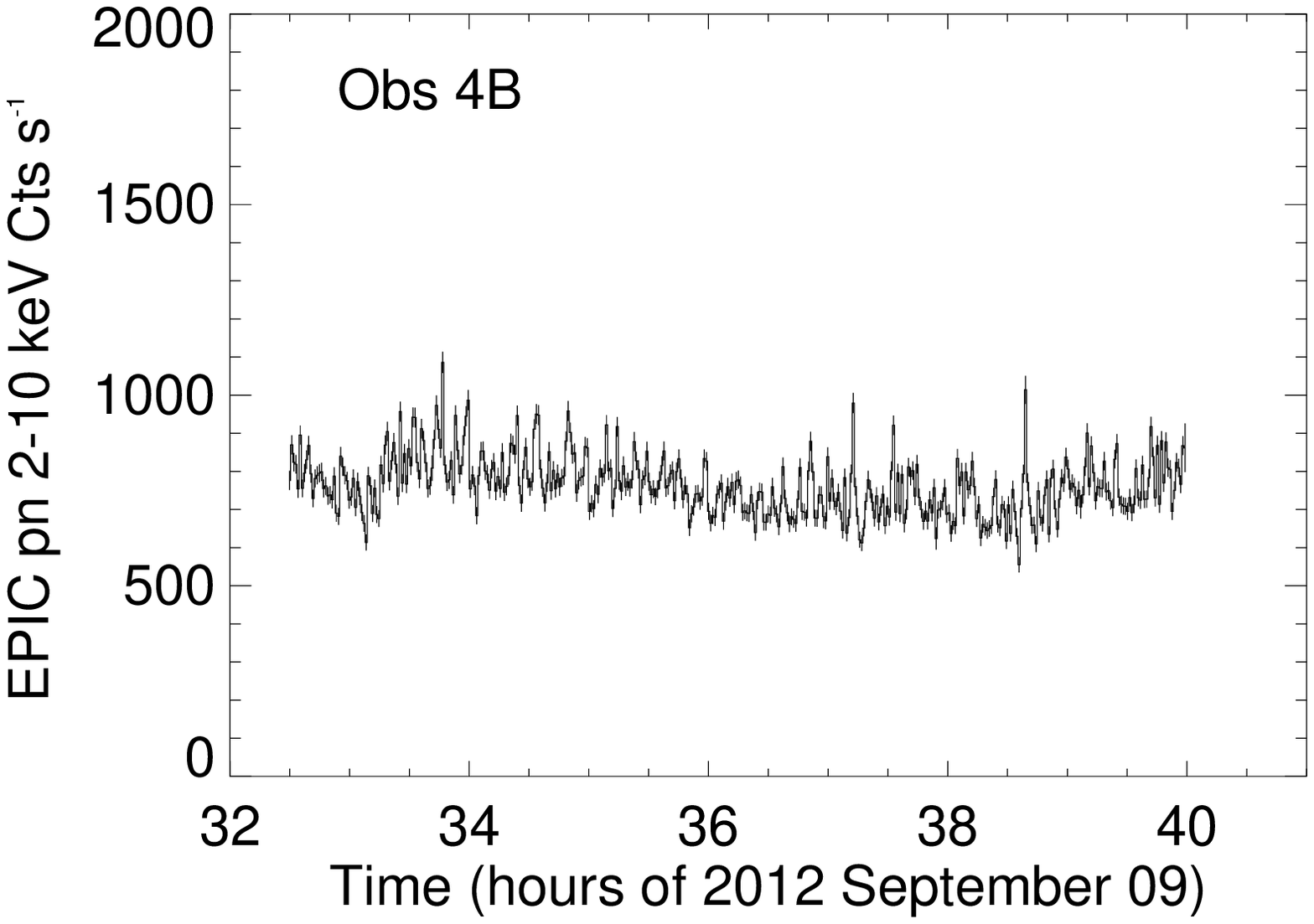}
\includegraphics[angle=0.0,width=0.25\textheight]{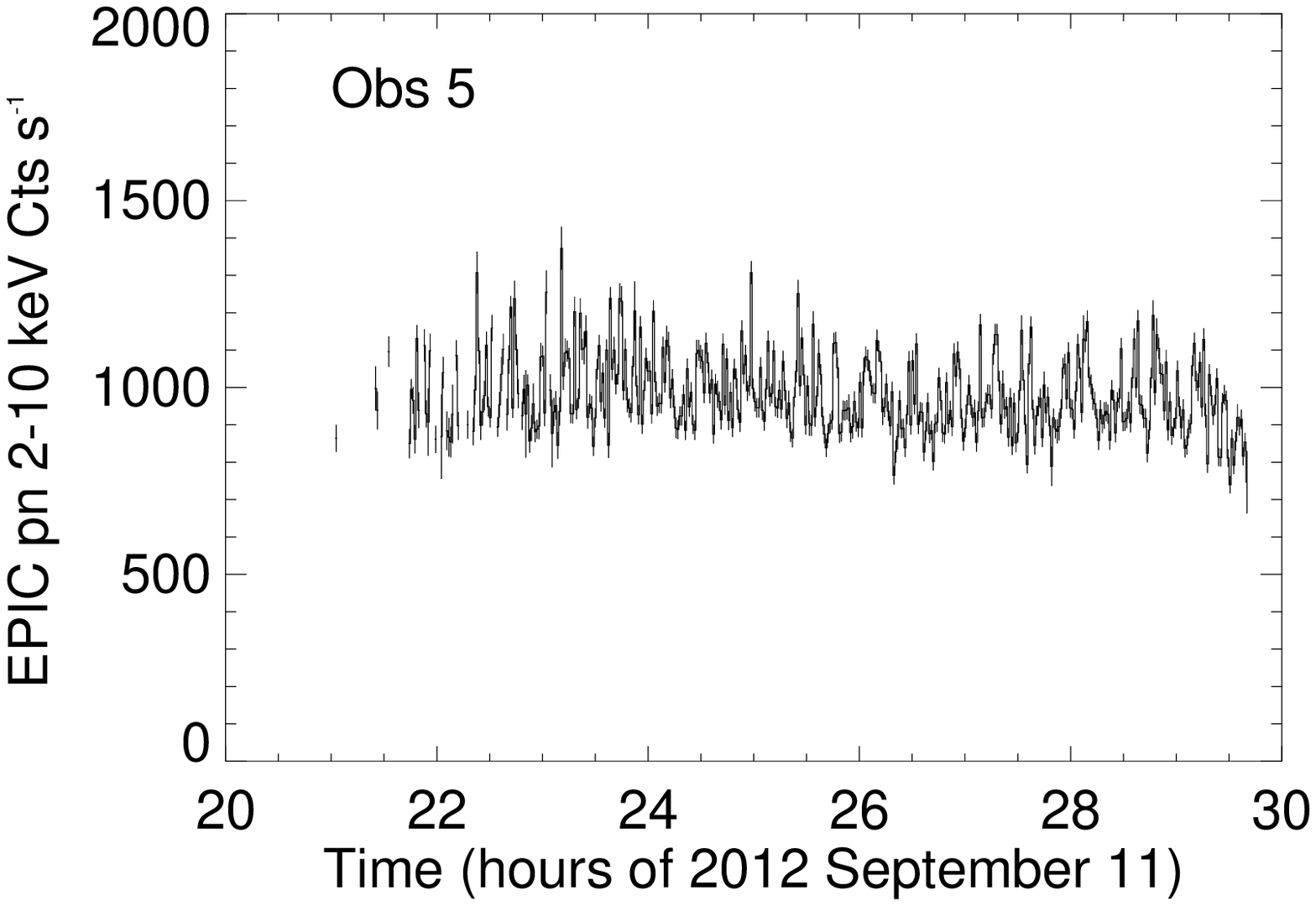}
\includegraphics[angle=0.0,width=0.25\textheight]{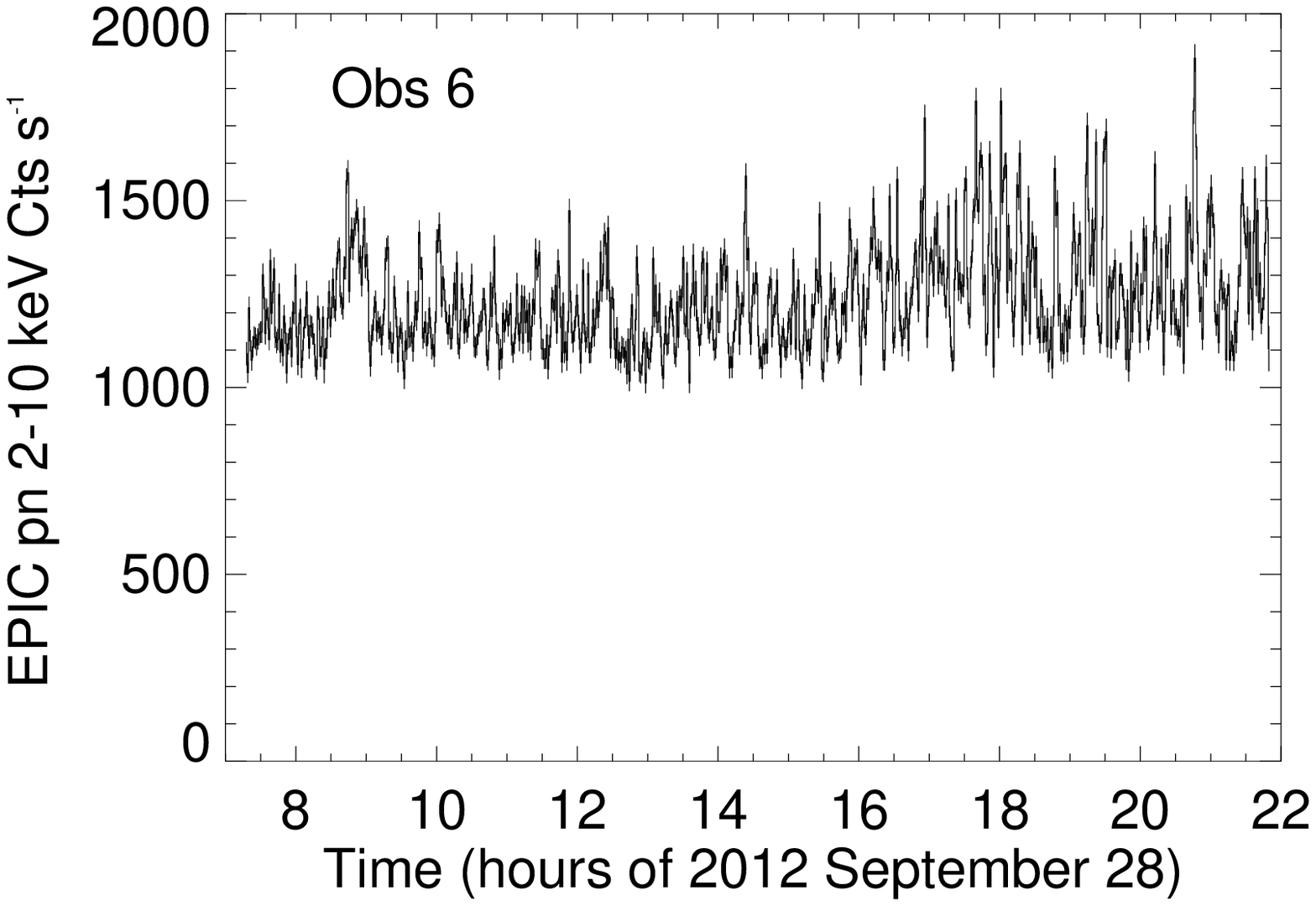}
\caption{2--10 keV EPIC pn light curves with a
binning of 64~s for obs 1--6. Time is shown in hours since the beginning of the
observation. The y-axis scale is the same for all observations. Note that obs~4 is split into two parts: 4T was stopped 
shortly before 08:00 UT on September 10 and 4B was started immediately
after (see text for more details).}
\label{fig:lightcurves}
\end{figure*}

Fig.~\ref{fig:lightcurves} shows the 2--10~keV EPIC pn light curves of the six \xmm\ observations of \src\ with a binning of 64~s.
The average count rate increases from one observation to the next. While the change is small for the first three observations, with count rate values between 450 and 550
counts s$^{-1}$, the last three observations show a significant increase, with values up to 1900 counts s$^{-1}$ in obs~6. In addition, the count rate variability increases significantly from obs~4 to 6.  

\begin{figure}[!ht]\centerline{\hspace{-1cm}\includegraphics[angle=0,width=0.5\textwidth]{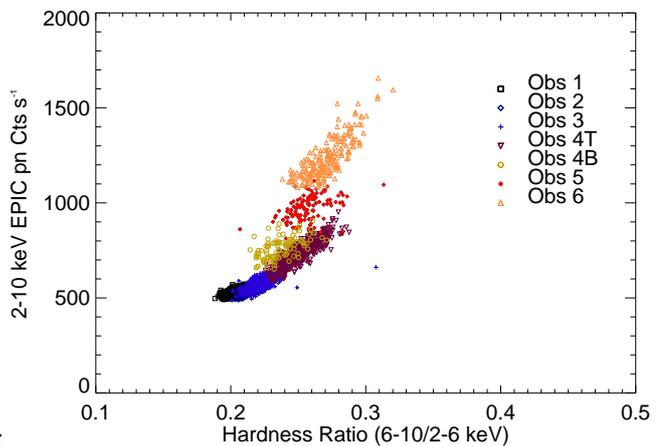}}
\caption{Hardness ratio (6--10~keV/2--6~keV counts) versus
2--10~keV EPIC pn count rate for all the observations. 
Each point corresponds to a binning of 64~s (256~s) for obs~1--4T (4B--6). 
} \label{fig:hr}
\end{figure}

Fig.~\ref{fig:hr} shows the hardness ratio (counts in the 6--10 keV
band divided by those between 2--6 keV) as a function of
the 2--10 keV count rate for all observations. In these observations, the count rate and the hardness ratio
are positively correlated.

\section{X-ray spectra}
\label{sec:spectra}

We extracted EPIC pn and RGS spectra for each observation. We rebinned the EPIC pn spectra to over-sample the full width
at half maximum (FWHM) of the energy resolution by a factor 3 and to have a minimum of 25 counts per bin, to allow 
the use of the \chisq\ statistic. To account for systematic uncertainties in the assumed effective area, we added 1.2 and 1.0\% uncertainty to each spectral bin after rebinning in the 2--3 and 3--10~keV spectral ranges (see Appendix~A for a description of the derivation of the systematic uncertainties). We rebinned the RGS spectra to oversample the $FWHM$ of the energy resolution by a factor 3 to be
sensitive to narrow features and we used the C-statistic \citep{cash79apj}. We performed spectral analysis using XSPEC \citep{arnaud96conf}, version 12.8.1. Since there were no photons in the RGS spectra below
$\sim$1.4~keV, we used the RGS spectra in the energy interval 1.4--2 keV. We used the pn spectra between 2 and 10 keV to exclude energy bins
for which we expect the spectrum to be affected by background (see Sect. 2.2) and
extended the energy range of the spectral fit to 0.01--200 keV when using convolution models. To account for absorption by neutral gas we
used the {\tt tbabs} XSPEC model with solar abundances \citep{anders89}. 
Spectral uncertainties are given at 90\% 
confidence ($\Delta$\chisq\,= 2.71 for one interesting parameter).

\subsection{EPIC-pn spectral analysis}
\label{subsec:pn}

\subsubsection{Phenomenological model}
\label{model1}

We first fitted the EPIC pn spectra with a model consisting of a disc blackbody 
 modified by neutral material for each observation separately. 
A disc blackbody described adequately the spectra of obs~1--5. In contrast, obs~6 showed some curvature in the residuals, pointing to the existence of a second component. 
Therefore, for this observation we used a combination of a disc blackbody and a power law for the continuum. All the observations showed an emission feature at $\sim$\,2.3~keV, 
most likely due to residual calibration uncertainties around the Au edge. We modeled those features with a Gaussian emission component and do not discuss them further. 
However, we note that these features, if due to residual uncertainties of the CTI correction, indicate that the energy gain could be compromised in the whole energy band.

\begin{figure*}[!ht]{
\includegraphics[angle=0,width=0.31\textwidth]{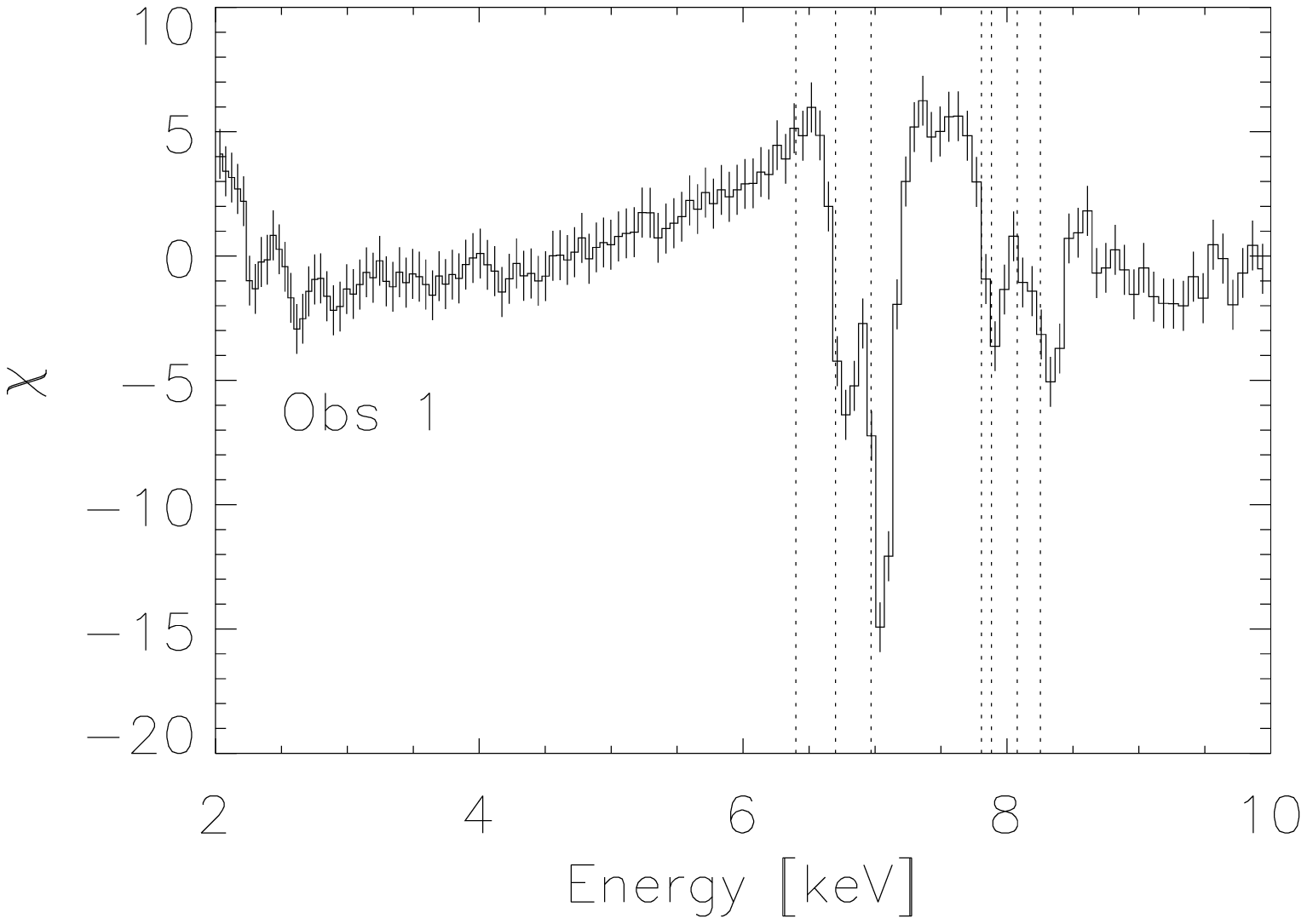}
\includegraphics[angle=0,width=0.31\textwidth]{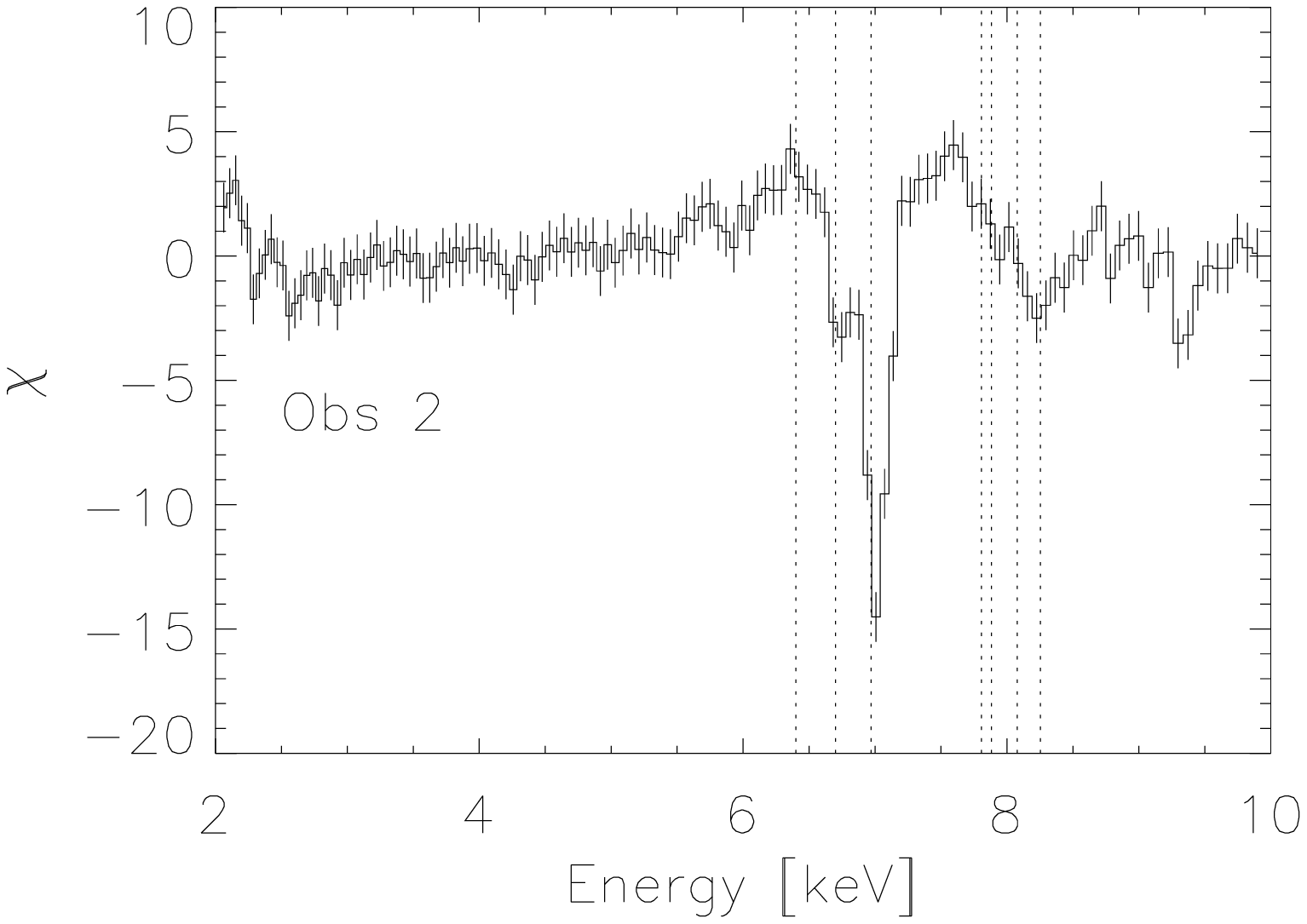}
\includegraphics[angle=0,width=0.31\textwidth]{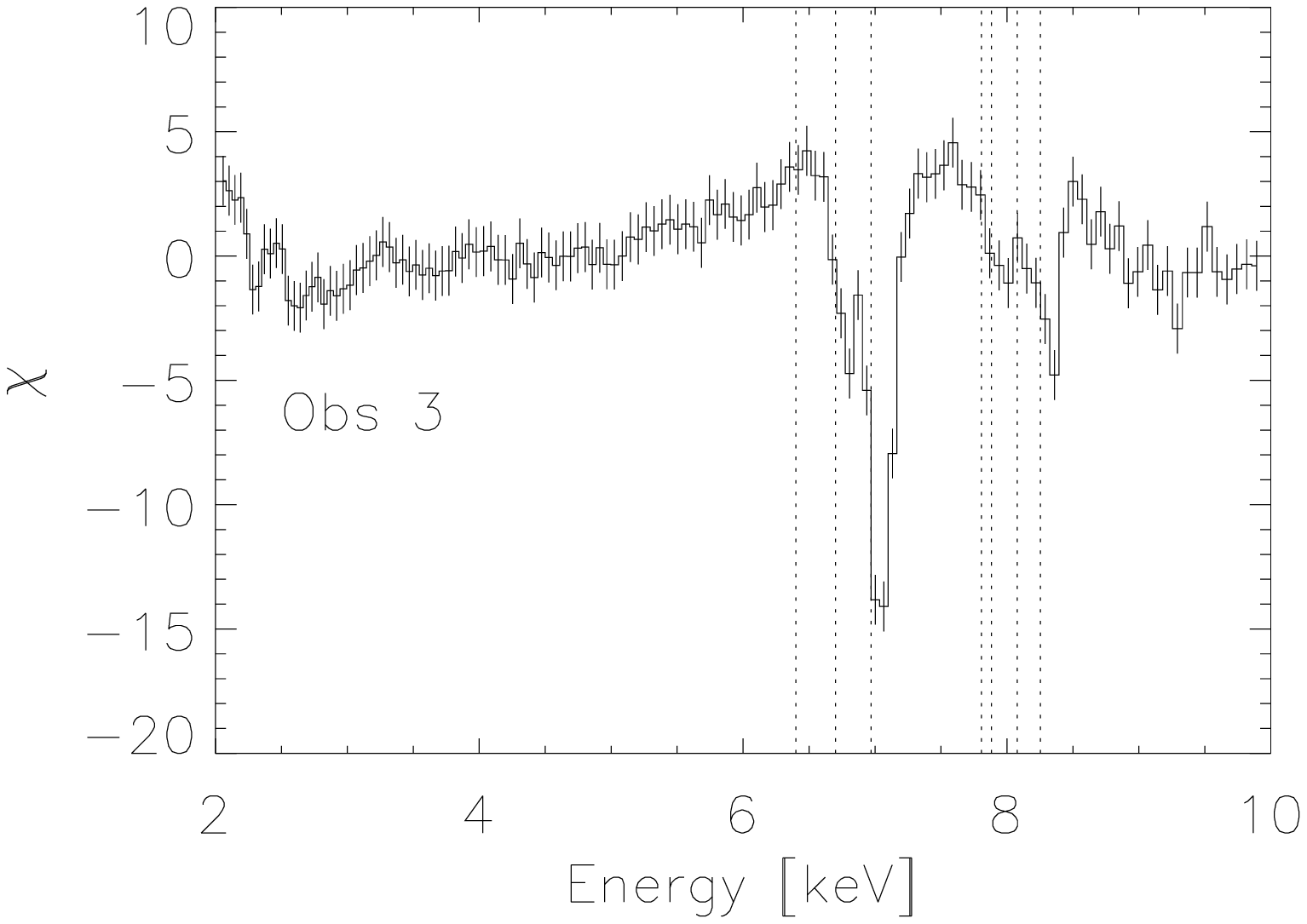}
\includegraphics[angle=0,width=0.31\textwidth]{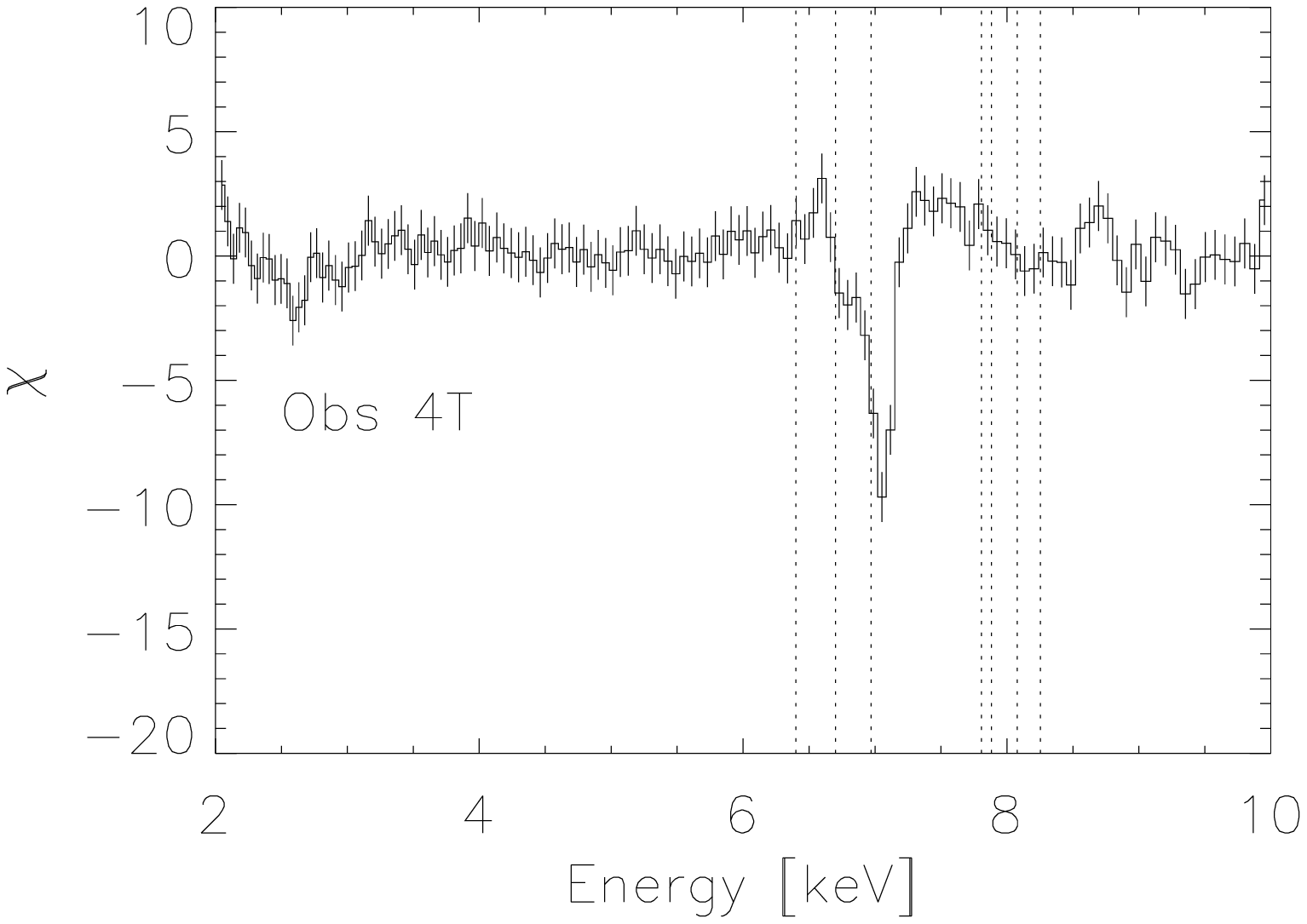}
\includegraphics[angle=0,width=0.31\textwidth]{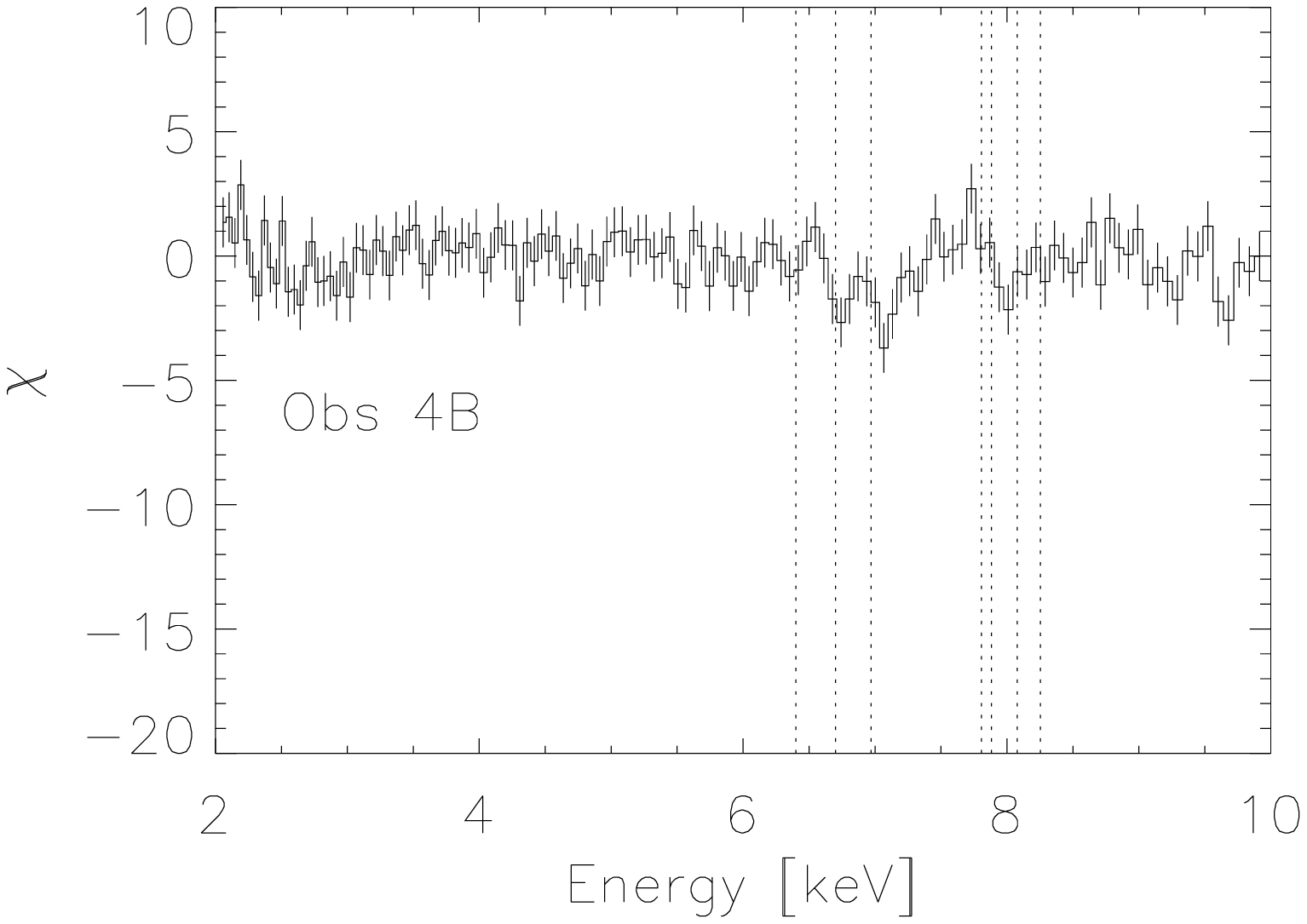}
\includegraphics[angle=0,width=0.31\textwidth]{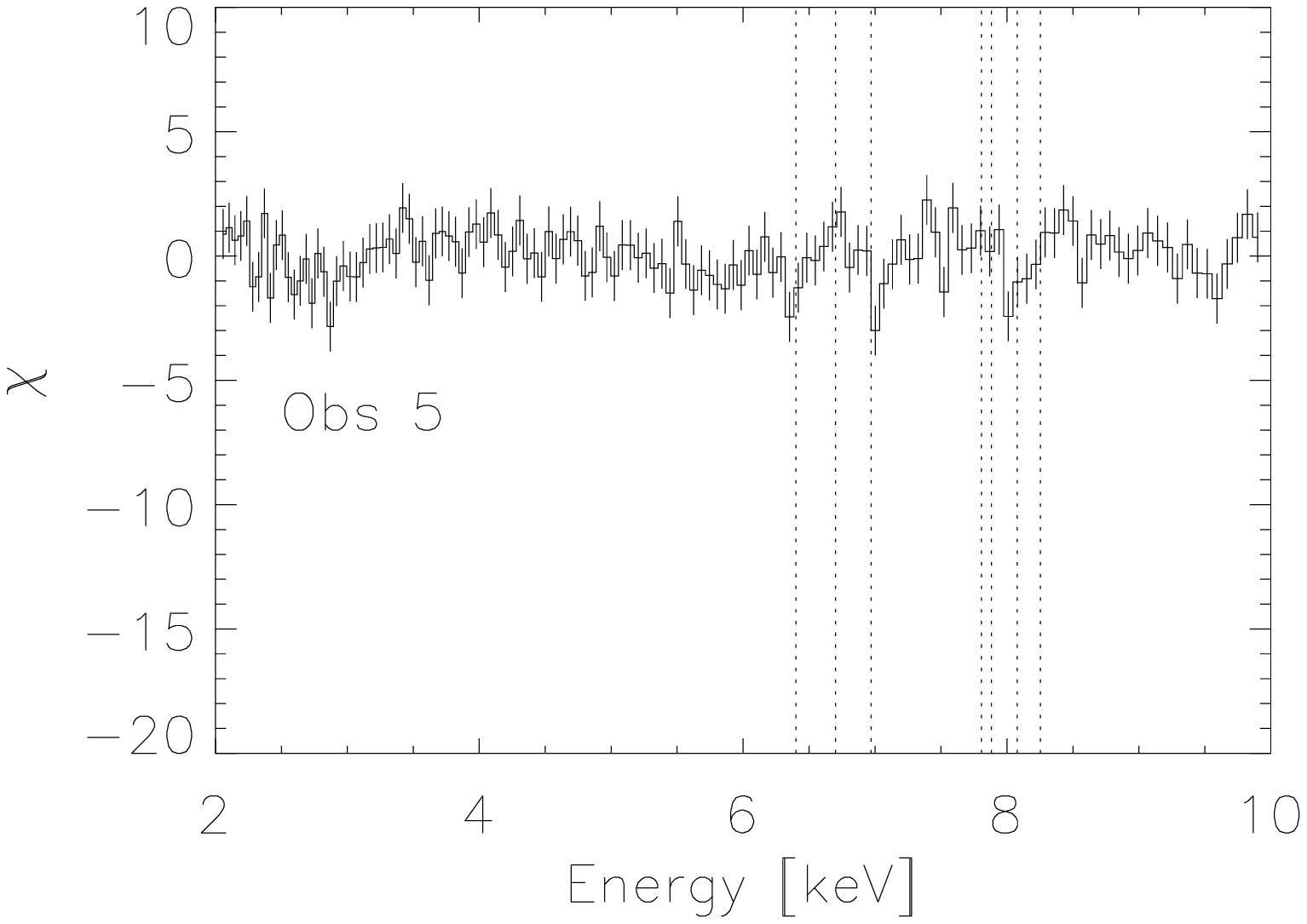}
\includegraphics[angle=0,width=0.31\textwidth]{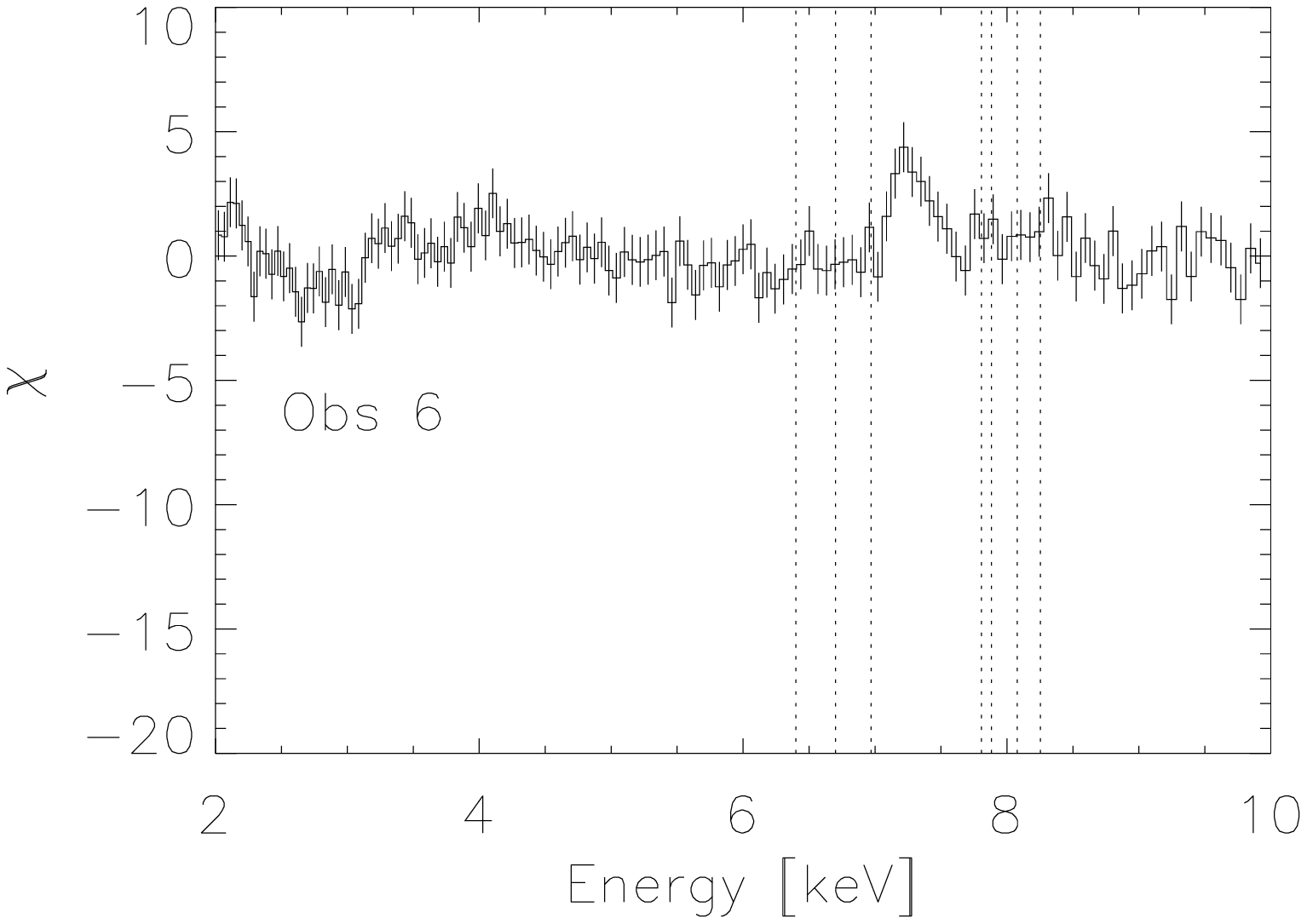}
}
\caption{Data residuals in units of standard deviation (\chisq) from the best-fit continuum model for obs~1--6 (see text). The dotted vertical lines indicate from left to right the rest energy of the transitions of neutral Fe, \fetfive\ \ka, \fetsix\ \ka, \nitseven\ \ka, \fetfive\ \kb, \niteight\ \ka\ and \fetsix\ \kb.} \label{fig:pndel}
\end{figure*}

\begin{figure*}[!ht]{
\includegraphics[angle=0,width=0.31\textwidth]{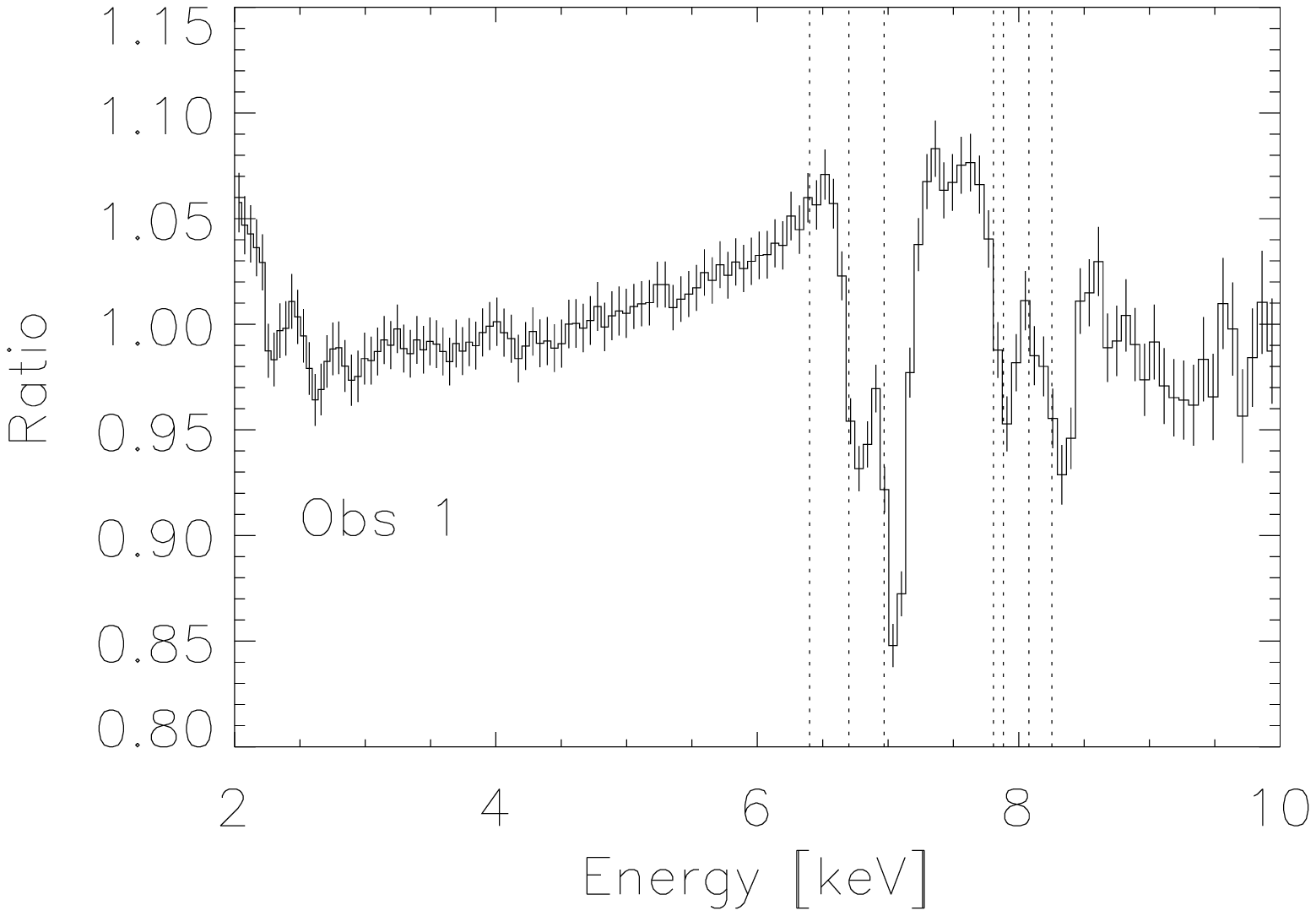}
\includegraphics[angle=0,width=0.31\textwidth]{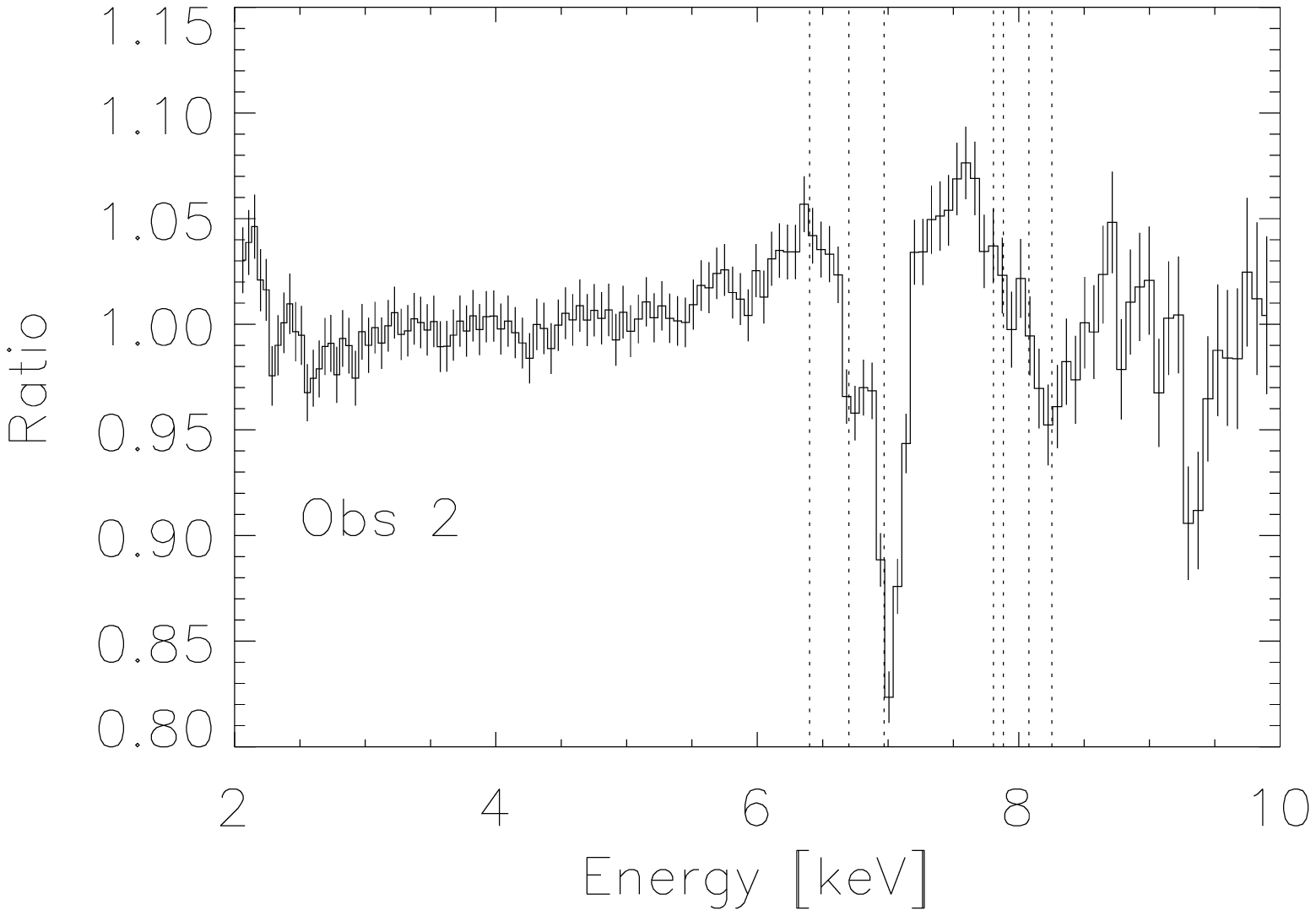}
\includegraphics[angle=0,width=0.31\textwidth]{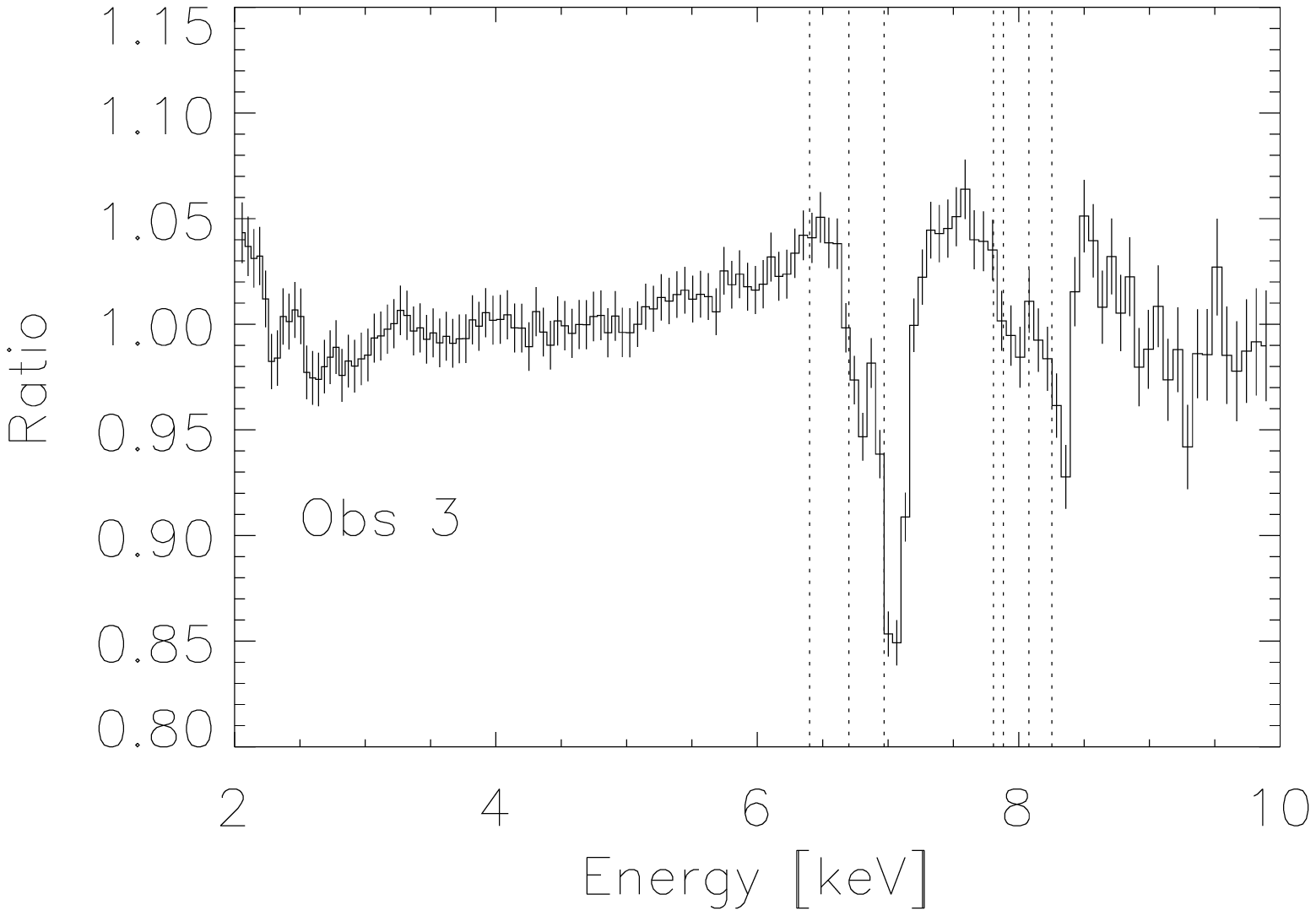}
\includegraphics[angle=0,width=0.31\textwidth]{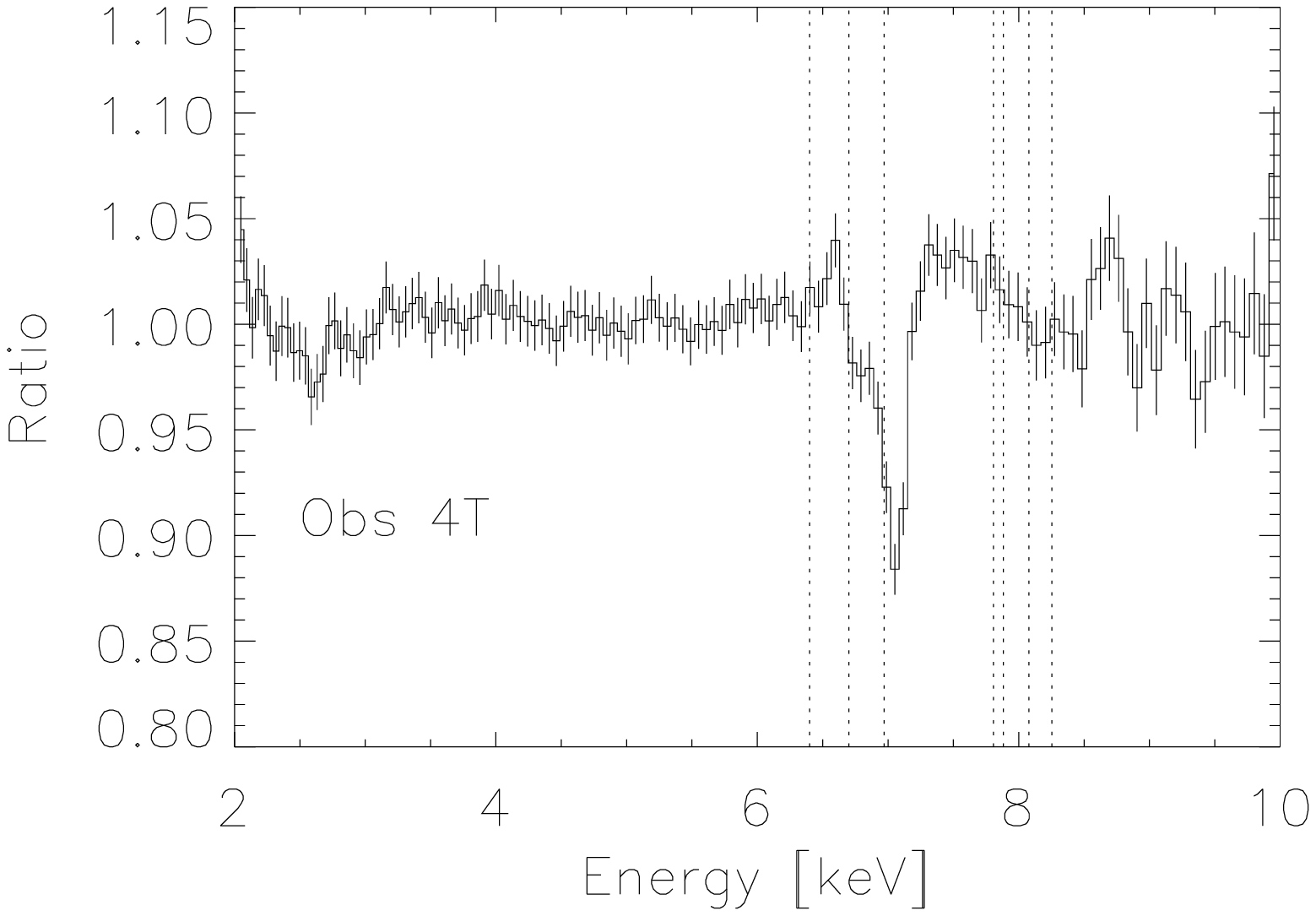}
\includegraphics[angle=0,width=0.31\textwidth]{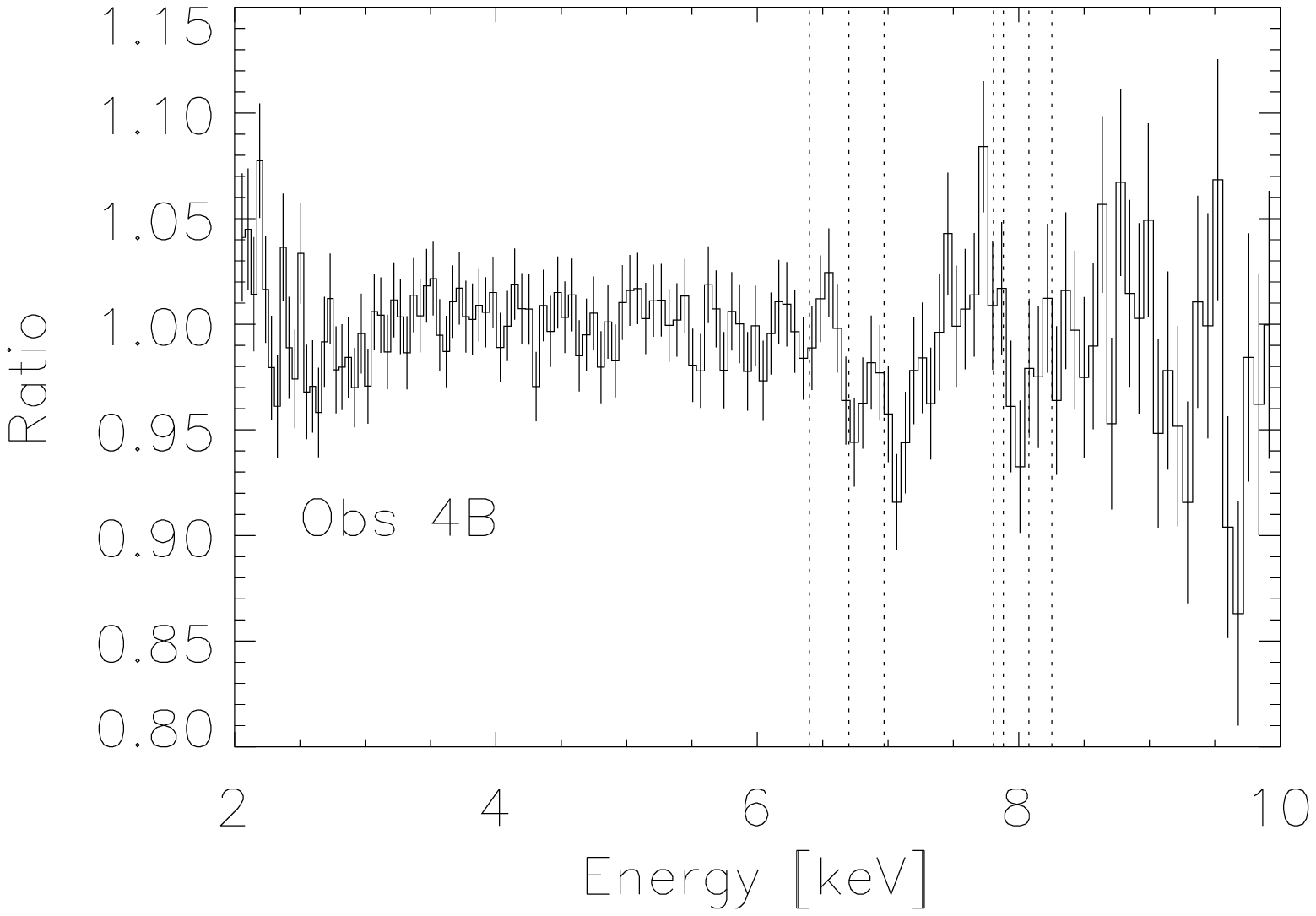}
\includegraphics[angle=0,width=0.31\textwidth]{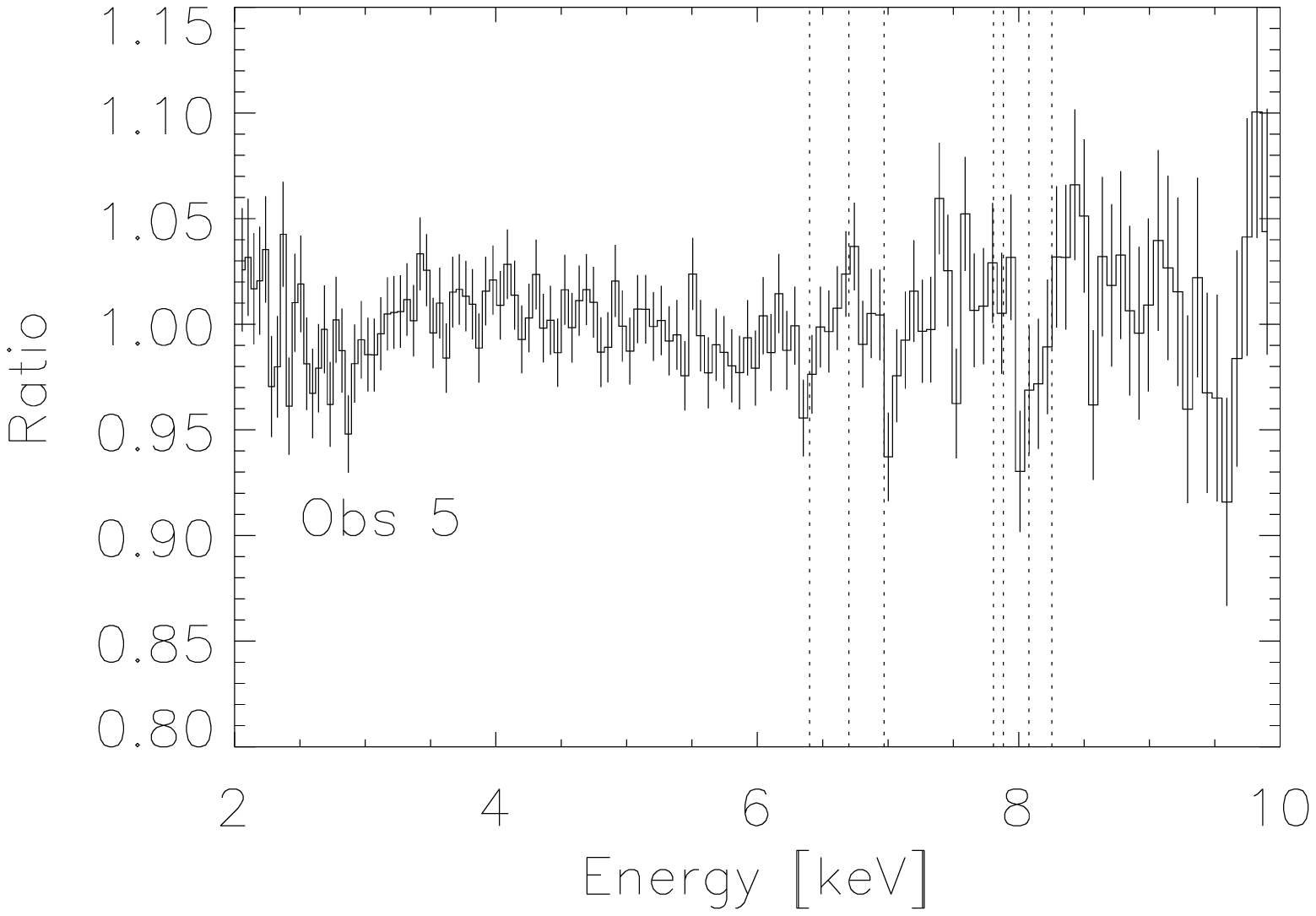}
\includegraphics[angle=0,width=0.31\textwidth]{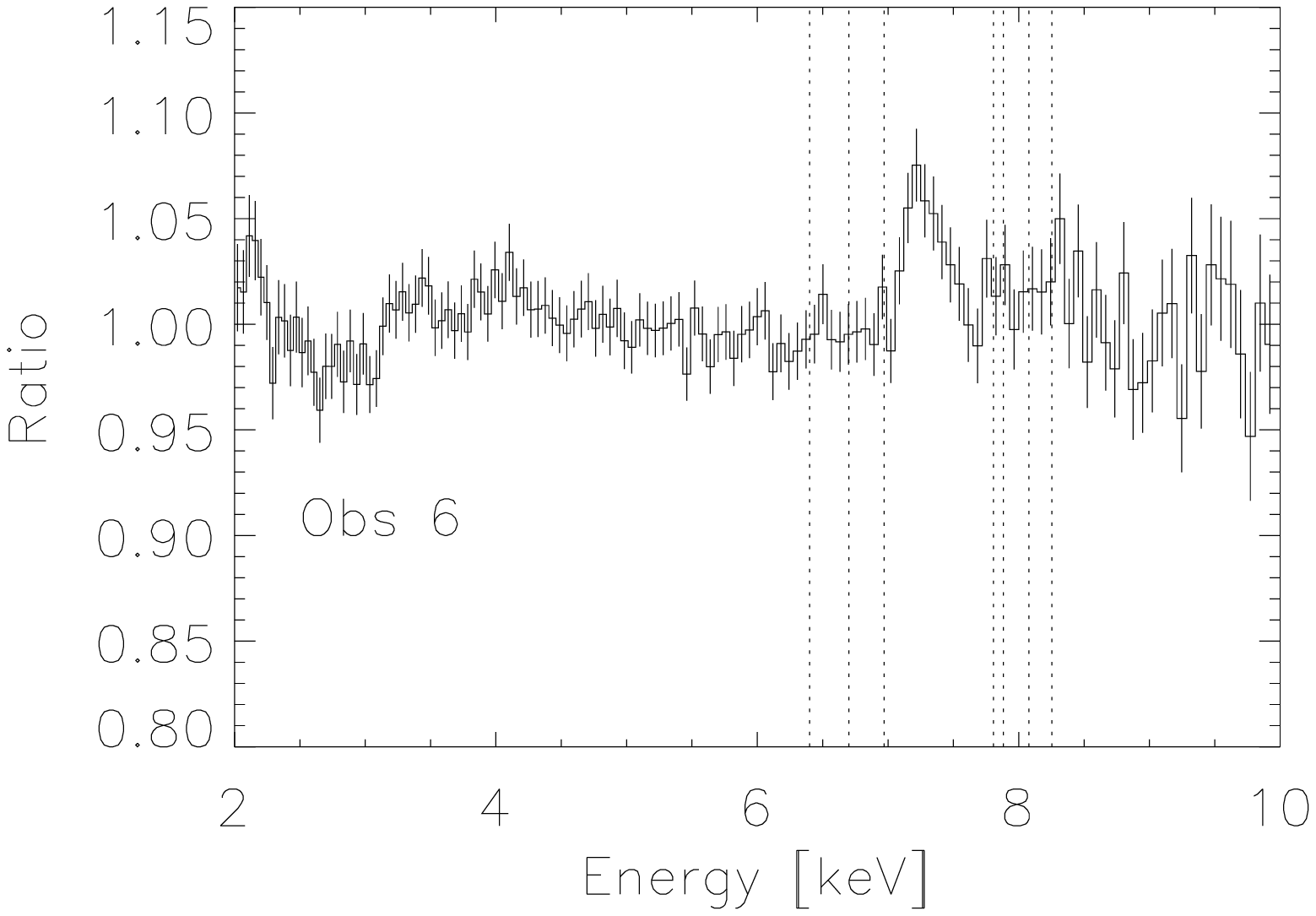}
}
\caption{Ratio of the data to the continuum model for obs~1--6. The dotted vertical lines indicate from left to right the rest energy of the transitions of neutral Fe, \fetfive\ \ka, \fetsix\ \ka, \nitseven\ \ka, \fetfive\ \kb, \niteight\ \ka\ and \fetsix\ \kb.} \label{fig:pnrat}
\end{figure*}

Detailed plots of the residuals from the best-fit continuum model around the Fe~K~region are 
shown in Figures~\ref{fig:pndel} and \ref{fig:pnrat}. 
Absorption features from highly ionised species of iron, such as 
\fetfive\ and \fetsix\ \ka\ and \kb\ and a broad iron
emission line are evident in obs~1--4. Both the absorption and emission features show significant
variations between observations. In contrast, obs~5 shows no significant
absorption or emission features at the Fe~K~region and obs~6 shows a prominent narrow emission line 
above 7~keV.  

\begin{figure*}[!ht]\centerline{\hspace{-1cm}
\includegraphics[angle=0,width=9.0cm,height=6cm]{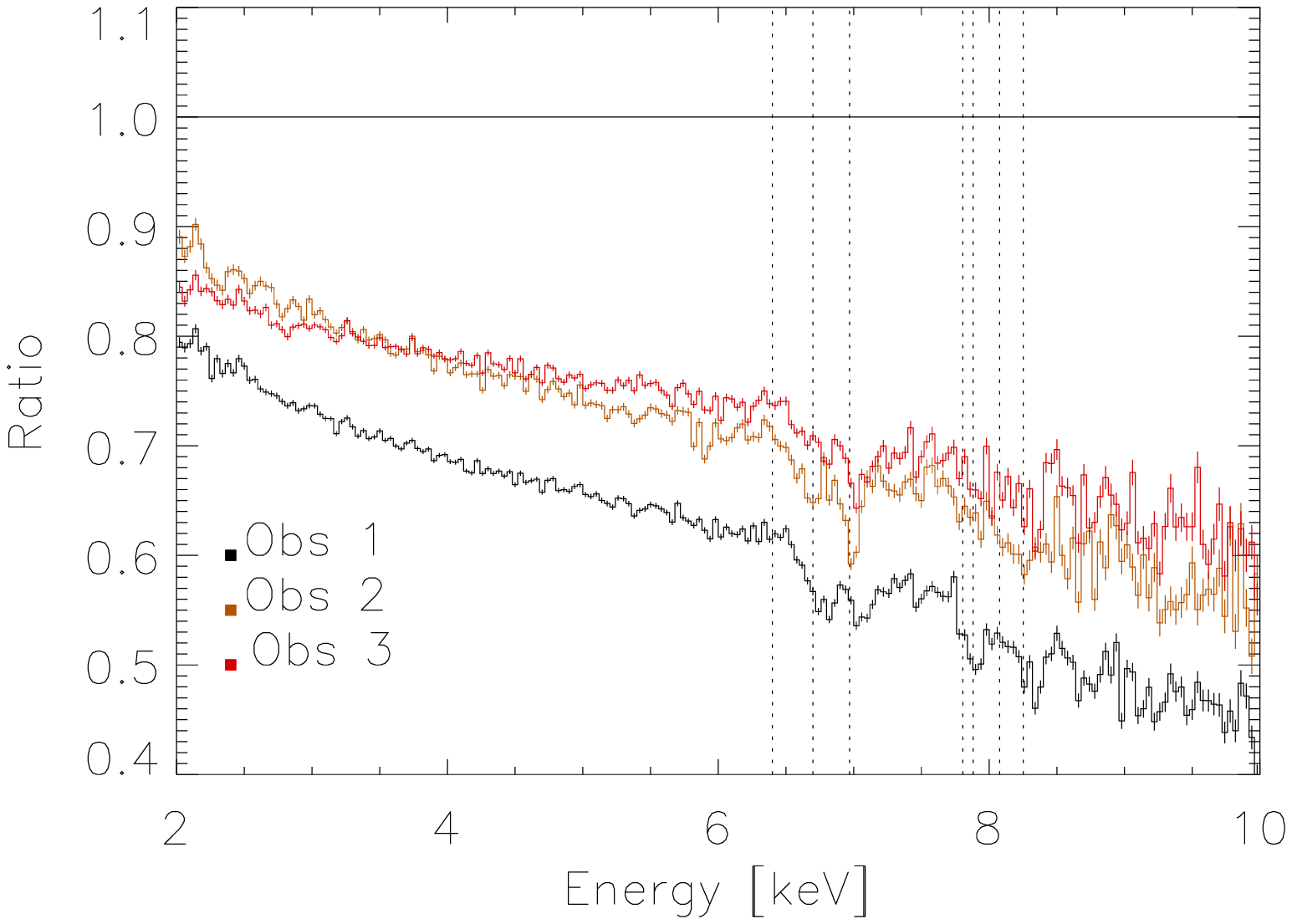}
\includegraphics[angle=0,width=9.0cm,height=6cm]{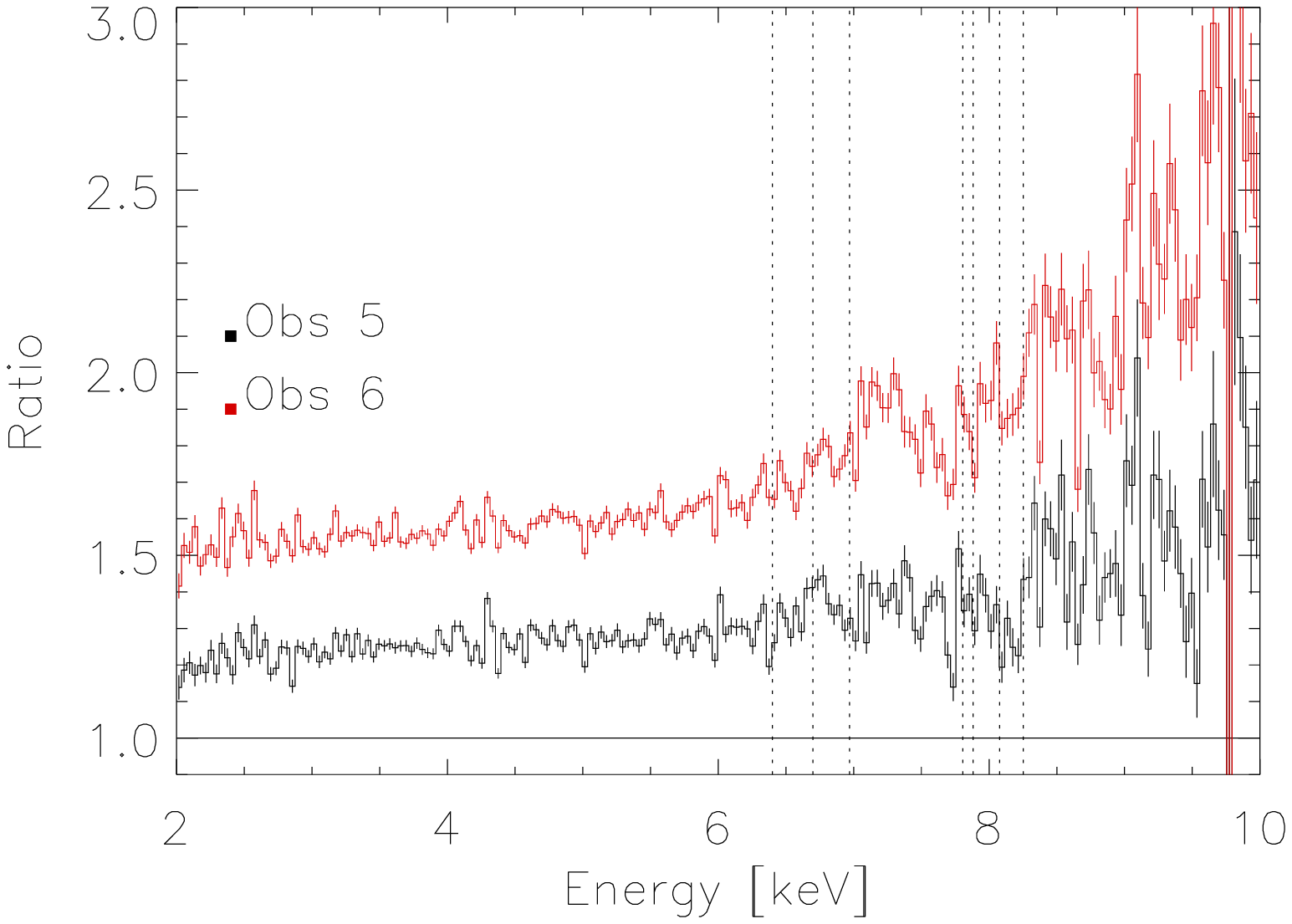}
}
\caption{{\it Left:} Ratio of EPIC pn \src\ timing mode spectra from obs~1, 2 and 3 with respect to 
obs~4. {\it Right:} Ratio of EPIC pn \src\ burst mode spectra from obs~5 and 6 with respect to 
obs~4. The dotted vertical lines indicate the same transitions as in Fig.~\ref{fig:pnrat}. } \label{fig:ratio}
\end{figure*}

We first evaluated the effect of the continuum on the absorption and emission features in a model-independent manner 
calculating the ratio among the spectra for all observations with respect to obs~4 (see Fig.~\ref{fig:ratio}). 
By obtaining the ratios between
obs~1--3 and obs~4T and between obs~5--6 and obs~4B, we can rule out changes in the spectrum due to cross-calibration
uncertainties between the timing and the burst modes. It is clear that the spectrum has systematically hardened and brightened from obs~1 to
obs~6. This is accompanied by a weakening of the absorption lines from obs~1--4 (see Fig.~\ref{fig:ratio}, left panel). Obs~6 shows for first time a prominent emission feature above 7~keV which is absent in obs~1--5 (Fig.~\ref{fig:ratio}, right panel).

\begin{table*}\begin{center}\caption[]{Best fits to the 2--10 keV EPIC pn spectra for all observations using Models 1a (obs~1--4), 1b (obs~5) and 1c (obs~6) (see text).  \nhabs\ is the column
  density for the neutral absorber. \kdbb, $k_{po}$ and
  \kgau\ are the normalizations of the disc blackbody component, power law and Gaussian features, respectively.
  \ktdbb\ is the temperature of the disc blackbody component and $\Gamma$ the index of the power law. \egau, $\sigma$ and \ew\ represent the energy, width and equivalent width of
  the Gaussian features. $E_{edge}$ and $\tau$ are the energy and optical depth of the absorption edges. $F$ is the unabsorbed
  flux of the continuum between 2 and 10~keV. {\it f} indicates that a parameter was fixed. The width of the absorption features was constrained to be $\approxlt$\,0.1~keV and of the broad emission feature ({\tt gau$_{e1}$}) to $\approxlt$\,1~keV. {\it p} indicates that the error of a parameter pegged at the limit imposed.}
\small\addtolength{\tabcolsep}{-3pt}
\scalebox{0.9}{
\begin{tabular}{lcccccccc}
\hline \hline\noalign{\smallskip}
Observation No. & & 1 & 2 & 3 & 4T & 4B & 5 & 6 \\
\noalign{\smallskip\hrule\smallskip}& Comp. & & & & \\
Parameter & & & & & \\
& {\tt tbabs} & & & & \\
\multicolumn{2}{l}{\nhabs$_{pn}$ {\small($10^{22}$ cm$^{-2}$)}} & 7.64\,$\pm$\,0.05 & 7.58\,$\pm$\,0.07 & 7.69\,$\pm$\,0.06 & 7.72\,$\pm$\,0.06 &  8.19\,$\pm$\,0.08 & 8.22\,$\pm$\,0.07 & 8.38\,$\pm$\,0.09 \\
& {\tt diskbb} & & & \\
\multicolumn{2}{l}{\ktdbb\ {\small(keV)}} & 1.41\,$\pm$\,0.01 & 1.44\,$\pm$\,0.02 &1.49\,$\pm$\,0.01 & 1.59\,$\pm$\,0.01 & 1.74\,$\pm$\,0.01 & 1.81\,$\pm$\,0.01 & 1.78\,$\pm$\,0.03 \\
\multicolumn{2}{l}{\kdbb\ {\small[(R$_{in}$/D$_{10}$)$^{2}$ cos$\theta$]}} & 197\,$^{+5}_{-7}$ & 200\,$^{+8}_{-10}$ & 181\,$\pm$\,7 & 175\,$\pm$\,5 & 100\,$\pm$\,4 & 107\,$\pm$\,3 & 118\,$\pm$\,5 \\
& {\tt po} & & & \\
\multicolumn{2}{l}{$\Gamma$} & -- &-- &-- & -- & -- & -- & 2 (f)\\ 
\multicolumn{2}{l}{$k_{po}$ {\small[ph keV$^{-1}$cm$^{-2}s^{-1}$ at 1 keV]}} &-- &-- &-- &-- &-- & -- & 1.76\,$\pm$\,0.25 \\ 
& {\tt gau$_{e1}$} & & & \\
\multicolumn{2}{l}{ \egau\ {\small(keV)}} & 6.40\,$^{+0.11}_{-0.0p}$ & 6.55\,$\pm$\,0.24 & 6.42\,$^{+0.30}_{-0.02p}$ & 6.97\,$^{+0p}_{-0.14}$ & -- & -- & -- \\
\multicolumn{2}{l}{ $\sigma$ {\small(keV)}} & 1\,$^{+0p}_{-0.14}$ & 1\,$^{+0p}_{-0.36}$ & 0.99\,$^{+0.01p}_{-0.48}$ & 0.33\,$^{+0.41}_{-0.18}$ & -- & --& --\\
\multicolumn{2}{l}{ \kgau\ {\small(10$^{-3}$ ph cm$^{-2}$ s$^{-1}$)}} & 12\,$\pm$\,3 & 10\,$^{+3}_{-5}$  & 7\,$\pm$\,5 & 3\,$^{+7}_{-2}$ & -- & --& --\\
 \multicolumn{2}{l}{ \ew\ (eV)} & 161\,$\pm$\,40 & 123\,$^{+37}_{-62}$ & 76\,$\pm$\,54 & 34\,$^{+79}_{-23}$ & -- & --& --\\
& {\tt gau$_{e2}$} & & & \\
\multicolumn{2}{l}{ \egau\ {\small(keV)}} &-- &-- &-- &-- &-- & --& 7.26\,$\pm$\,0.04 \\
\multicolumn{2}{l}{ $\sigma$ {\small(keV)}} &-- &-- &-- &-- &-- &-- & 0.10\,$\pm$\,0.06 \\
\multicolumn{2}{l}{ \kgau\ {\small(10$^{-3}$ ph cm$^{-2}$ s$^{-1}$)}} &-- &-- &-- &-- &-- &-- & 3.2\,$\pm$\,1.0 \\
 \multicolumn{2}{l}{ \ew\ (eV)} &-- &-- &-- &-- &-- &-- & 23\,$\pm$\,7 \\
& {\tt gau$_{a1}$} & & & \\
\multicolumn{2}{l}{ \egau\ {\small(keV)}} & 2.62\,$\pm$\,0.04 & 2.59\,$\pm$\,0.05 & 2.61\,$\pm$\,0.06 & 2.58\,$\pm$\,0.05 & -- & --&-- \\
\multicolumn{2}{l}{ $\sigma$ {\small(keV)}} & 0.08$^{+0.02p}_{-0.08p}$ & 0.1$^{+0.0p}_{-0.1p}$ &0.1\,$^{+0.0p}_{-0.1p}$ & 0.1\,$^{+0p}_{-0.06}$ &  -- & --& --\\
\multicolumn{2}{l}{ \kgau\ {\small(10$^{-3}$ ph cm$^{-2}$ s$^{-1}$)}} & 8.4\,$\pm$\,3.8  & 10\,$^{+6}_{-4}$  & 9.8\,$^{+3.1}_{-5.2}$ & 13\,$\pm$\,5 & -- & --& --\\
\multicolumn{2}{l}{ \ew\ (eV)} & 11\,$\pm$\,5 & 13\,$^{+5}_{-8}$ & 12\,$^{+4}_{-6}$ & 13\,$\pm$\,5 & -- & -- & -- \\
 & {\tt gau$_{a2}$} & & & \\
 \multicolumn{2}{l}{ \egau\ {\small(keV)}} & 2.94\,$\pm$\,0.08 & 2.92\,$^{+0.10}_{-0.13}$ & 2.92\,$\pm$\,0.09 & 2.94\,$^{+0.07}_{-0.10}$ & -- & --&-- \\
\multicolumn{2}{l}{ $\sigma$ {\small(keV)}} & 0.1$^{+0.0p}_{-0.07}$ & 0.1$^{+0p}_{-0.1p}$ &0.1\,$^{+0p}_{-0.1p}$ & 0.06\,$^{+0.04p}_{-0.06p}$ &  -- & --& --\\
\multicolumn{2}{l}{ \kgau\ {\small(10$^{-3}$ ph cm$^{-2}$ s$^{-1}$)}} & 5.5\,$^{+2.2}_{-2.5}$  & 5\,$\pm$\,3  & 5.5\,$\pm$\,2.9 & 4.1\,$^{+3.6}_{-2.7}$ & -- & --& --\\
\multicolumn{2}{l}{ \ew\ (eV)} & 9\,$\pm$\,4 & 7\,$\pm$\,5 & 8\,$\pm$\,4 & 5\,$^{+4}_{-3}$ & -- & -- & -- \\
 & {\tt gau$_{a3}$} & & & \\
\multicolumn{2}{l}{ \egau\ {\small(keV)}} & 6.78\,$\pm$\,0.01 & 6.73\,$\pm$\,0.03 & 6.78\,$\pm$\,0.02 & 6.79\,$\pm$\,0.04  & 6.75\,$^{+0.05}_{-0.21}$ & -- & --\\
\multicolumn{2}{l}{ $\sigma$ {\small(keV)}} & $<$\,0.07 & 0$^{+0.1p}_{-0p}$  &  $<$\,0.07 & 0.0\,$^{+0.1p}_{-0.0p}$ & 0.0\,$^{+0.1p}_{-0.0p}$ & --& -- \\
\multicolumn{2}{l}{ \kgau\ {\small(10$^{-3}$ ph cm$^{-2}$ s$^{-1}$)}} & 2.0\,$\pm$\,0.3 & 1.4\,$\pm$\,0.3 & 1.3\,$\pm$\,0.2 & 1.4\,$^{+0.2}_{-0.5}$ & 1.1\,$\pm$\,0.5 & -- & -- \\
\multicolumn{2}{l}{ \ew\ (eV)} & 32\,$\pm$\,5  & 19\,$\pm$\,4 & 17\,$\pm$\,3 & 13\,$^{+2}_{-5}$ & 10\,$\pm$\,5 & -- & -- \\
& {\tt gau$_{a4}$} & & &  \\
\multicolumn{2}{l}{ \egau\ {\small(keV)}} & 7.05\,$\pm$\,0.02 & 7.01\,$\pm$\,0.01 & 7.04\,$\pm$\,0.01 & 7.04\,$^{+0.01}_{-0.04}$ & 7.08\,$\pm$\,0.04 & -- &  -- \\
\multicolumn{2}{l}{ $\sigma$ {\small(keV)}} & 0.03\,$\pm$\,0.02 & 0.05\,$\pm$\,0.02 & 0.04\,$\pm$\,0.02 & 0.05\,$\pm$\,0.05p & 0.0\,$^{+0.1p}_{-0.0p}$ & -- & -- \\
\multicolumn{2}{l}{ \kgau\ {\small(10$^{-3}$ ph cm$^{-2}$ s$^{-1}$)}} & 2.4\,$\pm$\,0.3 & 3.2\,$\pm$\,0.4 & 2.9\,$\pm$\,0.4 & 3.4\,$^{+10.3}_{-0.9}$ & 1.3\,$\pm$\,0.5 & -- & --\\
\multicolumn{2}{l}{ \ew\ (eV)} & 48\,$\pm$\,6 & 52\,$\pm$\,6 & 47\,$\pm$\,6 & 38\,$^{+115}_{-10}$ & 16\,$\pm$\,6 & -- & --\\
& {\tt gau$_{a5}$} & & \\
\multicolumn{2}{l}{ \egau\ {\small(keV)}}  & 7.90\,$\pm$\,0.02 & 7.92\,$\pm$\,0.09 & 7.97\,$\pm$\,0.05 & 8.14\,$^{+0.12}_{-0.07}$ & -- & --& -- \\
\multicolumn{2}{l}{ $\sigma$ {\small(keV)}} &  0\,$^{+0.1p}_{-0p}$ & 0\,$^{+0.1p}_{-0p}$ & 0.1\,$^{+0.0p}_{-0.1p}$ & 0.05\,$^{+0.05p}_{-0.05p}$ & -- & -- & --\\
\multicolumn{2}{l}{ \kgau\ {\small(10$^{-3}$ ph cm$^{-2}$ s$^{-1}$)}} & 0.6\,$\pm$\,0.1 & 0.3\,$\pm$\,0.2  & 0.7\,$\pm$\,0.3 & $<$\,0.7 & -- & -- & --\\
\multicolumn{2}{l}{ \ew\ (eV)} & 20\,$\pm$\,3 & 7\,$\pm$\,5 & 17\,$\pm$\,7 & $<$\,16 & -- & -- & --\\
& {\tt gau$_{a6}$} \\
\multicolumn{2}{l}{ \egau\ {\small(keV)}} & 8.29\,$\pm$\,0.02 & 8.25\,$\pm$\,0.05 & 8.31\,$^{+0.03}_{-0.01}$  & 8.43\,$^{+0.11}_{-0.08}$ & -- & -- & --\\
\multicolumn{2}{l}{ $\sigma$ {\small(keV)}} & 0.1\,$^{+0.0p}_{-0.04}$ & 0.1\,$^{+0.0p}_{-0.04}$ & 0\,$^{+0.1p}_{-0p}$ & 0.0\,$^{+0.1p}_{-0p}$ & -- & -- & --\\
\multicolumn{2}{l}{ \kgau\ {\small(10$^{-3}$ ph cm$^{-2}$ s$^{-1}$)}} & 0.8\,$\pm$\,0.2 & 0.8\,$\pm$\,0.2 & 0.6\,$\pm$\,0.1 & 0.3\,$\pm$\,0.2 & -- & -- & --\\
\multicolumn{2}{l}{ \ew\ (eV)} & 33\,$\pm$\,8 & 28\,$\pm$\,7 & 20\,$\pm$\,3 & 6\,$\pm$\,4 & -- & -- & --\\
& {\tt edge$_1$} & & & \\
\multicolumn{2}{l}{ $E_{edge}$ {\small(keV)}} & 8.83 (f) & 8.83 (f) & 8.83 (f) & 8.83 (f) & 8.83 (f) & -- & --\\ 
\multicolumn{2}{l}{ $\tau$ } & 0.05\,$^{+0.03}_{-0.01}$ & 0.04\,$\pm$\,0.02  & 0.06\,$\pm$\,0.03 & 0.03\,$\pm$\,0.02 & $<$\,0.06 & -- & --\\
& {\tt edge$_2$} & & & \\
\multicolumn{2}{l}{ $E_{edge}$ {\small(keV)}} & 9.28 (f) & 9.28 (f) & 9.28 (f) & 9.28 (f) & -- & --& --\\
\multicolumn{2}{l}{ $\tau$ } & $<$\,0.02 & $<$\,0.08 & $<$\,0.03 & $<$\,0.03 &  -- & --& --\\
\hline\noalign{\smallskip}
\multicolumn{2}{l}{F$_{2-10 keV}$ {\small(10$^{-8}$ erg cm$^{-2}$ s$^{-1}$)}} & 1.01 & 1.14 & 1.16 & 1.51 & 1.29 & 1.63 & 2.12 \\
\noalign {\smallskip}
\hline\noalign {\smallskip}
\multicolumn{2}{l}{\rchisq (d.o.f.)} & 0.49 (107) & 0.63 (106) & 0.47 (106) & 0.46 (107) & 0.97 (122) & 1.10 (129) & 1.08 (126) \\
\noalign{\smallskip\hrule\smallskip}
\noalign{\smallskip\hrule\smallskip}
\label{tab:bestfit-gau}
\end{tabular}
}
\end{center}
\end{table*} 

To quantify the relation between the significance of the
emission and absorption features and the changes in the continuum, we next included in the model Gaussian absorption
and emission features and absorption edges to account for
the residuals near 7 keV. 
For obs~1--4, the total model consisted of a disc
blackbody component, one Gaussian emission
feature at $\sim$6.6 keV, and five Gaussian absorption features
at the energies of $\ssixteen$ (2.62~keV), $\fetfive$ \ka\ (6.70~keV), $\fetsix$ \ka\ (6.97~keV), $\fetfive$ \kb\ (7.88~keV) and 
$\fetsix$ \kb\ (8.25~keV) modified by photo-electric absorption from neutral material and
by absorption edges at the energies of \fetfive\ (8.83~keV) and \fetsix\ (9.28~keV). We also added a sixth absorption feature at $\sim$2.9~keV, to
account for residuals at that energy, although we could not identify a known transition. Therefore, the final model was 
({\tt tbabs*edge$_1$*edge$_2$*(diskbb+gau$_{e1}$+gau$_{a1}$+gau$_{a2}$+gau$_{a3}$ +gau$_{a4}$+gau$_{a5}$+gau$_{a6}$}), hereafter Model 1a). The fits with this model were
acceptable, with \rchisq\ $\sim$ 0.4--1.0 for 106--122 d.o.f., for obs 1--4. 
For obs~5 the total model consisted of a disc
blackbody, modified by photo-electric absorption from neutral material ({\tt tbabs*(diskbb)}, hereafter Model 1b). This model gave an acceptable fit, with \rchisq\ = 1.10 for 129 d.o.f. For obs~6, we included a power-law component, in addition to the
disc blackbody, to account for the curvature observed in the residuals when only one continuum component was used. A further motivation for inclusion of the
power-law component is that the hard X-ray flux was significantly higher in obs~5--6 compared to obs~1--4, as observed by MAXI/ASM and Swift/BAT. In particular,
the SWIFT/BAT 15--50 keV count rate increased from 0.009\,$\pm$\,0.001 s$^{-1}$ at the time of obs~4 to 0.019\,$\pm$\,0.001 s$^{-1}$ and 0.036\,$\pm$\,0.002 s$^{-1}$ at the time of obs~5 and 6, respectively. 
We note that inclusion of a power-law component for obs~5 did not improve the fit significantly and therefore we used only the initial disc blackbody.
For obs~6, after fitting the spectra with a disc blackbody and a power law modified by photo-electric absorption from neutral material, narrow emission features remained at $\sim$\,4.1 and 7.3 and 8.2~keV. Inclusion of the most significant feature at 7.3~keV improved the fit from \rchisq\ = 1.43 (129) to 1.08 (126). Therefore the best-fit model for obs~6 was ({\tt tbabs*(diskbb+po+gau$_{e2}$)}, hereafter Model 1c). 
The parameters of the best fit with models 1a, 1b and 1c are given in Table~\ref{tab:bestfit-gau}.

The parameters of the continuum and the neutral absorber appear to change significantly between obs~4T and 4B. The different value of neutral absorption translates to a different value of unabsorbed flux, which decreases from obs~4T to obs~4B. However, the light curves indicate a clear increase of count rate between both parts of obs~4. Therefore, we attribute the differences of the continuum between obs~4T and 4B to cross-calibration differences between timing and burst mode. We can estimate the systematic error on the temperature of the disc by modifying the energy-scale to minimize the residuals at the Au edge and assuming that the energy scale error is the same at the Au edge and at the Fe region (but note that this assumption has to be confirmed yet). We obtain that the disc temperature decreases from 1.59 and 1.74 keV to 1.52 and 1.65 keV for obs~4T and 4B, respectively. Therefore, we obtain a significant increase in the disc temperature between obs~4T and 4B even after considering energy-scale calibration deficiencies. While the timing mode observations have been recently subject to a major recalibration, which should yield smaller systematic uncertainties compared to the burst mode, it should be borne in mind that we are discarding the better-calibrated columns at the centre of the PSF for the timing mode observations due to pile-up, and therefore it is not clear whether the calibration should be worse for the burst mode than for the timing mode at the wings of the PSF. Clearly, one should be cautious when interpreting absolute values of the continuum such as the inner disc radius. In contrast, we can examine relative changes for observations taken in the same mode. However, we caution that even in this case, the continuum modelling may change significantly the value of the disc radius. For example, substituting the power law by the more physical {\tt simpl} component (available in XSPEC) in obs~6, the quality of the fit improves from \rchisq\ = 1.08 (126) to 1.02 (126) and the inner disc radius derived from the disc normalisation increases from 17 to 19 km. This highlights the difficulty of identifying and disentangling effects due to calibration or to a deficient modeling when comparing observations (see also the different relative changes on the parameters of obs~4T and 4B between models~3 and 4 in Sect.~\ref{sec:model3} below). 

Fig.~\ref{fig:lines2} shows the depth of the \fetfive\ \ka\ and \kb\ and \fetsix\ \kb\ narrow absorption lines as a function of the disc unabsorbed flux. The constant value of the \fetsix\ \ka\ line in obs~1--3 as opposed to the evolution of the \kb\  line indicates that the \fetsix\ \ka\ is probably saturated. This is a signature of the high column density of the plasma. In contrast, for obs~4, the \fetfive\ \ka\ line is very weak and we can only set upper limits to the existence of the \fetfive\ \kb\ line. The \fetsix\ lines have also become significantly smaller. Clearly, the lines become weaker as the flux of the disc increases, indicating that the ionising flux has a direct effect on the column density and/or ionisation of the plasma. Moreover, we cannot detect any significant absorption line in obs~5--6 (although a weak \fetsix\ \ka\ line may be present in obs~5, see also Sect.~\ref{sec:wind-hss}), indicating that the wind has ``disappeared''.
Since obs~5 and obs~6 show a significantly different spectrum compared to obs~1--4, with no signature of the presence of a wind, we do not explore these observations in this work any further and refer to \citet{1630:diaz13nat} for details. We discuss the interpretation of these two observations with respect to obs~1--4 in Sect.~\ref{sec:wind-hss} and \ref{subsec:wind-jet}.

\subsubsection{Photoionised plasma model}
\label{sec:model2}

The strength of the absorption features shown in Table~\ref{tab:bestfit-gau} indicates that the continuum may be significantly 
affected by the absorbing plasma for obs~1--4. Therefore, to quantify the above changes in a more physical manner, we substituted the absorption features
and edges by the component {\tt warmabs}, which models the absorption due to a photoionised
plasma in the line of sight. This component not only accounts
for the narrow absorption features evident near 7~keV but also modifies
the overall continuum shape at regions where the spectral resolution is not enough to resolve
individual features. We note that the {\tt warmabs} model
does not include Compton scattering, in contrast to the {\tt xabs} model in the SPEX package \citep{kaastra96}\footnote[7]{We used XSPEC instead of SPEX for spectral fitting because the reflection model {\tt rfxconv} (see next section) is not available in the latter spectral package.}.
The contribution of Compton scattering is expected to be significant when the column density of the absorber is $\approxgt$\,10$^{22}$~cm$^{-2}$\citep{1323:boirin05aa, ionabs:diaz06aa}. Therefore 
we added the component {\tt cabs} to the model, which accounts for non-relativistic, optically-thin Compton scattering. We forced
the column density of the {\tt cabs} model to be equal to the column density of the {\tt warmabs} component multiplied by a factor of 1.21. The latter
factor accounts for the number of electrons per hydrogen atom for a material of solar abundances \citep{herx1:stelzer99}. The non-relativistic approximation
of {\tt cabs} will overestimate the scattering fraction at energies $\approxgt$10~keV \citep{1323:boirin05aa}, and therefore it should
not be used when broad band energy data are available. Moreover, the column density of the {\tt warmabs} component sets only 
a lower limit to the amount of Compton scattering since there may be fully ionised material, which cannot be identified via line absorption but still
contributes to Compton scattering.

\begin{figure}[!ht]
\includegraphics[angle=0,width=0.24\textwidth]{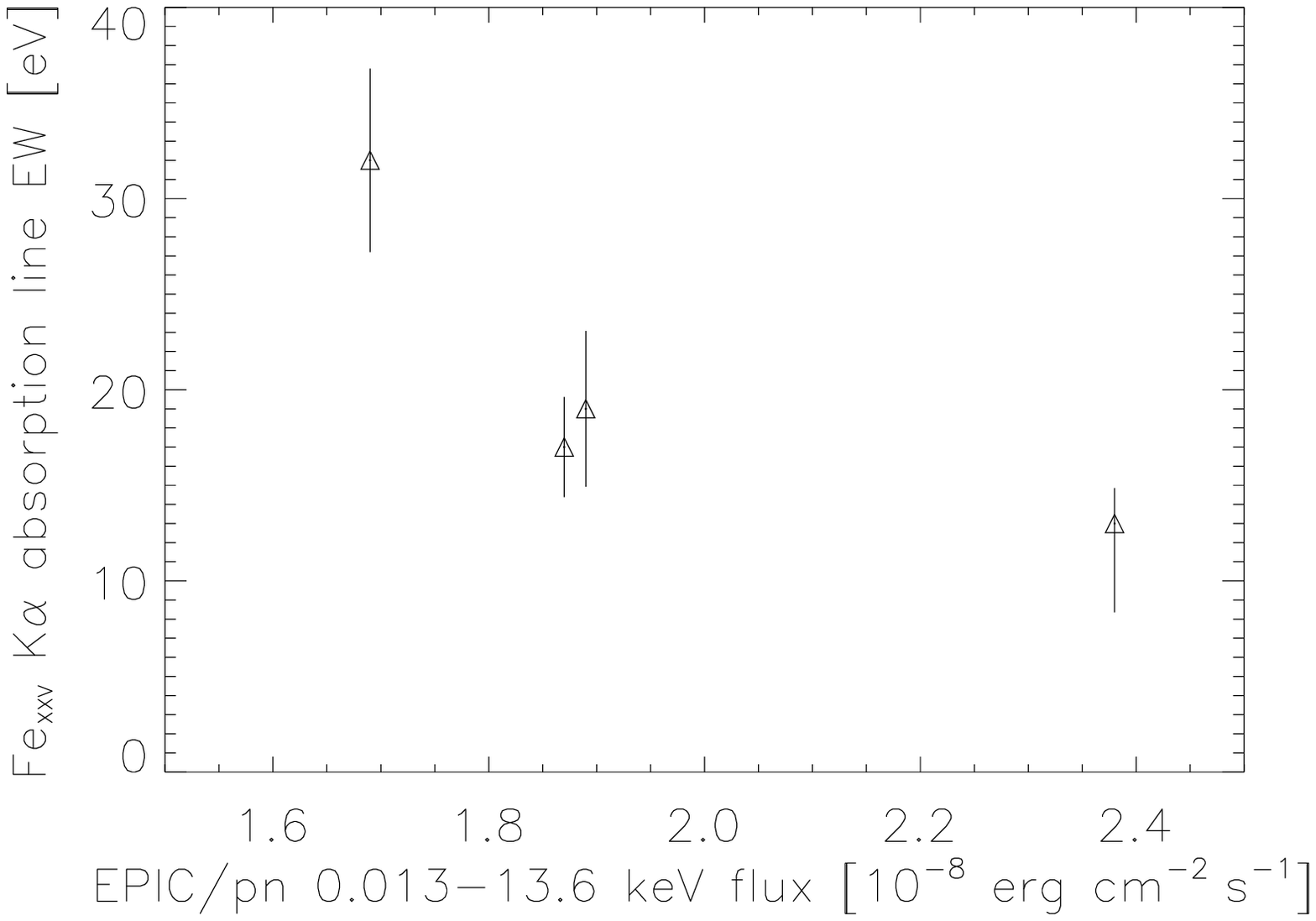}
\includegraphics[angle=0,width=0.24\textwidth]{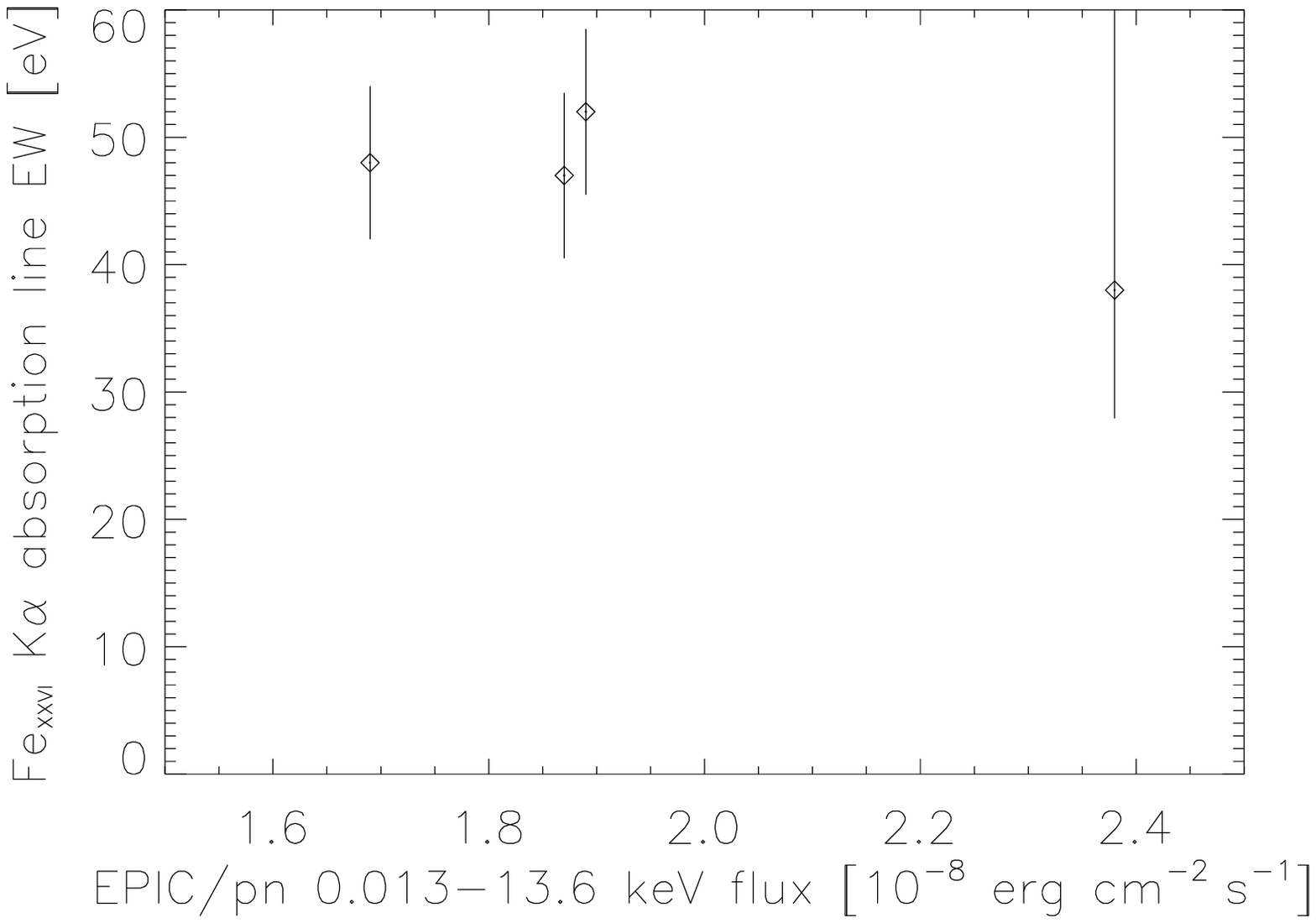}
\includegraphics[angle=0,width=0.24\textwidth]{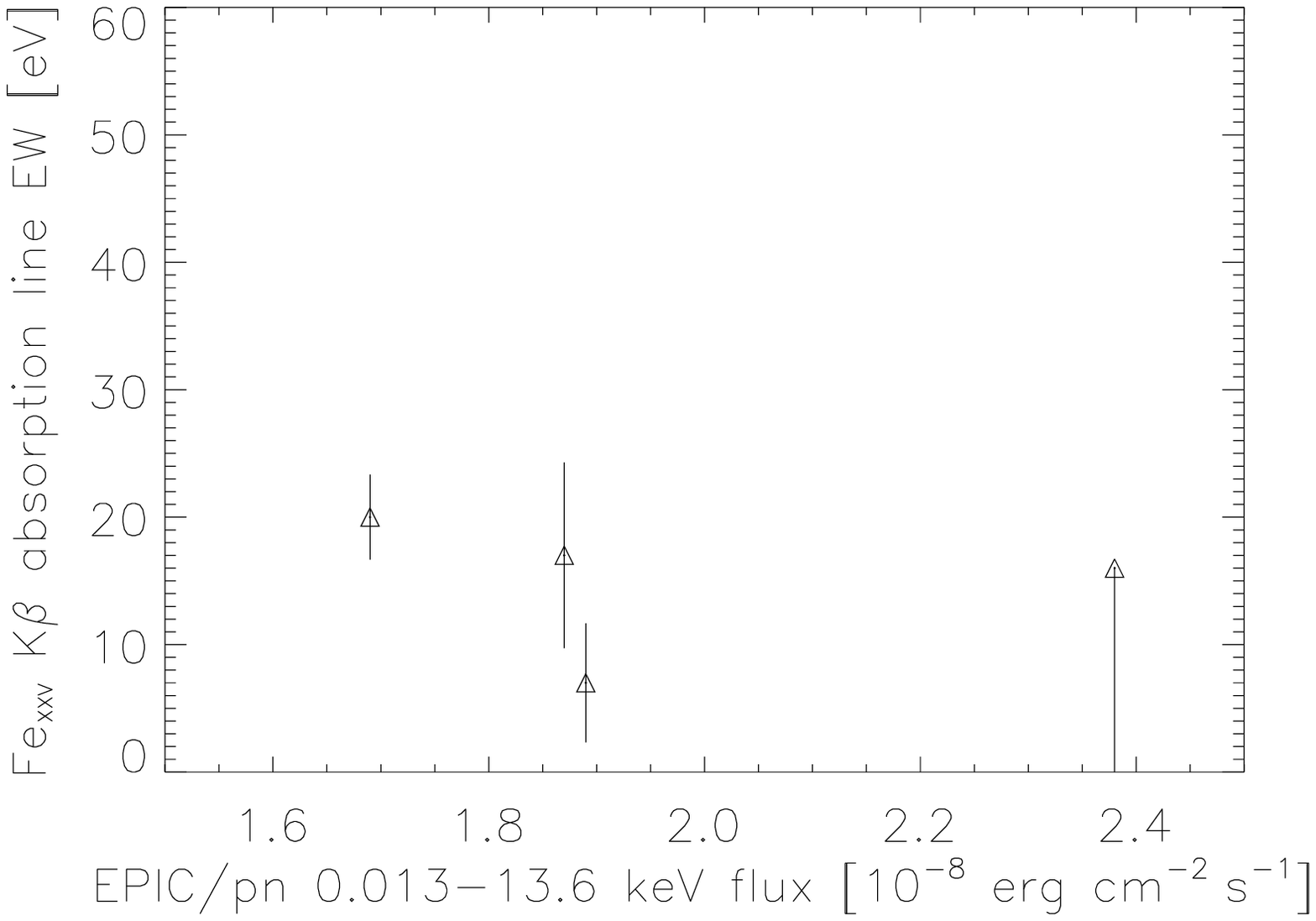}
\includegraphics[angle=0,width=0.24\textwidth]{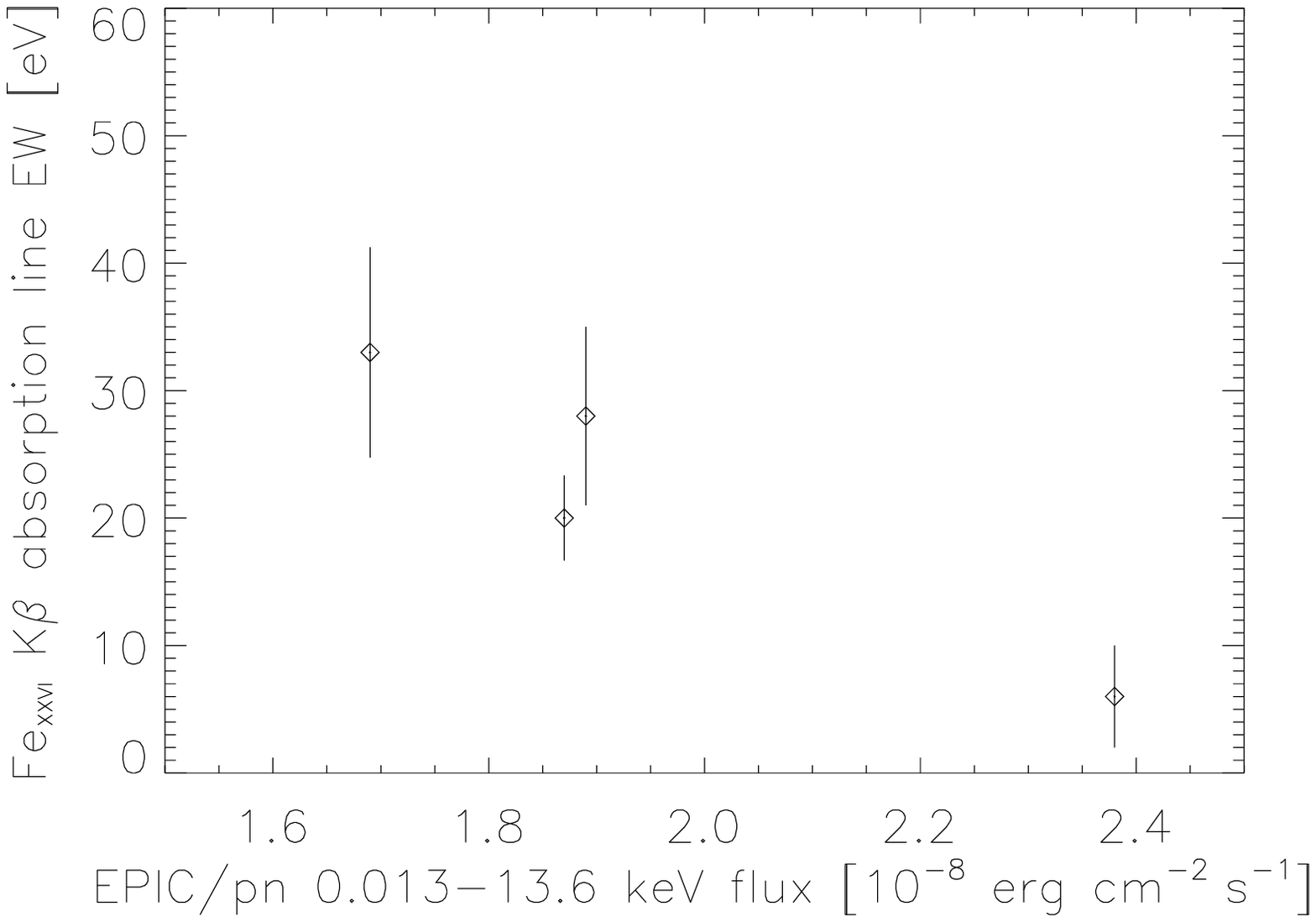}
\caption{\ew\ of the narrow absorption lines of \fetfive\ \ka\ (upper-left), 
\fetsix\ \ka\ (upper-right), \fetfive\ \kb\ (lower-left) and \fetsix\ \kb\ (lower-right) with respect to the 0.013--13.6 keV unabsorbed flux of the disc blackbody component.} 
\label{fig:lines2}
\end{figure}

Our final model consisted of a disc blackbody component and
one Gaussian emission feature at $\sim$6.6~keV, modified by photo-electric 
absorption from neutral and ionised material
({\tt tbabs*cabs*warmabs*(diskbb+gau)}, hereafter Model~2).

We considered that the illumination of the photoionised absorber was only due to the disc. Therefore, for each observation, we used the best-fit 
disc continuum (i.e. not including the broad emission line or the photo-ionised absorber itself) as the
ionising continuum for the photoionised absorber. The density of the plasma
was set to be $n_e$ = 10$^{12}$ cm$^{-3}$ following \citet{1630:kubota07pasj}. The
luminosity was defined in the 0.013--13.6 keV band for each
spectrum.
The fits with Model~2 were acceptable for obs~1--4, with \rchisq\ of 0.6--0.9
for 121--124 d.o.f.. The parameters of the best fit with
this model are given in Table~\ref{tab:bestfit-warm}.

We observe a trend of decreasing plasma column density from obs~1 to obs~4T.
At the same time, the disc temperature increases significantly, while the disc radius stays roughly constant at $\sim$22--24~km, consistent with the evolution expected for an optically thick and geometrically thin disc during the HSS. 
 
 We detect an overabundance of Sulphur of $\sim$\,13--21 with respect to solar abundances. This might indicate a deficiency in the model used, such as the absence of a reprocessed wind component, which would include resonance line emission and fill in a different proportion of saturated and non-saturated ions. Alternatively, it could be related to the shape of the illumination continuum not being realistic (e.g. if there is a power-law component above 10~keV) since we do not observe a significant overabundance when a power--law illumination continuum is used. As an example, for obs~4T, we obtain a S abundance of 3.7\,$^{+2.7}_{-1.5}$ and a ionisation parameter of 3.63\,$^{+0.07}_{-0.01}$ when using a power law of index 2 as ionising continuum (all the other parameters of the fit remain constant within the errors with respect to those shown in Table~\ref{tab:bestfit-warm}. 

We also observe a significant decrease of the flux and \ew\  of the broad Fe line from obs~1 to 4. 

\begin{figure}[!ht]
\includegraphics[angle=0,width=0.24\textwidth]{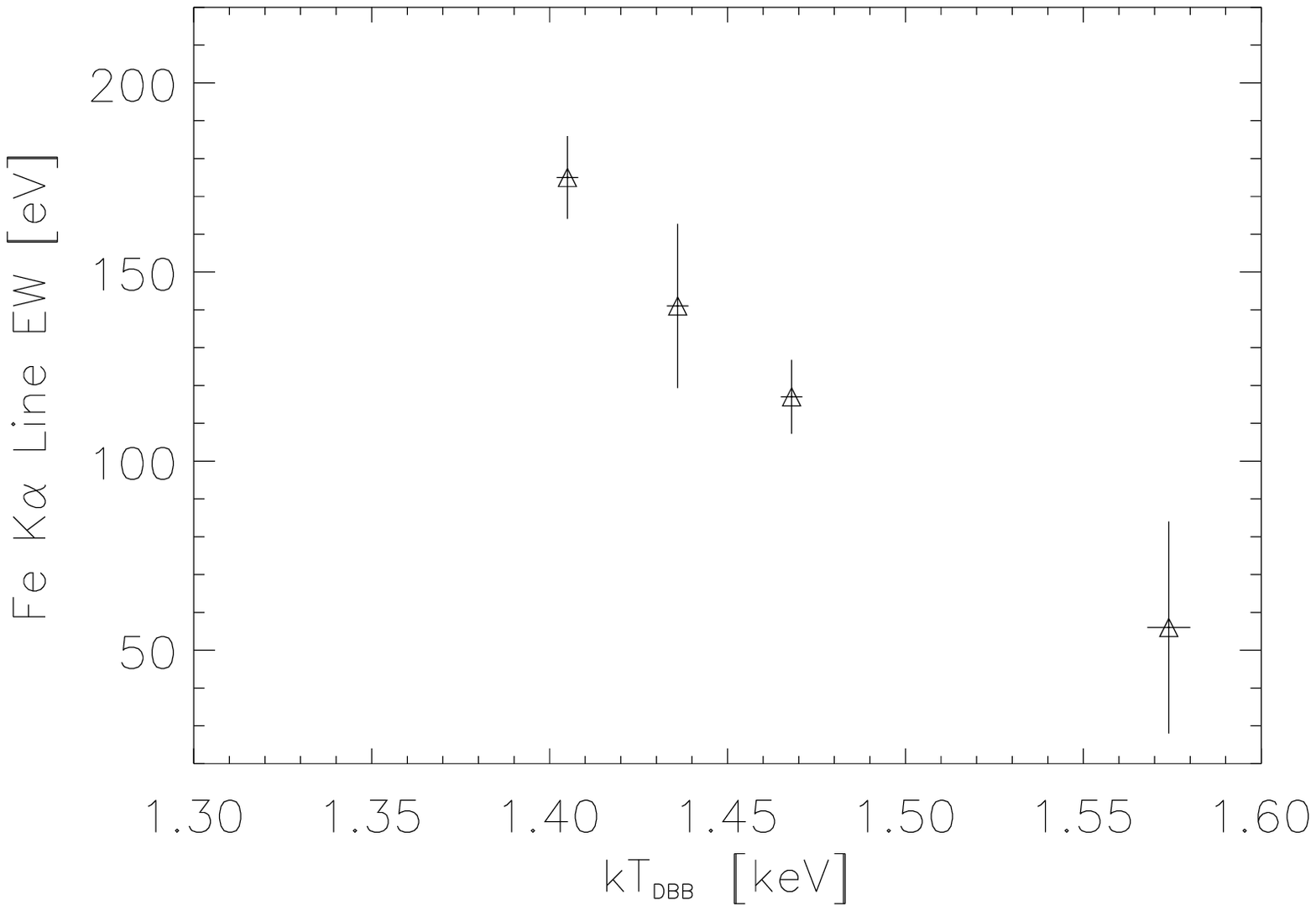}
\includegraphics[angle=0,width=0.24\textwidth]{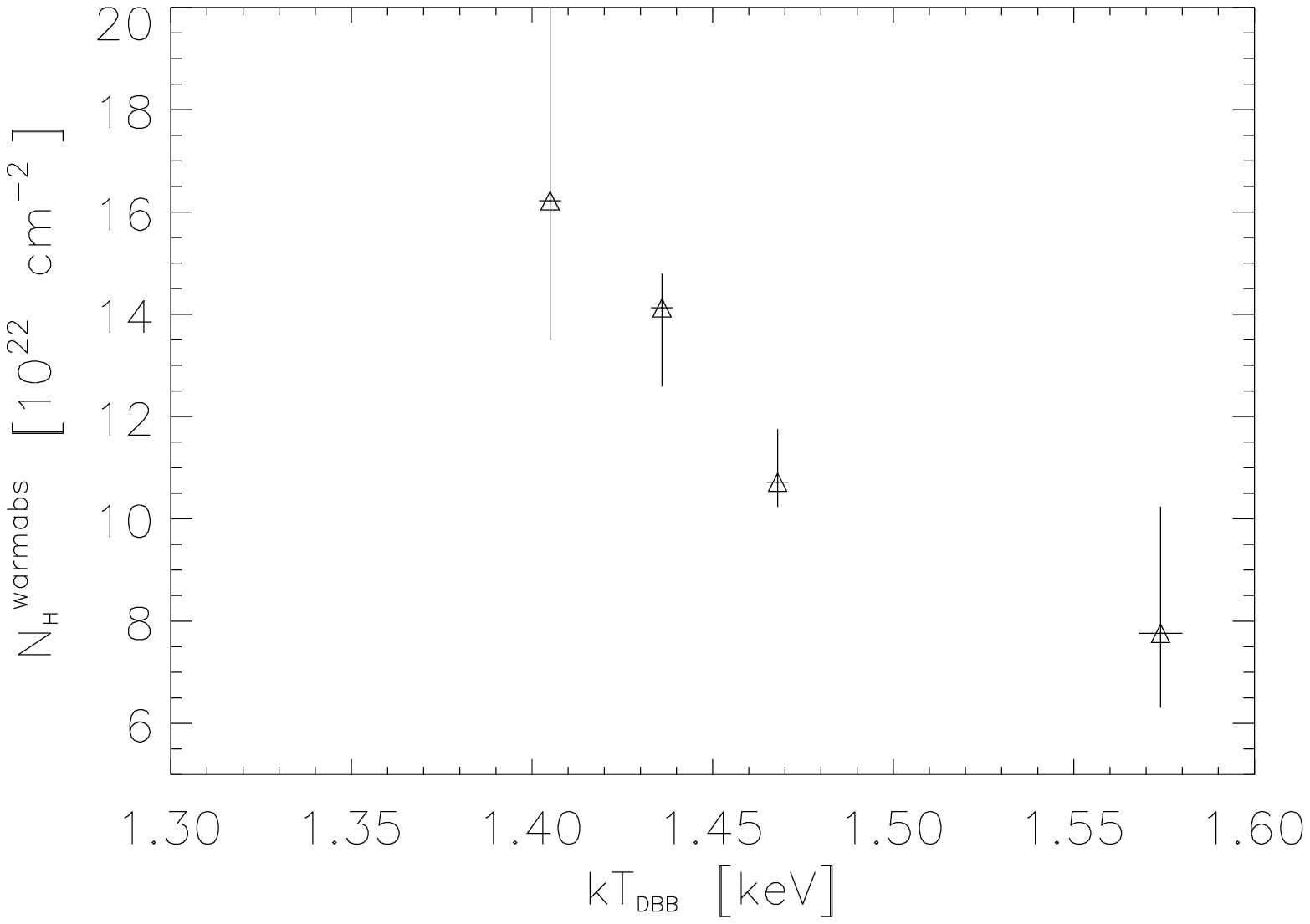}
\includegraphics[angle=0,width=0.24\textwidth]{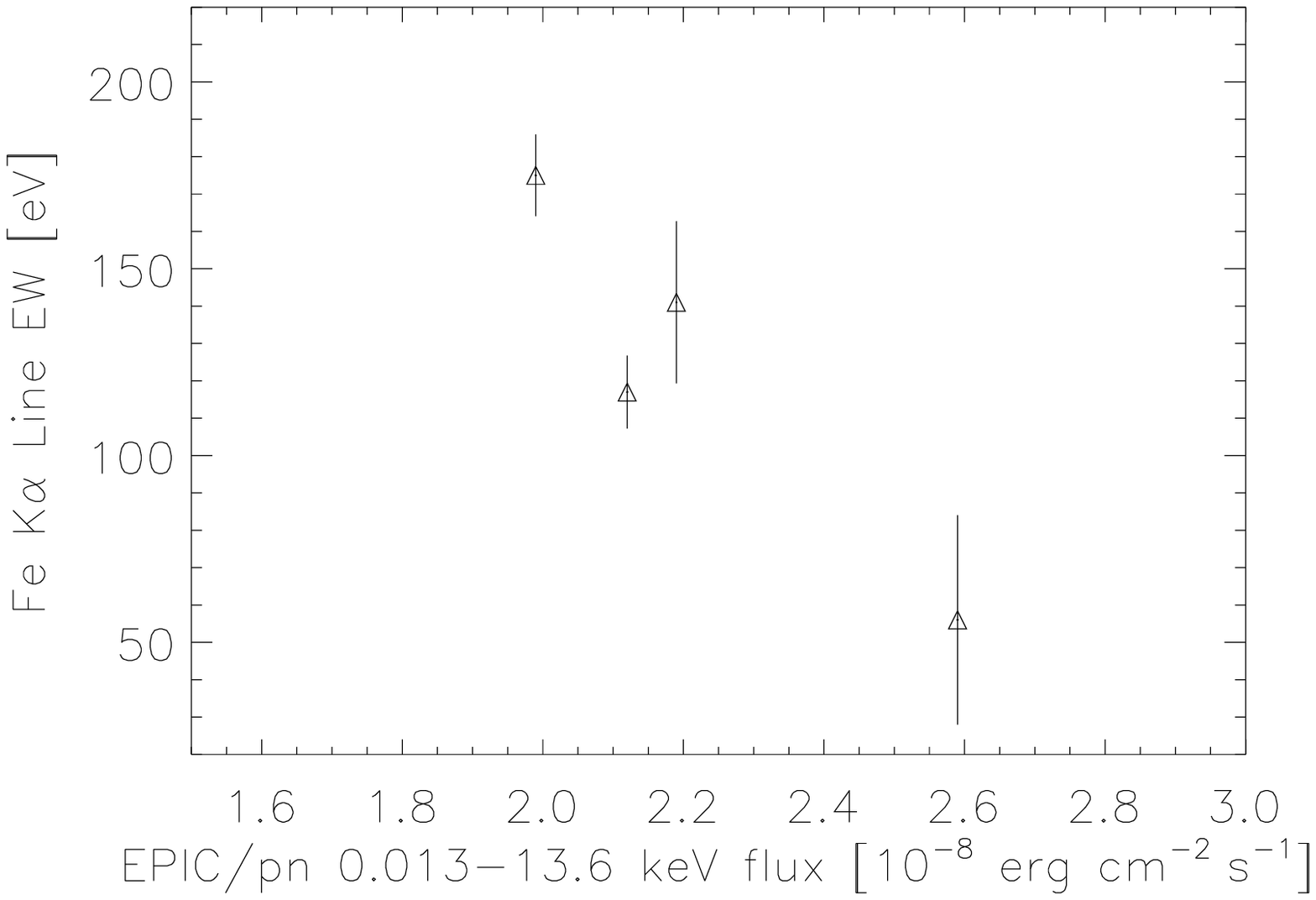}
\includegraphics[angle=0,width=0.24\textwidth]{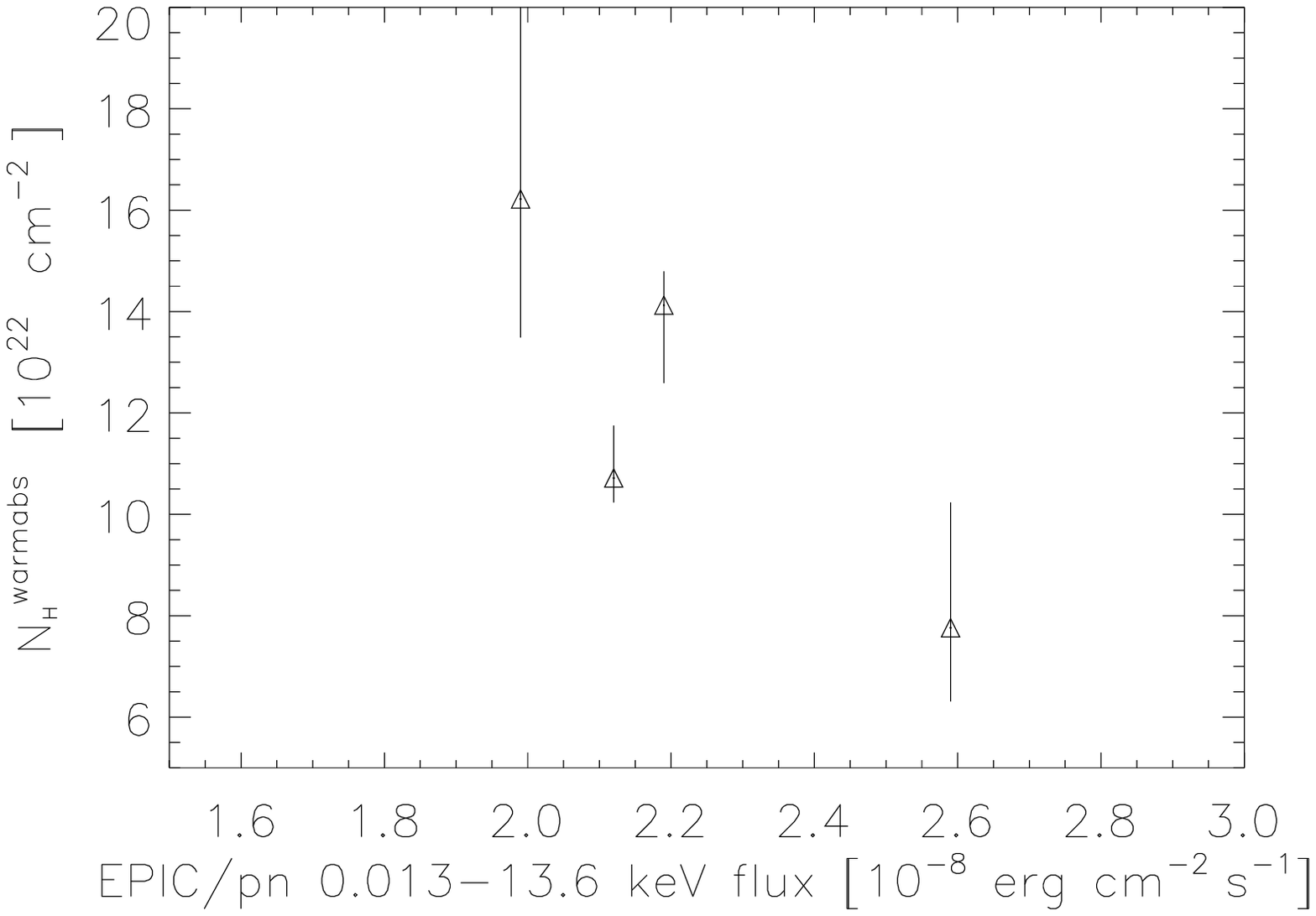}
\caption{From left to right: \ew\ of the broad emission line of Fe \ka\ and column density of the warm absorber with respect to temperature (top) and 0.013-13.6 keV unabsorbed flux of the disc blackbody component (bottom).
} 
\label{fig:line_warm_cont}
\end{figure}

\begin{figure}[!ht]
\includegraphics[angle=0,width=0.24\textwidth]{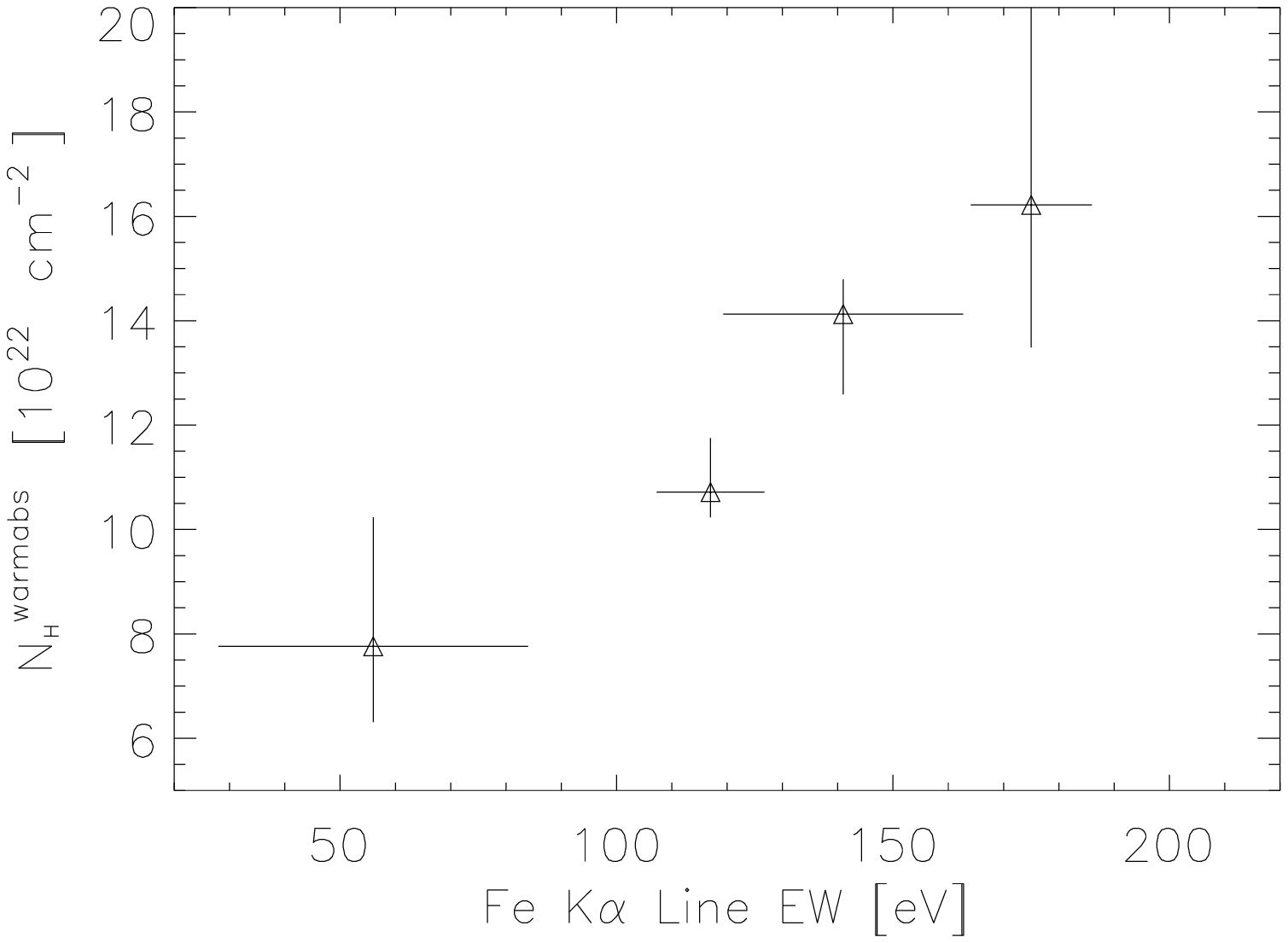}
\includegraphics[angle=0,width=0.24\textwidth]{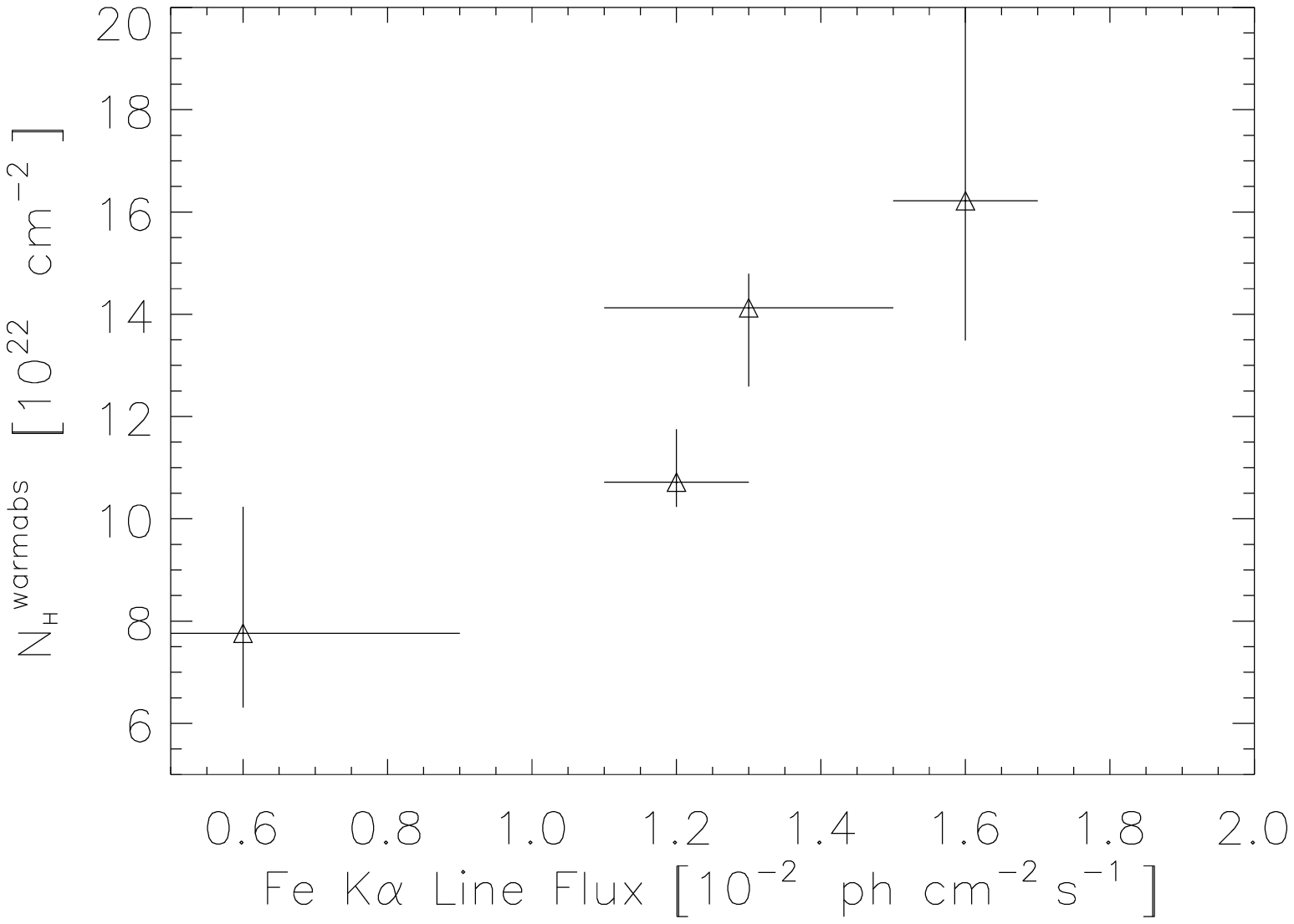}
\caption{Column density of the warm absorber with respect to the \ew\ and flux of the broad Fe emission line. } 
\label{fig:line_warm}
\end{figure}

Fig.~\ref{fig:line_warm_cont} shows the \ew\ of the broad Fe line (left) and the column density of the warm absorber (right) with respect to the temperature of the disc and the 0.013--13.6~keV unabsorbed flux\footnote[8]{We do not include obs~4B in Fig.~\ref{fig:line_warm_cont} and \ref{fig:line_warm} due to the differences between the parameters of obs~4T and 4B possibly arising as a result of calibration differences between timing and burst modes (see previous section).}. Interestingly, both the broad line and the warm absorber show a change correlated to the temperature and flux of the disc. In contrast, this is not the case for the disc radius inferred from the disc normalisation (see Table~\ref{tab:bestfit-warm}). Fig.~\ref{fig:line_warm} shows the column density of the warm absorber with respect to the equivalent width and the flux of the broad Fe line: there seems to be a trend of the \ew\ and flux of the line decreasing as the column density of the warm absorber decreases.

\begin{table*}\begin{center}\caption[]{Best fits to the 2--10 keV EPIC pn persistent spectra for all observations using Model~2 (see text). The column density
  of the {\tt cabs} component was tied to the column density of the photoionised
  absorber (see text). \kdbb\ and
  \kgau\ are the normalizations of the disc
  blackbody component and Gaussian emission feature, respectively.
  \ktdbb\ is the temperature of the disc blackbody component. \egau, $\sigma$ and \ew\ represent the energy, width and equivalent width of
  the Gaussian feature. \nhabs\ and \nhwarmabs\ are the column
  densities for the neutral and ionised absorbers,
  respectively. \xil, \sigmav, $v$ and $S$ are the ionisation parameter
  (in units of erg cm s$^{-1}$), the turbulent velocity broadening, the average systematic velocity shift  (negative
  values indicate blueshifts) and the Sulphur abundance of the absorber in units of solar abundances. $F$ is the unabsorbed
  flux of the disc continuum between 2 and 10~keV.
The width ($\sigma$) of the Gaussian emission line {\tt
 gau} was constrained to be $\le$1~keV in the fits. {\it p} indicates that the error of a parameter pegged at the limit imposed.
}
\begin{tabular}{lcccccc}
\hline \hline\noalign{\smallskip}
Observation No. & & 1 & 2 & 3 & 4T & 4B  \\
\noalign{\smallskip\hrule\smallskip}& Comp. & & & \\
Parameter & & & & \\
& {\tt tbabs} & & & \\
\multicolumn{2}{l}{\nhabs\ {\small($10^{22}$ cm$^{-2}$)}} & 7.70\,$\pm$\,0.02 & 7.67\,$\pm$\,0.02 & 7.80\,$\pm$\,0.02 & 7.81\,$\pm$\,0.03 & 8.17\,$\pm$\,0.07   \\
& {\tt diskbb} & & & \\
\multicolumn{2}{l}{\ktdbb\ {\small(keV)}} & 1.405\,$\pm$\,0.003 & 1.436\,$\pm$\,0.003 & 1.468\,$\pm$\,0.003 & 1.574\,$\pm$\,0.006 & 1.74\,$\pm$\,0.01  \\
\multicolumn{2}{l}{\kdbb\ {\small[(R$_{in}$/D$_{10}$)$^{2}$ cos$\theta$]}} & 237\,$\pm$\,3 & 237\,$\pm$\,3 & 212\,$\pm$\,2 &  196\,$\pm$\,3 & 103\,$\pm$\,3   \\
& {\tt warmabs} & & & \\
\multicolumn{2}{l}{\nhwarmabs\ {\small($10^{22}$ cm$^{-2}$)}}  & 16.2\,$^{+4.7}_{-2.7}$ & 14.1\,$^{+0.7}_{-1.5}$ & 10.7\,$^{+1.0}_{-0.5}$ & 7.8\,$^{+2.5}_{-1.5}$ & 2.6\,$^{+5.0}_{-0.9}$  \\
\multicolumn{2}{l}{\logxi (\xiunit)} & 4.34\,$^{+0.01}_{-0.08}$ & 4.43\,$^{+0.01}_{-0.07}$  & 4.28\,$^{+0.07}_{-0.01}$ & 4.34\,$^{+0.03}_{-0.16}$ & 3.9\,$^{+0.6}_{-0.2}$\\
\multicolumn{2}{l}{S} & 13\,$\pm$\,5 & 15\,$\pm$\,6 & 16\,$\pm$\,6 &  21\,$\pm$\,9 & 19\,$^{+29}_{-15}$ \\
\multicolumn{2}{l}{\sigmav {\small(km s$^{-1}$)}} & 1000\,$^{+115}_{-315}$ & 2580\,$^{+865}_{-660}$ & 1680\,$^{+725}_{-230}$ & 2160\,$^{+1260}_{-990}$ & 1060\,$^{+3085}_{-930}$  \\
\multicolumn{2}{l}{$v$ {\small(km s$^{-1}$)}} & -3690\,$^{+60}_{-30}$ & -2190\,$^{+120}_{-60}$ & -3180\,$^{+180}_{-60}$ & -3420\,$^{+210}_{-360}$ & -3660\,$\pm$\,1500 \\
& {\tt gau$_{e}$} \\
\multicolumn{2}{l}{ \egau\ {\small(keV)}} & 6.40\,$^{+0.05}_{-0p}$ & 6.59\,$\pm$\,0.09 & 6.47\,$^{+0.09}_{-0.07p}$ & 6.97\,$^{+0p}_{-0.08}$ & -- \\
\multicolumn{2}{l}{ $\sigma$ {\small(keV)}} & 1\,$^{+0p}_{-0.04}$ & 1\,$^{+0p}_{-0.07}$ & 1\,$^{+0p}_{-0.06}$ & 0.85\,$^{+0.15p}_{-0.16}$ & -- \\
\multicolumn{2}{l}{ \kgau\ {\small(10$^{-2}$ ph cm$^{-2}$ s$^{-1}$)}} & 1.6\,$\pm$\,0.1 & 1.3\,$\pm$\,0.2 & 1.2\,$\pm$\,0.1 & 0.6\,$\pm$\,0.3 & -- \\
\multicolumn{2}{l}{ \ew\ (eV)} & 175\,$\pm$\,11 & 141\,$\pm$\,22 & 117\,$\pm$\,10 & 56\,$\pm$\,28 & -- \\
& & & & \\
\noalign{\smallskip\hrule\smallskip}
\multicolumn{2}{l}{F$_{2-10 keV}$ {\small(10$^{-8}$ erg cm$^{-2}$ s$^{-1}$)}} & 1.19 & 1.31 & 1.30 & 1.64 & 1.33  \\
\noalign {\smallskip}
\noalign {\smallskip}
\hline\noalign {\smallskip}
\multicolumn{2}{l}{\rchisq (d.o.f.)} & 0.86 (122) &  0.86 (121) & 0.86 (121) & 0.67 (122) & 0.93 (124) \\
\noalign{\smallskip\hrule\smallskip}
\noalign{\smallskip\hrule\smallskip}
\label{tab:bestfit-warm}
\end{tabular}
\end{center}
\end{table*} 

\subsubsection{Photoionised plasma model and reflection component}
\label{sec:model3}

Next we attempted to model the broad Fe line with a more physical model. The contemporaneous evolution of the broad Fe line and the warm absorber (see previous sections) indicates that the broad Fe line could arise as reprocessed emission in the wind, as for the NS LMXB \gx\ \citep{gx13:diaz12aa}.
Unfortunately, models that include self-consistently the reprocessed emission in an absorbing plasma are not yet available for XRBs. Instead, since the reprocessed component of a wind should resemble a disc reflection component \citep[][]{sim10mnras} we substituted the broad Fe line by such a component and evaluated the extent to which the continuum is altered. Based on the results, we can also examine the validity of the interpretation of the broad Fe line as a BH spin indicator in the context of disc reflection models (see Sect.~\ref{sec:broadline}). 
We used the reflection model {\tt rfxconv} \citep{gx339:kolehmainen11mnras}, available as a local model in XSPEC, and convolved it with the relativistic smearing Green's function \citep{laor91apj} {\tt kdblur}. The advantage of this model with respect to other reflection models such as {\tt reflionx} \citep{ross05mnras} or {\tt relxill} \citep{garcia14apj} is that it can be used with any input continuum (as opposed to the power law of all alternative reflection models). Indeed, for the spectra modeled in this paper,  reflection from a power law yields significantly worse fits than reflection from a pure disc blackbody, especially for those observations for which the reflection fraction is large. As an example, the \rchisq\ of the best fit to obs~1 is 2.95 (120) when using  {\tt relxill} to model the broad iron line and only 1.01 (123) when {\tt rfxconv} is used (see Table~\ref{tab:bestfit-warm-rfxconv} below). 

\begin{figure*}[!ht]
\includegraphics[angle=0,width=0.21\textwidth]{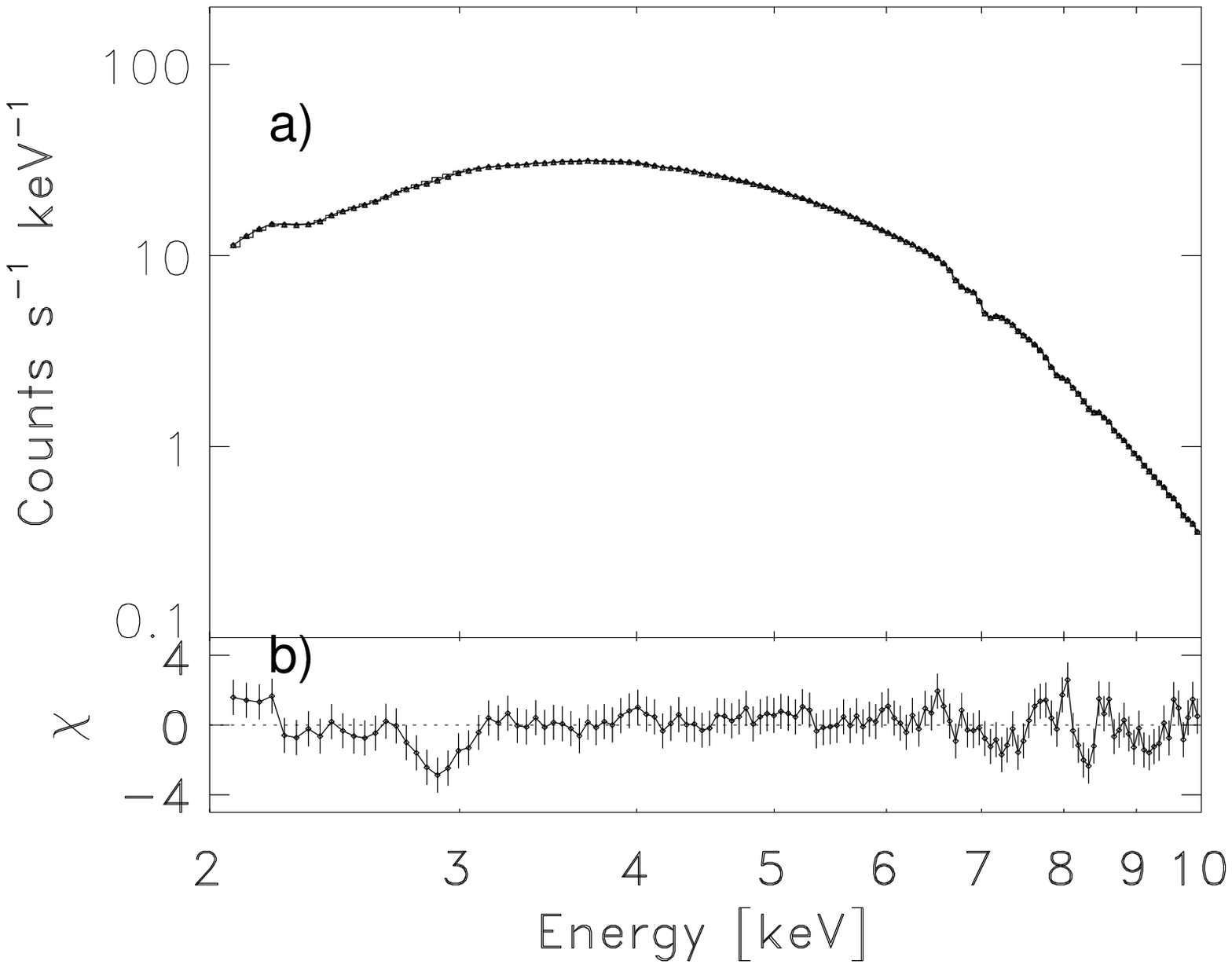}
\hspace{-0.4cm}
\includegraphics[angle=0,width=0.21\textwidth]{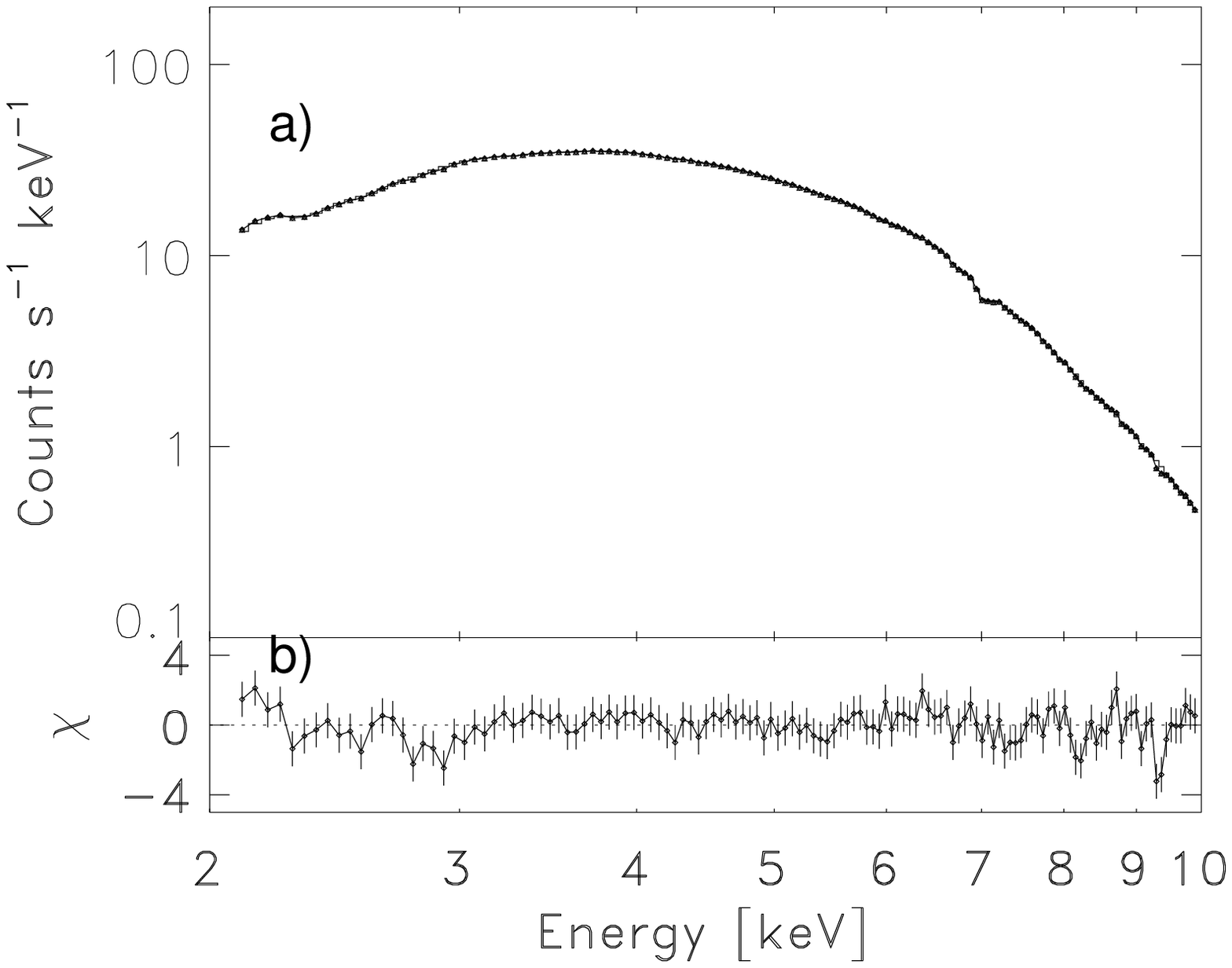}
\hspace{-0.4cm}
\includegraphics[angle=0,width=0.21\textwidth]{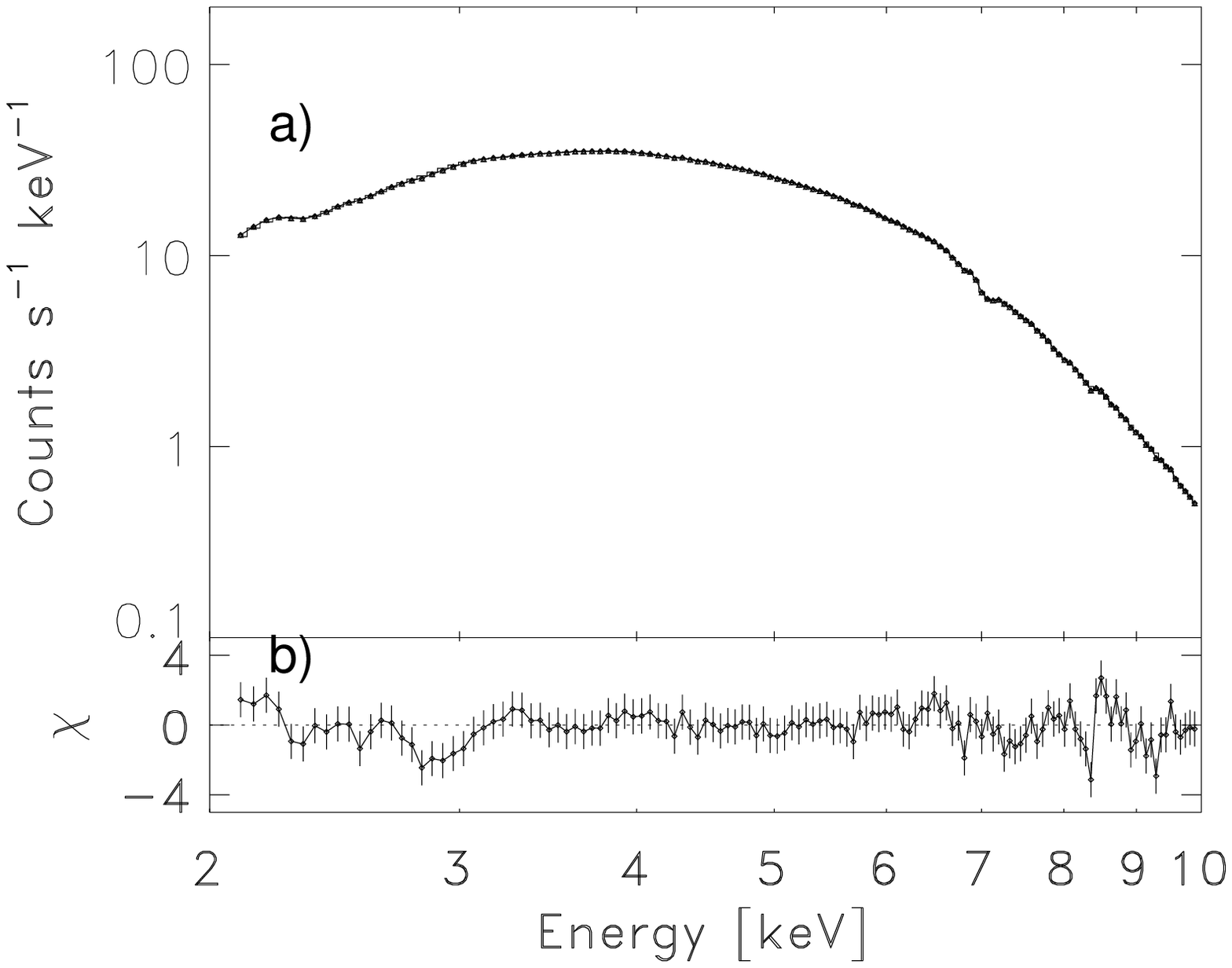}
\hspace{-0.4cm}
\includegraphics[angle=0,width=0.21\textwidth]{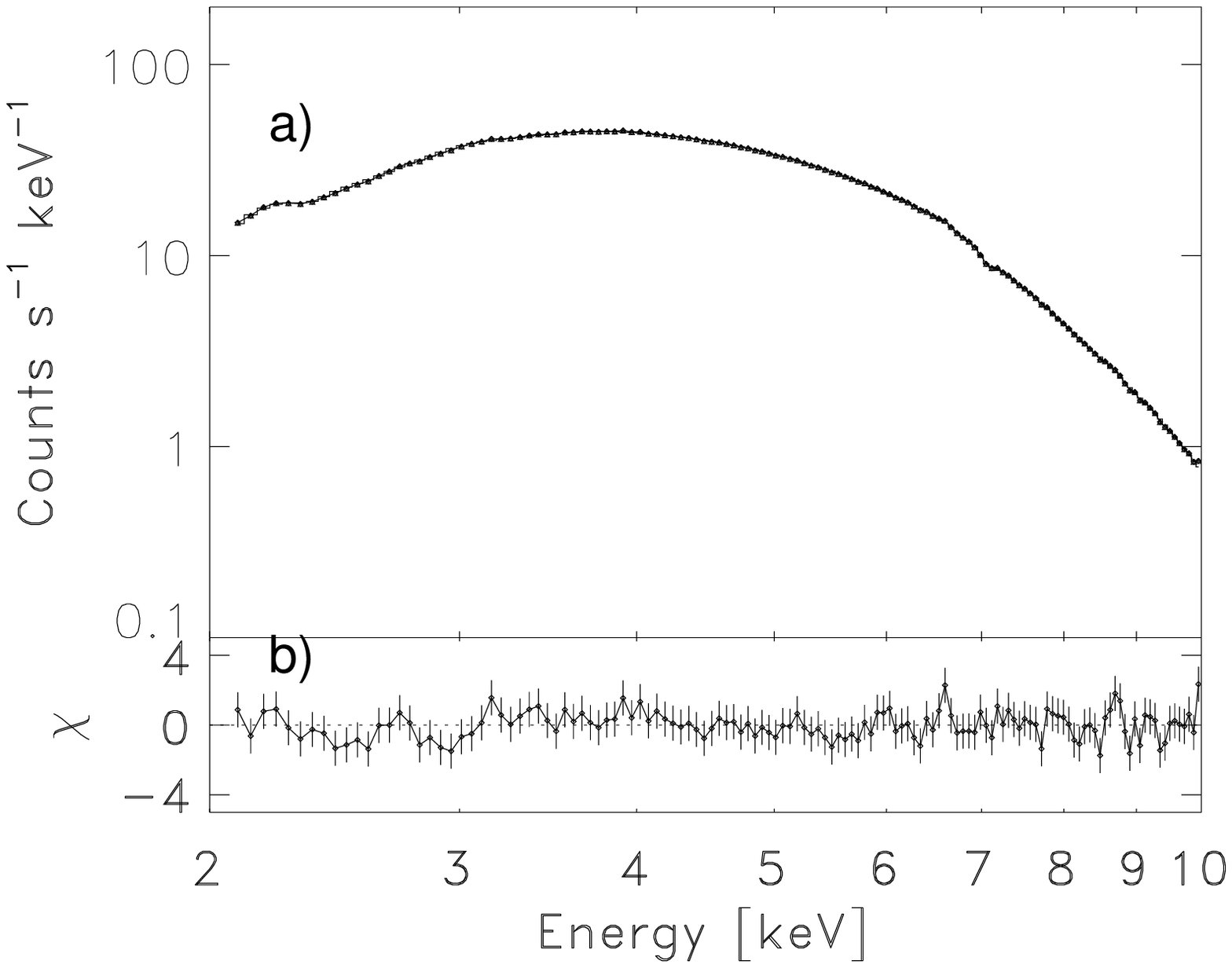}
\hspace{-0.4cm}
\includegraphics[angle=-0,width=0.21\textwidth]{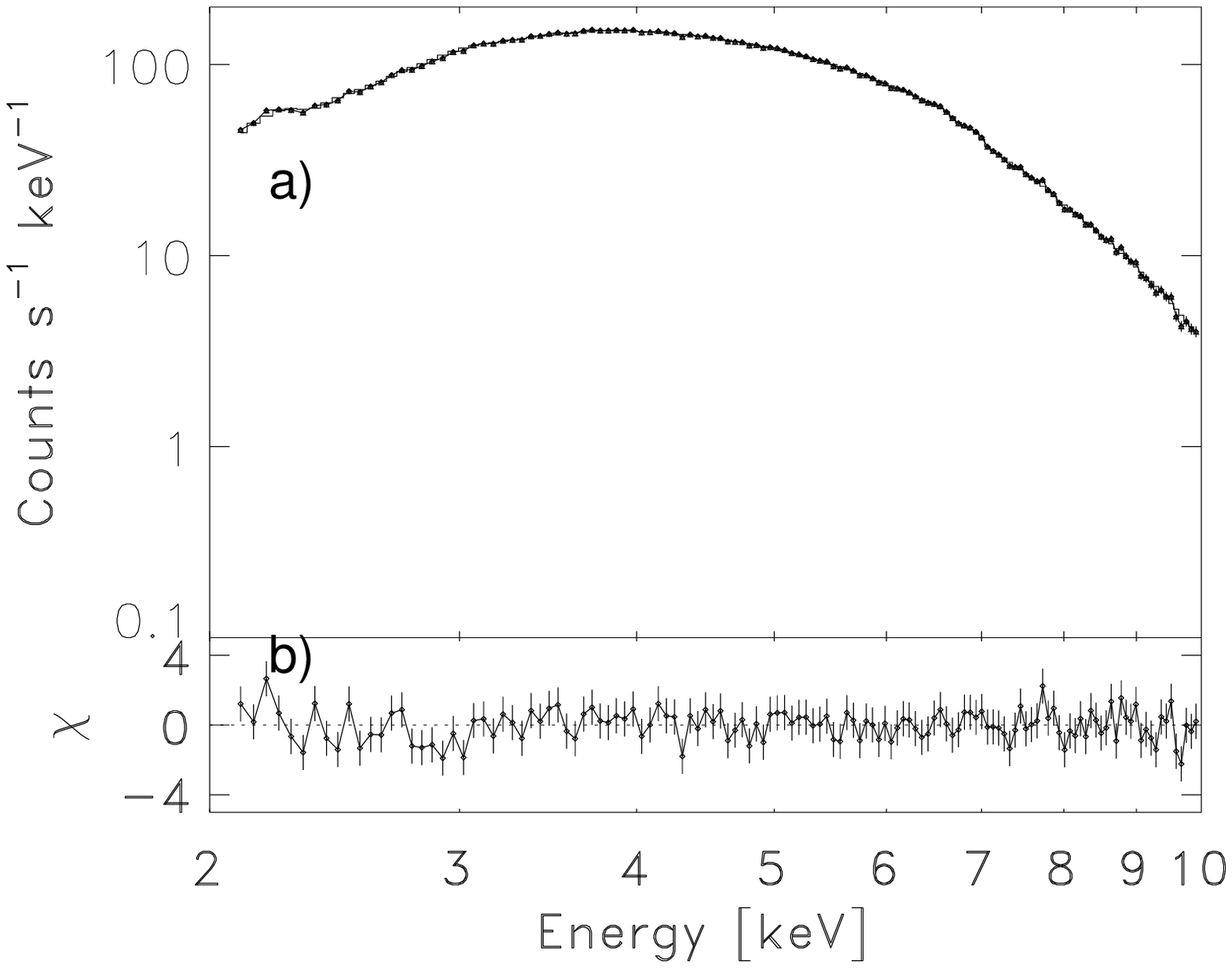}
\caption{Obs~1--4B are shown from left to right.  (a) 2--10 keV EPIC pn \src\ spectra fitted with Model 3 (see text). (b) Residuals in units of standard
deviation from the above model.} 
\label{fig:lddel_mod2}
\end{figure*}

\begin{figure*}[!ht]
\includegraphics[angle=0,width=0.21\textwidth]{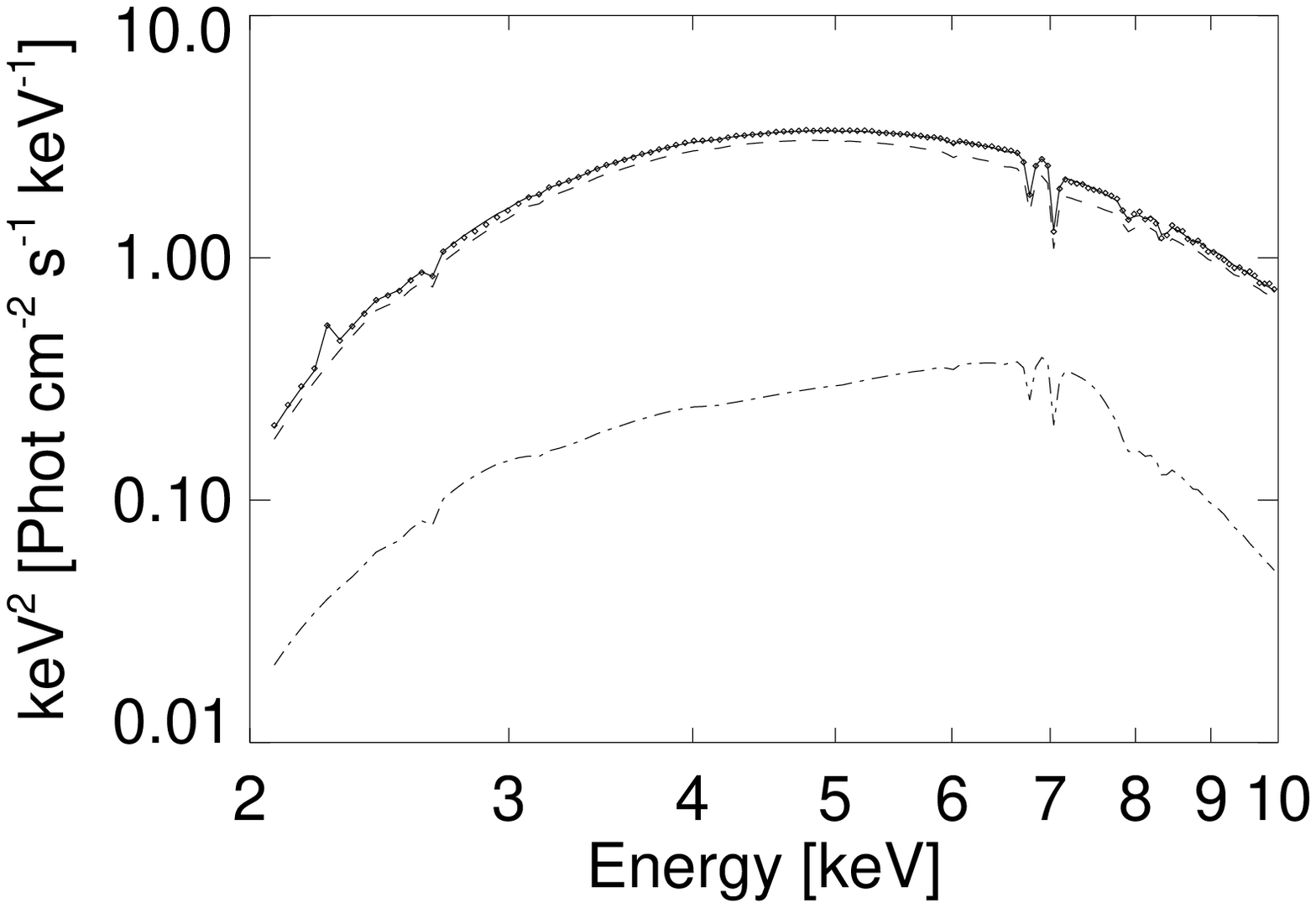}
\hspace{-0.4cm}
\includegraphics[angle=0,width=0.21\textwidth]{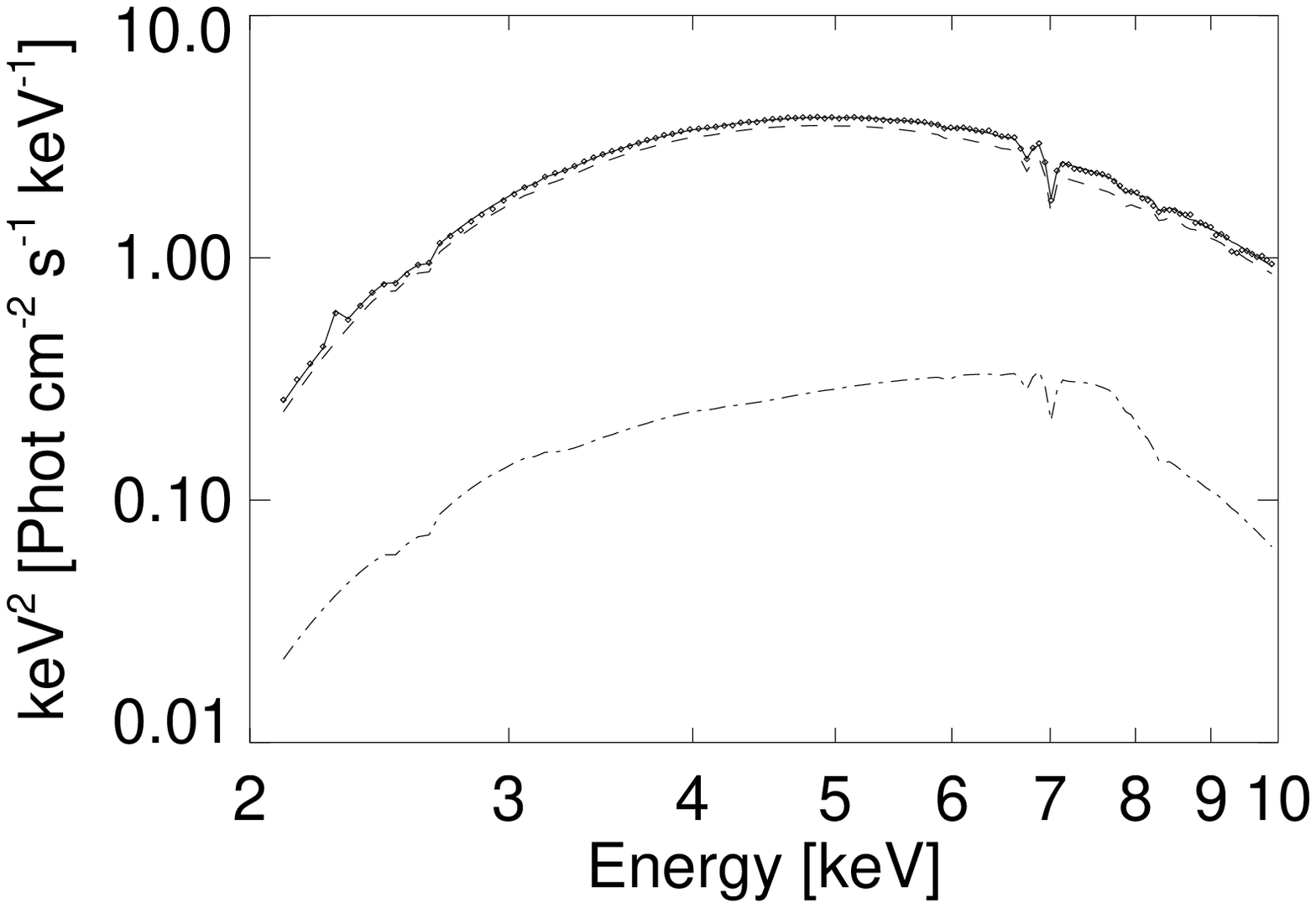}
\hspace{-0.4cm}
\includegraphics[angle=0,width=0.21\textwidth]{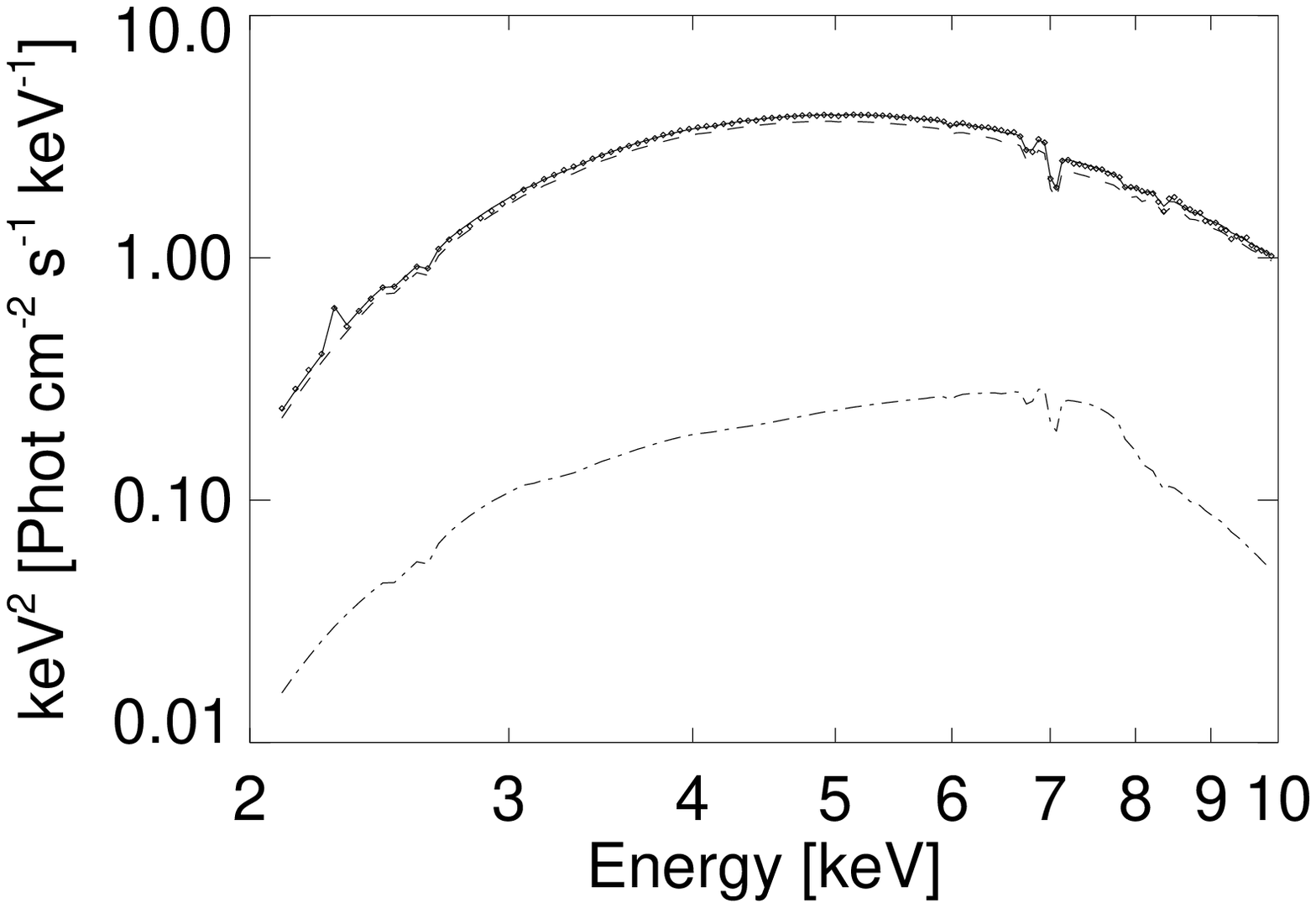}
\hspace{-0.4cm}
\includegraphics[angle=0,width=0.21\textwidth]{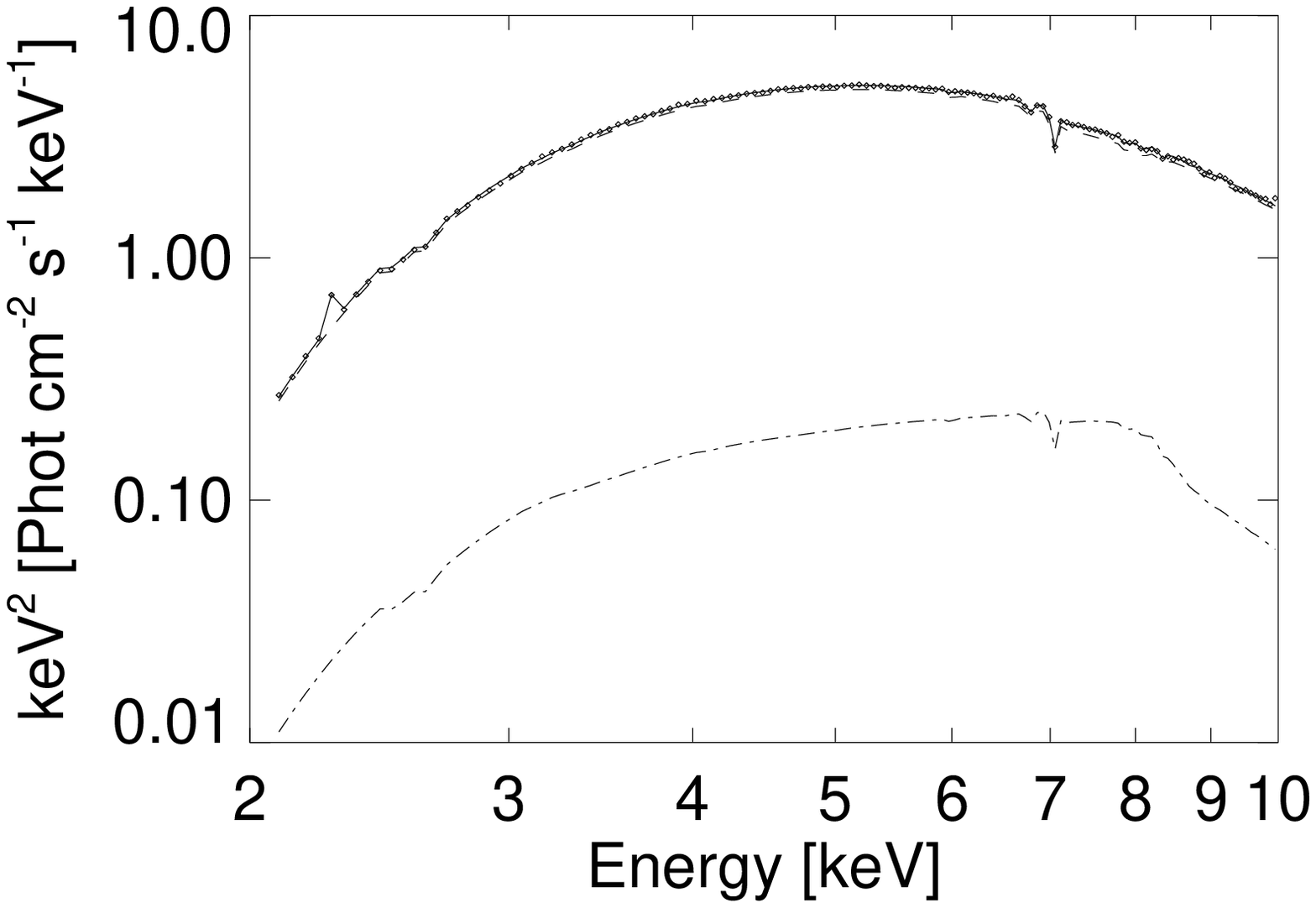}
\hspace{-0.4cm}
\includegraphics[angle=0,width=0.21\textwidth]{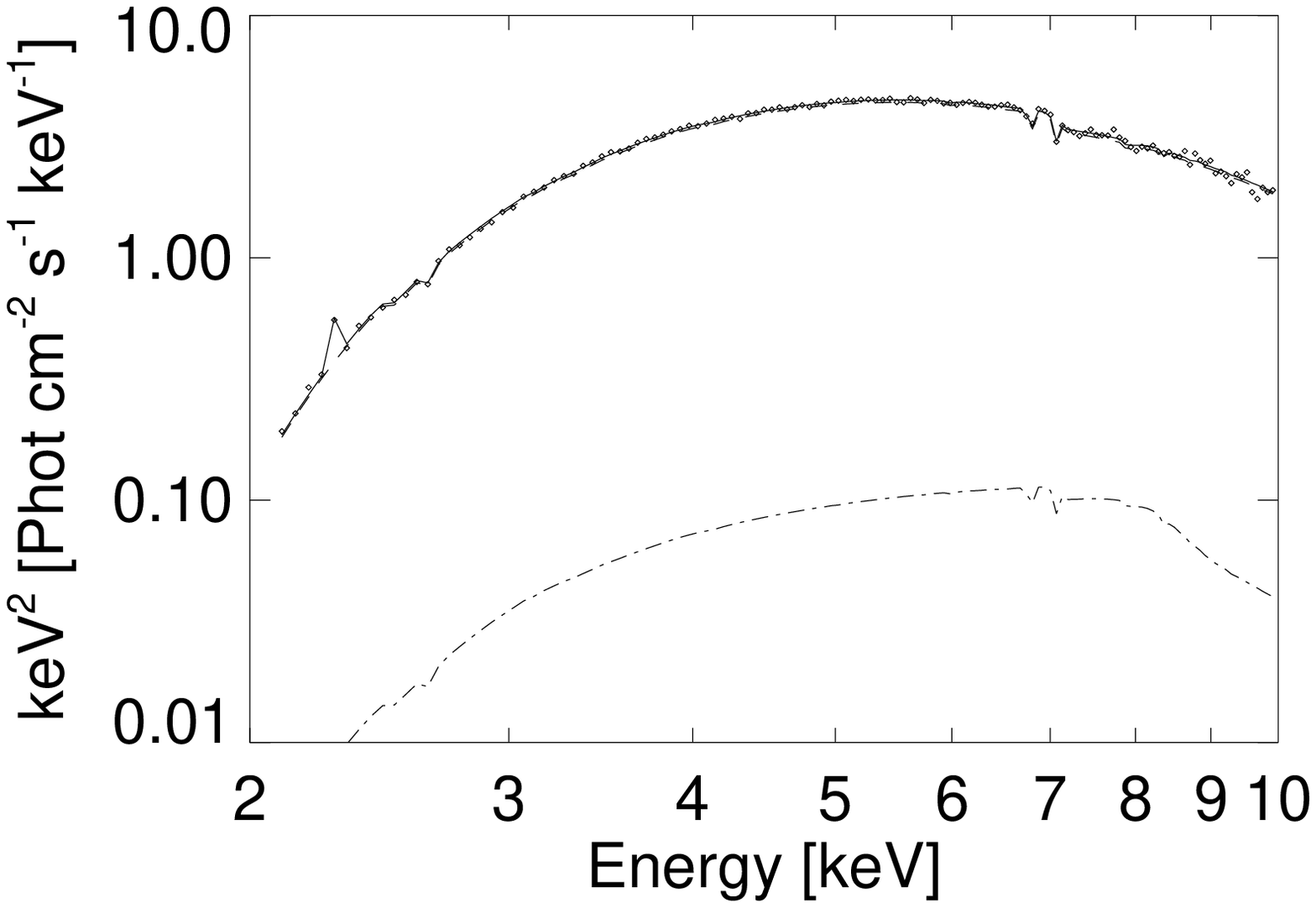}
\caption{2--10~keV unfolded spectra (obs~1--4B are shown from left to right). Solid, dashed and dashed-dotted lines represent
the contribution of the total model, warm absorber and reflection components, respectively. } 
\label{fig:eeuf_mod2}
\end{figure*}

Table~\ref{tab:bestfit-warm-rfxconv} shows the results of the fit to the final model ( {\tt tbabs*cabs*warmabs*(diskbb+kdblur*rfxconv*diskbb)}, hereafter Model~3).  The residuals of the fit and unfolded spectra are shown in Figs.~\ref{fig:lddel_mod2} and \ref{fig:eeuf_mod2}. 
Some residuals are still visible at $\sim$2.9~keV, especially for obs~1 and 3. This is expected, since we could not identify an atomic transition at this energy, and therefore this absorption feature is not accounted for by the warm absorber.  
Although the parameters of the warm absorber are consistent with those obtained with Model~2 within the errors (except for the column density in obs~1), the column density decreases at a lesser rate compared to Model~2. The values of neutral absorption and disc normalisation become also less variable for Model~3. The variation is now absorbed by the reflection component, which is strong in obs~1 and decreases steadily towards obs~4. For the reflection component, an ionisation parameter of \logxi\ $\sim$\,4 gave the best \rchisq\ for all observations. Therefore we fixed this value during the fits. However, we note that this is only a lower limit, since the ionisation parameter of {\tt rfxconv} is based on the tabulated values of {\tt reflionx} and there are no values calculated above 4. For this reason, the evolution of the ionisation of the reflection component cannot be directly observed and is only indirectly inferred from the evolution of the reflection fraction, which should be related to the ionisation of the reflecting material. In addition, we cannot compare directly the ionisation parameter of the reflection component and the warm absorber in Model 3, since different illumination continua are used for both components (for {\tt warmabs} we used the self-consistent disc blackbody ionising continuum while for {\tt rfxconv} the ionisation parameter is obtained from the pre-calculated tables of {\tt reflionx}, which are based on a power law illumination). Finally, the index of the blurring kernel, representing the power law dependence of the emissivity, was largely unconstrained, and therefore we fixed it to the default value of 3 during the fits.

Fig.~\ref{fig:rfl_warm} shows the reflection fraction with respect to the temperature and radius of the disc, the 0.013--13.6~keV unabsorbed flux and column density of the warm absorber\footnote[9]{Similarly to Figs.~\ref{fig:line_warm_cont} and \ref{fig:line_warm}, we do not include obs~4B in these plots due to the differences between the parameters of obs~4T and 4B possibly arising as a result of calibration differences between timing and burst modes.}. The reflection fraction decreases as the disc temperature and flux increase. The reflection fraction also seems to decrease with the column density of the warm absorber, but the errors on the column density are large. In contrast, no correlation is observed between the reflection fraction and the disc radius inferred from the disc normalisation. Similarly, the changes in the disc radius when derived from the disc normalisation and from blurring the reflection component with a relativistic kernel are not correlated. This is discussed in the context of black hole spin studies in Sect.~\ref{sec:broadline}.

Finally, we note that while Model~3 accounts for the blurring of the reflection component, the disc itself is not being blurred. Therefore, we investigated whether the blurring of the disc would make a significant difference to the parameters of the fit. Table~\ref{tab:bestfit-warm-rfxconv2} shows the results of the fit to the final model ( {\tt tbabs*cabs*warmabs*(kdblur*rfxconv*diskbb)}, hereafter Model~4). While the relative changes between the parameters of the warm absorber, the continuum and the reflection component follow the same trends as in Model 3 (see Fig.~\ref{fig:rfl_warm2} as compared to Fig.~\ref{fig:rfl_warm}), we also find some interesting differences.  The value of the radius for blurring the disc and the reflection component is now somewhat larger than for Model~3. In addition, the value of the disc temperature is now more different between obs~4T and 4B, while the difference  between the value of neutral \nhabs\ for those observations has decreased.

\begin{table*}\begin{center}\caption[]{Best fits to the 2--10 keV EPIC pn spectra for observations 1--4 using Model~3 (see caption of Table~\ref{tab:bestfit-warm} for definitions). r$_{in}$ is the inner disc radius at which the reflection component is located (in units of r$_g$) and rel$_{refl}$ the reflection fraction.
The inclination has been fixed to 65\deg\ both for the {\tt kdblur} and {\tt rfxconv} components and the ionisation parameter to \logxi~=~4 for the {\tt rfxconv} component. In addition, the power law emissivity index has been fixed to 3 and the outer radius to 400\,r$_g$ for the {\tt kdblur} component (see text). The flux has been calculated both including (F$^{tot}$) and excluding (F$^{disc}$) the reflection component.
}
\begin{tabular}{lcccccccc}
\hline \hline\noalign{\smallskip}
Observation No. & & 1 & 2 & 3 & 4T & 4B \\
\noalign{\smallskip\hrule\smallskip}& Comp. & & & \\
Parameter & & & & \\
& {\tt tbabs} & & & \\
\multicolumn{2}{l}{\nhabs\ {\small($10^{22}$ cm$^{-2}$)}} & 7.81\,$\pm$\,0.02 & 7.74\,$^{+0.04}_{-0.02}$ & 7.85\,$\pm$\,0.02 & 7.84\,$\pm$\,0.04 & 8.19\,$\pm$\,0.07  \\
& {\tt diskbb} & & & \\
\multicolumn{2}{l}{\ktdbb\ {\small(keV)}} & 1.403\,$\pm$\,0.002 & 1.435\,$\pm$\,0.003  & 1.471\,$\pm$\,0.003 & 1.572\,$\pm$\,0.004 & 1.73\,$\pm$\,0.01 \\
\multicolumn{2}{l}{\kdbb\ {\small[(R$_{in}$/D$_{10}$)$^{2}$ cos$\theta$]}} & 215\,$\pm$\,2 & 218\,$\pm$\,3  & 200\,$\pm$\,1 &  194\,$\pm$\,3 & 103\,$\pm$\,3 \\
& {\tt warmabs} & & & \\
\multicolumn{2}{l}{\nhwarmabs\ {\small($10^{22}$ cm$^{-2}$)}} & 13.5\,$\pm$\,0.6 & 12.0\,$^{+1.2}_{-1.0}$ & 11.0\,$\pm$\,0.8 & 9.3\,$^{+1.2}_{-2.7}$ & 3.4\,$^{+11.0}_{-1.5}$ \\
\multicolumn{2}{l}{\logxi (\xiunit)} & 4.27\,$^{+0.07}_{-0.01}$ & 4.37\,$^{+0.09}_{-0.01}$ & 4.33\,$^{+0.02}_{-0.08}$ &  4.40\,$^{+0.10}_{-0.15}$ & 4.02\,$^{+0.39}_{-0.13}$ \\
\multicolumn{2}{l}{S} & 15\,$\pm$\,5 & 16\,$\pm$\,7 & 16\,$\pm$\,6 & 21\,$^{+12}_{-10}$ & 25\,$^{+32}_{-20}$ \\
\multicolumn{2}{l}{\sigmav {\small(km s$^{-1}$)}} & 1565\,$^{+415}_{-175}$ & 2400\,$^{+770}_{-730}$ & 1880\,$^{+815}_{-400}$ &  2035\,$^{+1495}_{-1155}$ & 720\,$^{+4250}_{-520}$ \\
\multicolumn{2}{l}{$v$ {\small(km s$^{-1}$)}} & -3570\,$\pm$\,30 & -2100\,$\pm$\,120 & -3480\,$^{+30}_{-90}$ & -3480\,$^{+390}_{-570}$  & -3870\,$^{+1200}_{-1800}$ \\
&  {\tt kdblur} \\ 
 \multicolumn{2}{l}{$r_{in}$ ($r_g$)} & 19\,$^{+6}_{-3}$ & 11\,$\pm$\,2 & 14\,$\pm$\,3 &  $<$\,5 & $<$\,17 \\
& {\tt rfxconv} \\
\multicolumn{2}{l}{$rel_{refl}$} & 0.60\,$\pm$\,0.05 & 0.44\,$\pm$\,0.05 & 0.30\,$\pm$\,0.04 & 0.15\,$\pm$\,0.05 & 0.08\,$^{+0.15}_{-0.10}$ \\
& \\
\noalign{\smallskip\hrule\smallskip}
\multicolumn{2}{l}{F$^{disc}_{2-10 keV}$ {\small(10$^{-8}$ erg cm$^{-2}$ s$^{-1}$)}} & 1.07 & 1.20 & 1.23 & 1.60 & 1.30 \\
& \\
\multicolumn{2}{l}{F$^{tot}_{2-10 keV}$ {\small(10$^{-8}$ erg cm$^{-2}$ s$^{-1}$)}} & 1.18 & 1.31 & 1.32 &  1.67 & 1.34 \\
\noalign {\smallskip}
\noalign {\smallskip}
\hline\noalign {\smallskip}
\multicolumn{2}{l}{\rchisq (d.o.f.)} & 1.01 (123) &  0.87 (122) & 0.95 (122) & 0.70 (123) & 0.93 (122) \\
\noalign{\smallskip\hrule\smallskip}
\noalign{\smallskip\hrule\smallskip}
\label{tab:bestfit-warm-rfxconv}
\end{tabular}
\end{center}
\end{table*} 

\begin{table*}\begin{center}\caption[]{Best fits to the 2--10 keV EPIC pn spectra for observations 1--4 using Model~4 (see caption of Table~\ref{tab:bestfit-warm-rfxconv} for definitions).}
\begin{tabular}{lcccccccc}
\hline \hline\noalign{\smallskip}
Observation No. & & 1 & 2 & 3 & 4T & 4B \\
\noalign{\smallskip\hrule\smallskip}& Comp. & & & \\
Parameter & & & & \\
& {\tt tbabs} & & & \\
\multicolumn{2}{l}{\nhabs\ {\small($10^{22}$ cm$^{-2}$)}} & 7.87\,$\pm$\,0.02 & 7.89\,$\pm$\,0.04 & 7.92\,$\pm$\,0.03 & 7.95\,$\pm$\,0.03 & 8.18\,$^{+0.07}_{-0.04}$  \\
& {\tt diskbb} & & & \\
\multicolumn{2}{l}{\ktdbb\ {\small(keV)}} & 1.378\,$\pm$\,0.002 & 1.381\,$\pm$\,0.009  & 1.440\,$^{+0.003}_{-0.006}$ & 1.51\,$^{+0.01}_{-0.03}$ & 1.74\,$\pm$\,0.01 \\
\multicolumn{2}{l}{\kdbb\ {\small[(R$_{in}$/D$_{10}$)$^{2}$ cos$\theta$]}} & 226\,$\pm$\,3 & 256\,$\pm$\,6  & 217\,$^{+2}_{-6}$ &  222\,$^{+11}_{-5}$ & 104\,$\pm$\,3 \\
& {\tt warmabs} & & & \\
\multicolumn{2}{l}{\nhwarmabs\ {\small($10^{22}$ cm$^{-2}$)}} & 11.2\,$^{+0.3}_{-0.7}$ & 12.6\,$\pm$\,0.9 & 11.2\,$^{+1.7}_{-0.7}$ & 7.8\,$^{+1.8}_{-2.1}$ & 3.3\,$^{+11.5}_{-1.5}$ \\
\multicolumn{2}{l}{\logxi (\xiunit)} & 4.24\,$^{+0.05}_{-0.07}$ & 4.37\,$^{+0.05}_{-0.03}$ & 4.31\,$\pm$\,0.04 &  4.28\,$^{+0.12}_{-0.07}$ & 4.0\,$^{+0.4}_{-0.2}$ \\
\multicolumn{2}{l}{S} & 16\,$\pm$\,5 & 18\,$^{+7}_{-4}$ & 17\,$\pm$\,6 & 23\,$\pm$\,10 & 22\,$^{+37}_{-18}$ \\
\multicolumn{2}{l}{\sigmav {\small(km s$^{-1}$)}} & 2160\,$^{+430}_{-595}$ & 2540\,$^{+810}_{-870}$ & 2280\,$^{+700}_{-615}$ &  2055\,$^{+1500}_{-900}$ & 700\,$^{+2470}_{-560}$ \\
\multicolumn{2}{l}{$v$ {\small(km s$^{-1}$)}} & -3660\,$\pm$\,60 & -2130\,$^{+90}_{-30}$ & -3720\,$\pm$\,30 & -3750\,$^{+510}_{-240}$  & -3900\,$\pm$\,1500 \\
&  {\tt kdblur} \\ 
 \multicolumn{2}{l}{$r_{in}$ ($r_g$)} & 29\,$^{+2}_{-6}$ & 13\,$\pm$\,2 & 24\,$\pm$\,3 & 12\,$\pm$\,3  & $>$\,25 \\
& {\tt rfxconv} \\
\multicolumn{2}{l}{$rel_{refl}$} & 0.73\,$^{+0.06}_{-0.04}$ & 0.66\,$\pm$\,0.07 & 0.39\,$\pm$\,0.05 & 0.21\,$\pm$\,0.04 & $<$\,0.07 \\
& \\
\noalign{\smallskip\hrule\smallskip}
\multicolumn{2}{l}{F$^{disc}_{2-10 keV}$ {\small(10$^{-8}$ erg cm$^{-2}$ s$^{-1}$)}} & 1.04 & 1.18 & 1.22 & 1.56 & 1.33 \\
& \\
\multicolumn{2}{l}{F$^{tot}_{2-10 keV}$ {\small(10$^{-8}$ erg cm$^{-2}$ s$^{-1}$)}} & 1.17 & 1.34 & 1.32 &  1.67 & 1.34 \\
\noalign {\smallskip}
\noalign {\smallskip}
\hline\noalign {\smallskip}
\multicolumn{2}{l}{\rchisq (d.o.f.)} & 1.20 (123) &  1.05 (122) & 1.13 (122) & 0.75 (123) & 0.94 (122) \\
\noalign{\smallskip\hrule\smallskip}
\noalign{\smallskip\hrule\smallskip}
\label{tab:bestfit-warm-rfxconv2}
\end{tabular}
\end{center}
\end{table*} 

\subsection{RGS spectral analysis}
\label{subsec:RGS}

We examined the 1.4--2.0~keV first order RGS spectra to constrain the \nh\ in the direction of the source and to
search for the signature of narrow absorption and emission features.  We note that the second order spectra have
too few counts to add any value to the fits.

We could fit the RGS spectra of all the observations with a continuum consisting of a disc blackbody modified by
photo-electric absorption from neutral material. We obtained a C-statistic of between 
170 and 440 for $\sim$162--330 d.o.f. for obs~1 to 4. The values of \nh\ in units of 10$^{22}$\,cm$^{-2}$ were 
7.4\,$\pm$\,0.4, 7.8\,$^{+0.8}_{-0.6}$, 7.3\,$^{+0.6}_{-0.2}$ and 7.2\,$^{+0.3}_{-0.5}$ for obs~1 to 4, respectively.
None of the observations showed significant narrow features. 

In summary, we obtained consistent values of \nh\ for all the observations within the errors. However, the errors are larger than when using the pn data alone. This is expected due to the combination of limited energy band of the RGS and low statistics of these observations below 2~keV due to the high interstellar absorption of this source. 
Therefore, we conclude that the RGS spectra alone cannot constrain the value of \nh\ for these observations due to the poor statistics and the limited bandwidth, necessary to determine the continuum.

\section{Discussion}
\label{sec:discussion}

The black hole XRB \src\ entered a new outburst in December 2011 (Nakahira et al. 2011, ATel \#3830). This unusually long outburst lasted for about two years. We triggered six observations with \xmm\ in March and September 2012 to investigate the variability of the disc wind with accretion state and the potential connection between the presence of winds and jets in this source.

The first five observations can be fitted with a continuum consistent with a disc blackbody. The disc temperature and 2--10~keV luminosity increase progressively from obs~1 to 5. Obs~6 shows an additional continuum component, which can be fitted with a power law of index $\Gamma$~$\sim$~2, although the index is very poorly constrained. The appearance of this component is consistent with the re-brightening in hard X-rays reported by INTEGRAL (Bodaghee et al. 2012, ATel \#4360) and confirmed by MAXI/ASM and Swift/BAT. 

\citet{1630:abe05pasj} studied the accretion states of \src\ in five outbursts from 1996 to 2004. They identified three different states during the high state, which have also been identified during outbursts of \gro\ \citep{1655:kubota01apjl} and \xtefifteen\ \citep{1550:kubota04apj}. The first one is consistent with the classical HSS and the concept of a standard Shakura-Sunyaev accretion disc, which follows the $L \propto T^4$ relation at a constant disc inner radius. The second state is characterised by a dominant hard component and a disc radius apparently becoming very small. This ``anomalous'' state is also called ``very high'' state (VHS) or ``steep power-law'' state (SPL), due to its steeper power-law of $\Gamma$~=~2.5--3. 
In this state, the disc structure is still standard, but the ambient hot electron corona that comptonises the disc emission is formed and evolved. The high temperatures and unphysically small inner radii of the disc are then attributed to electron scattering opacity in the inner regions of the accretion disc \citep{1655:kubota01apjl,1550:kubota04apj}. Optically thin radio flares are associated with this state in \gro, \xtefifteen\ and \src, pointing to a link between the comptonising plasma and the radio emission \citep{1630:abe05pasj}. 
The third state follows a $L \propto T^2$ relation, consistent with an optically--thick and advection--dominated slim disc. A transition from the first to the second or third states occurs when the total X-ray luminosity rises above a given critical luminosity, but it is not clear what triggers the transition to one state or the other.

\begin{figure}[!ht]
\includegraphics[angle=0,width=0.24\textwidth]{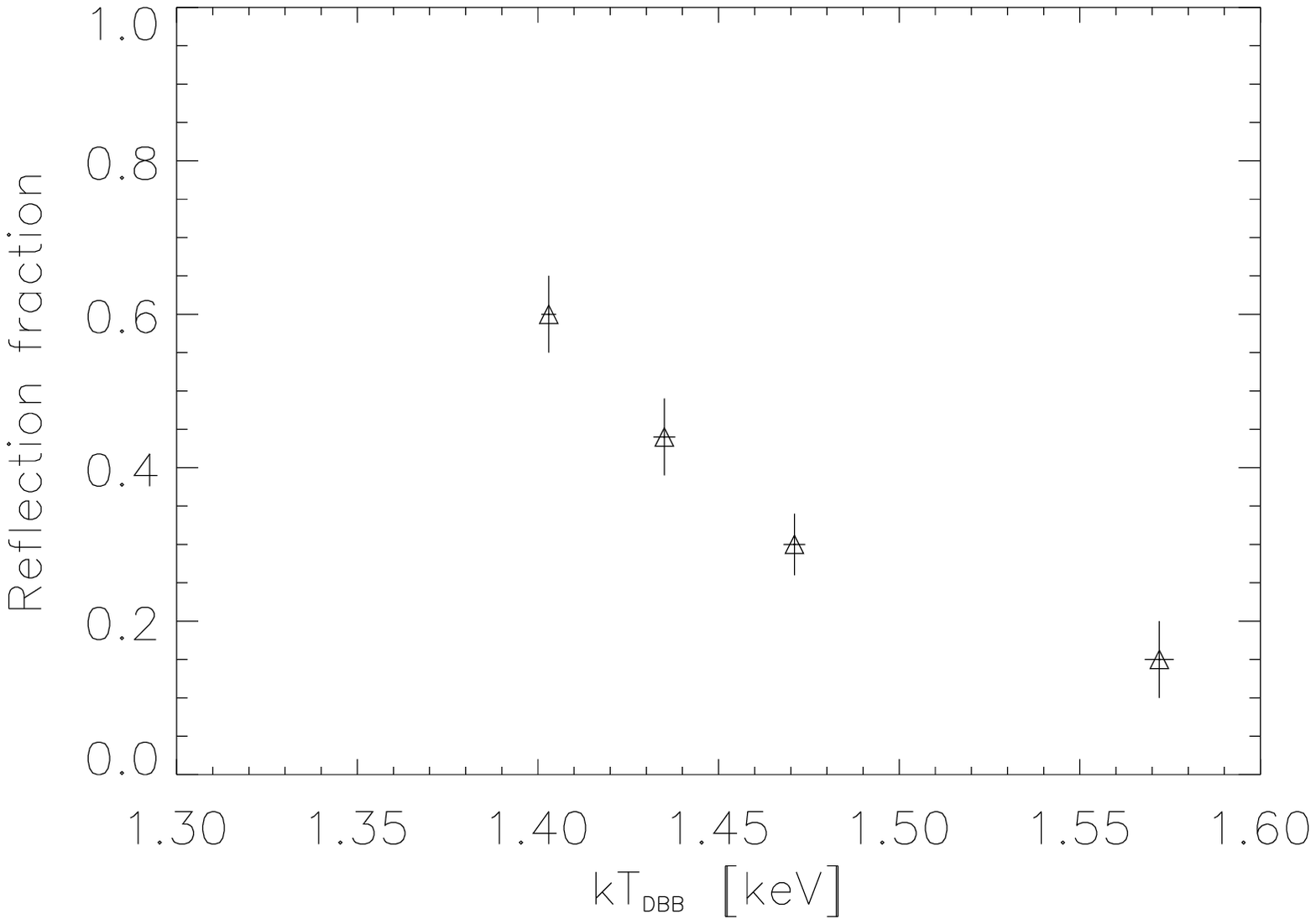}
\includegraphics[angle=0,width=0.24\textwidth]{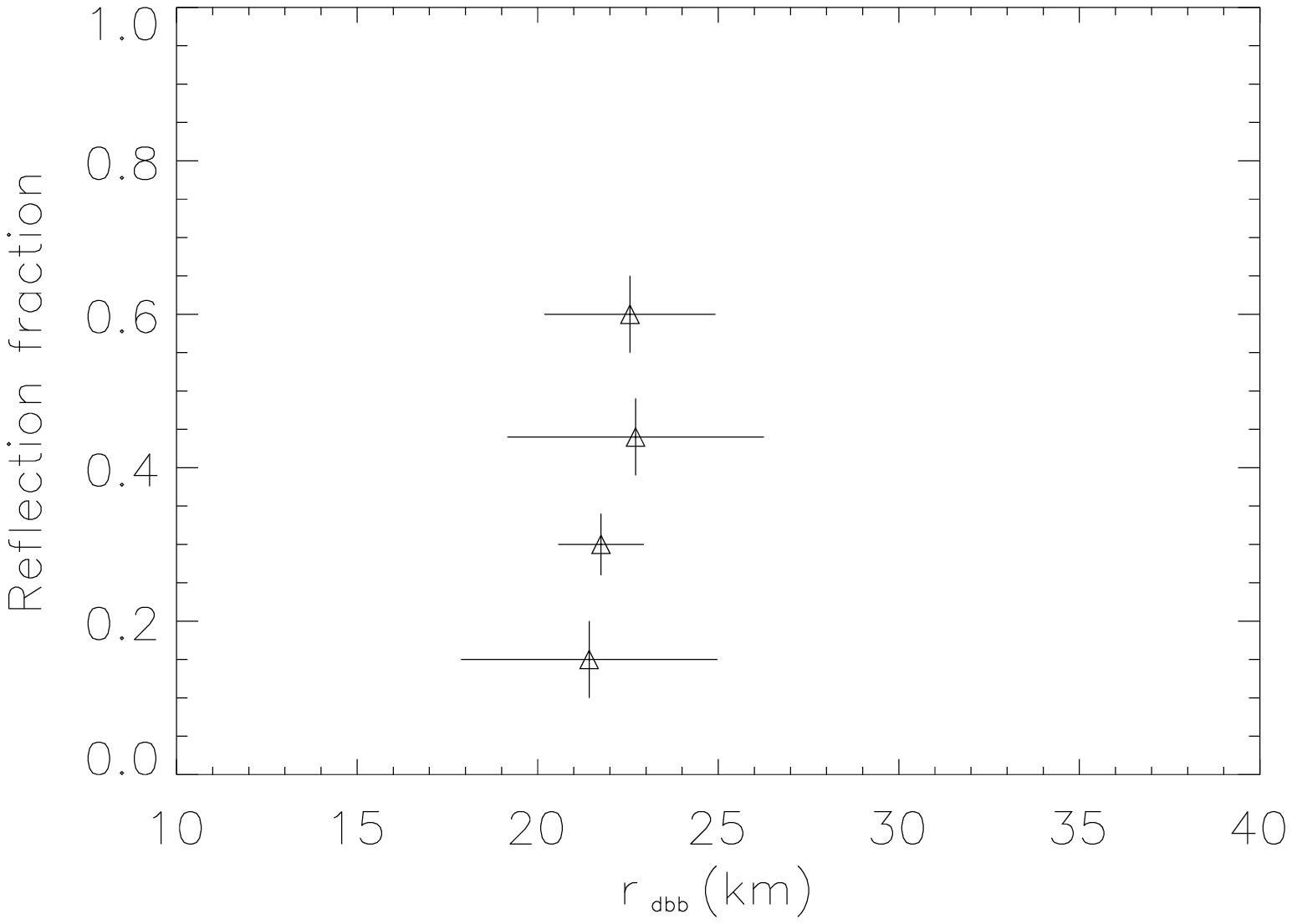}
\includegraphics[angle=0,width=0.24\textwidth]{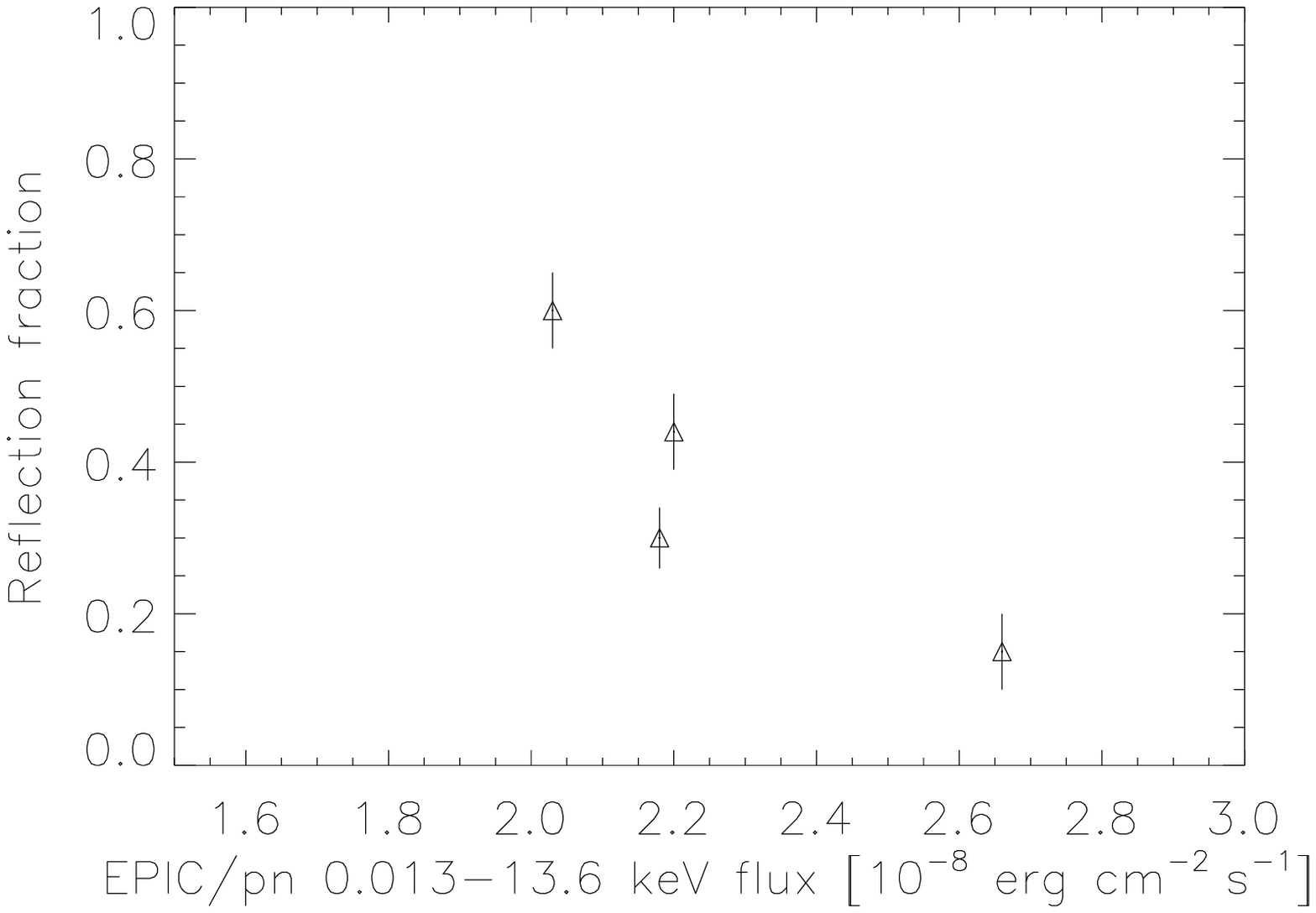}
\includegraphics[angle=0,width=0.24\textwidth]{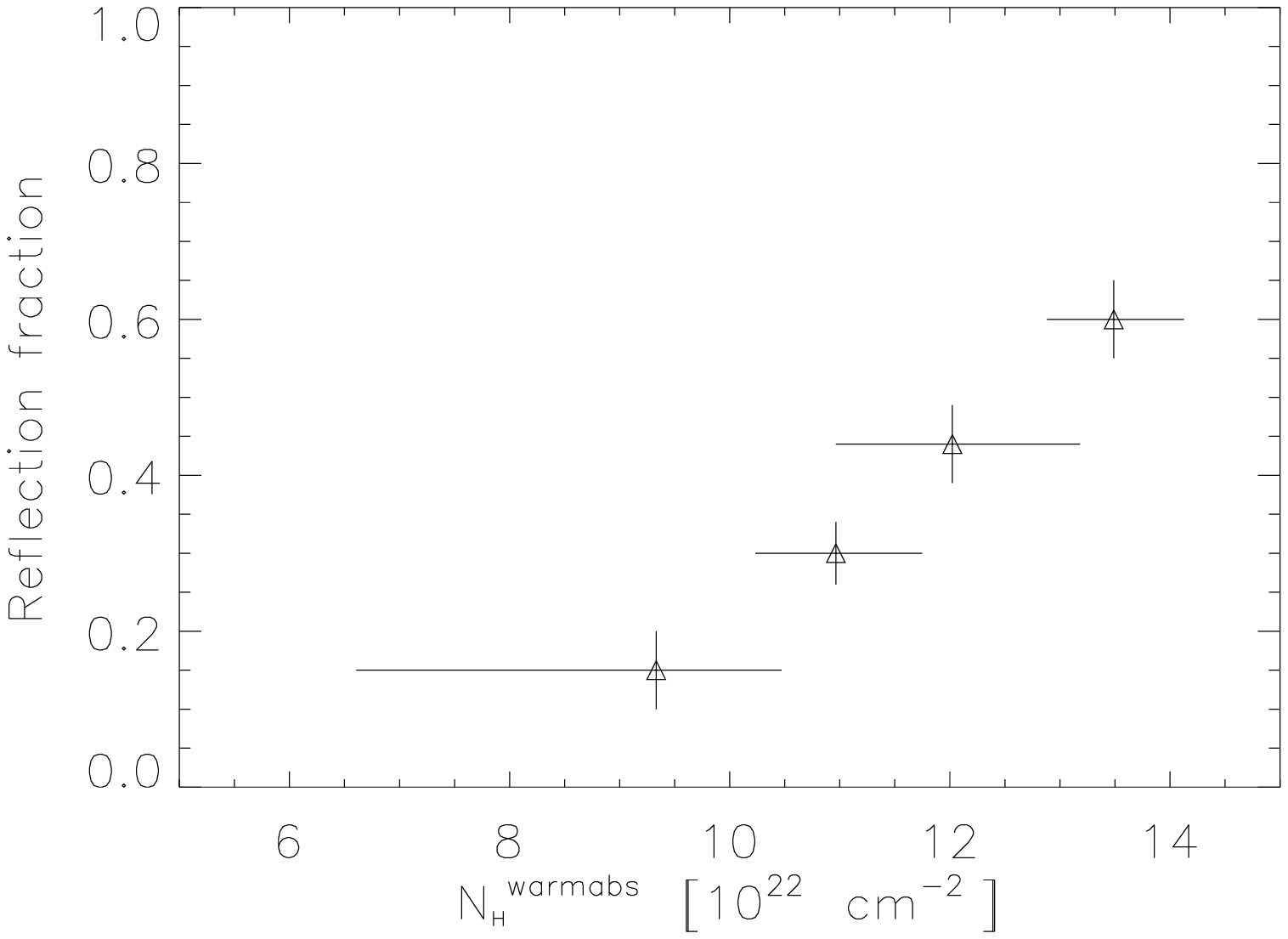}
\caption{Reflection fraction with respect to the temperature (upper left), inner radius (upper right) and 0.013-13.6 keV unabsorbed flux (lower left) of the disc blackbody component and column density of the warm absorber (lower right), for Model 3.} 
\label{fig:rfl_warm}
\end{figure}

\begin{figure}[!ht]
\includegraphics[angle=0,width=0.24\textwidth]{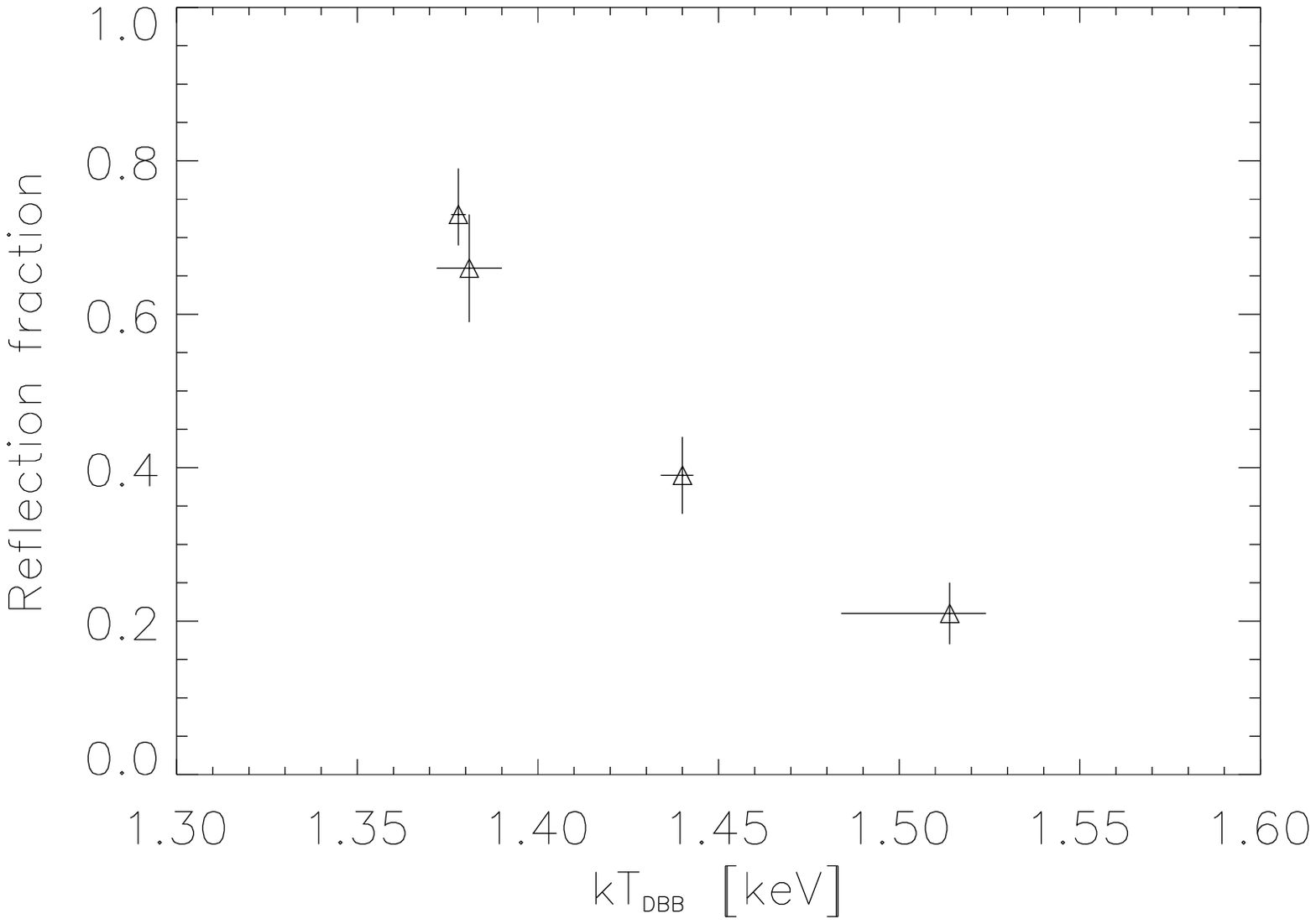}
\includegraphics[angle=0,width=0.24\textwidth]{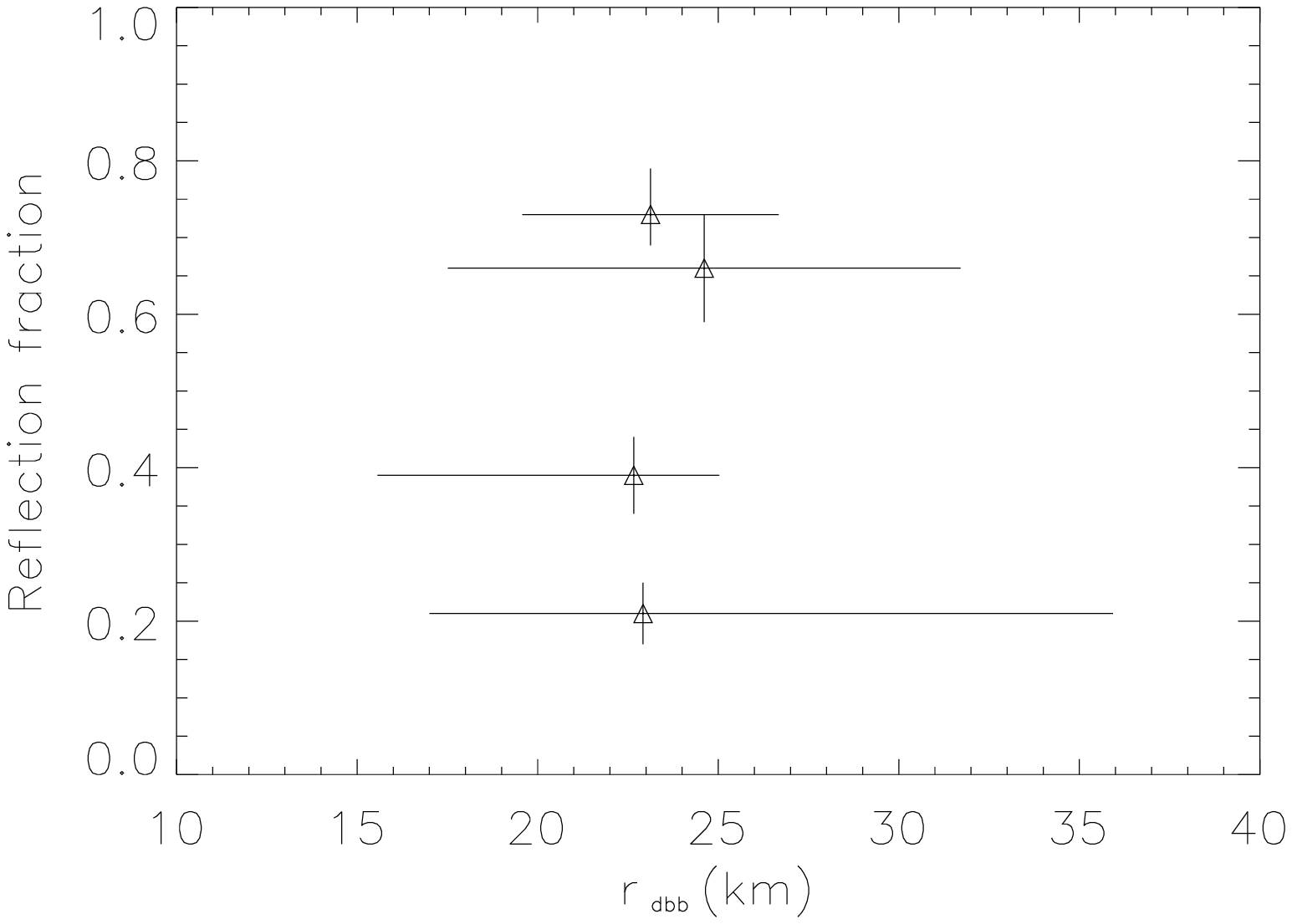}
\includegraphics[angle=0,width=0.24\textwidth]{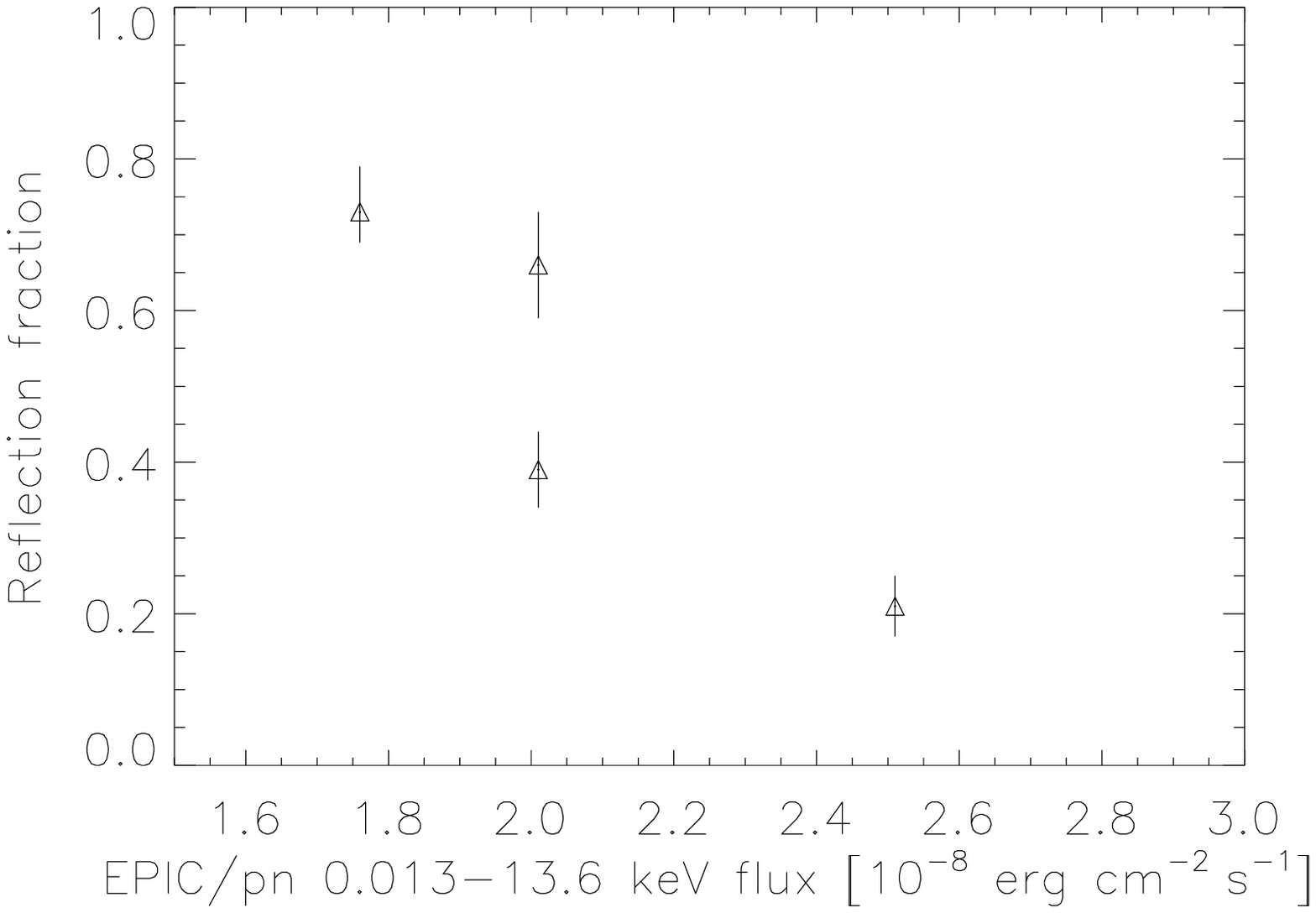}
\includegraphics[angle=0,width=0.24\textwidth]{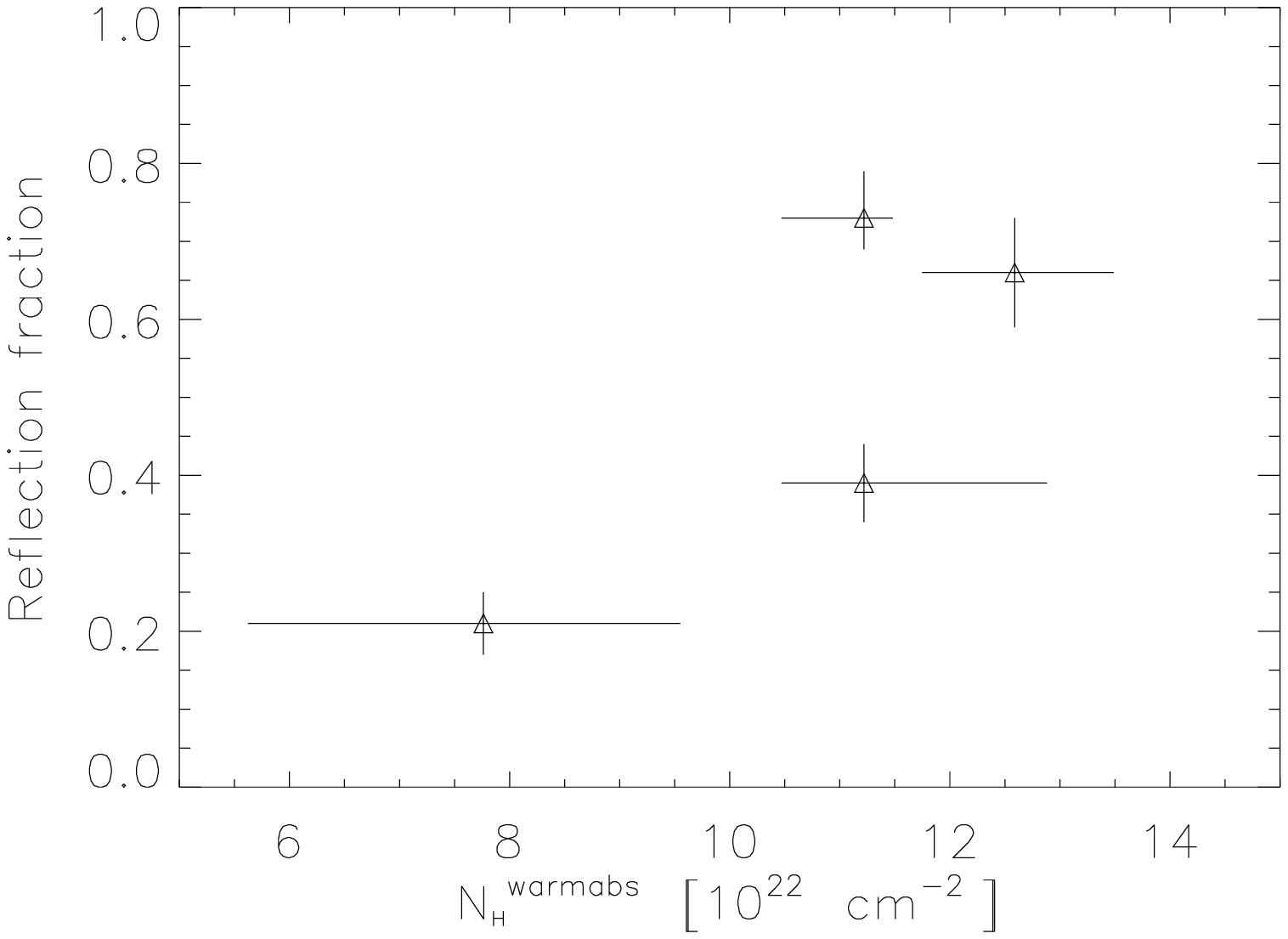}
\caption{Reflection fraction with respect to the temperature (upper left), inner radius (upper right) and 0.013-13.6 keV unabsorbed flux (lower left) of the disc blackbody component and column density of the warm absorber (lower right), for Model 4.}
\label{fig:rfl_warm2}
\end{figure}

For \src, a critical luminosity of 2.5--3\,$\times$\,10$^{38}$~erg/s for a distance of 10~kpc is derived \citep{1630:abe05pasj, 1630:tomsick05apj}, corresponding to an Eddington fraction of 0.7-0.8 (0.2-0.3) for a 3 (10) $\Msun$ black hole. Interestingly, the value of the total X-ray luminosity in obs~5 is $\sim$~2.9\,$\times$\,10$^{38}$~erg/s and increases in obs~6 to $\sim$~4.9\,$\times$\,10$^{38}$~erg/s, when a continuum of disc blackbody and power law is considered, and to $\sim$~3.8\,$\times$\,10$^{38}$~erg/s when an alternative continuum that physically models the cutoff at lower energies such as {\tt simpl*diskbb} is used.  

Therefore, we identify obs~1--5 with a HSS and interpret the spectral changes in obs~6 as the transition from a HSS to the ``anomalous'' state, described by \citet{1630:abe05pasj}. The high variability and spectral characteristics of obs~6 allow also its classification in the ``flaring intermediate''  state described by \citet{1630:tomsick05apj} or the ``anomalous'' state of \citet{belloni10lnp}. We note that it is difficult to obtain conclusive results about the absolute disc temperature at which the transition occurs, since the absolute value could be affected by calibration uncertainties. However, we can firmly conclude the existence of a transition as signaled by changes between observations, such as an increase of hard X-ray luminosity, disc temperature and variability. Since we are comparing observations of one source and taken with the same instrument, residual calibration effects or properties of the source such as inclination should not affect our conclusions. 

We observe significant changes in the spectra related to the state transition: firstly, the disc wind becomes invisible in obs~5, likely due to its very high ionisation after the wind has become weaker and more ionised from obs~1 to obs~4 (see Sect.~\ref{sec:wind-hss}). Secondly, in obs~6 we detect for the first time optically thin radio emission, that we interpreted as the appearance of a jet. Optically thin radio emission was also detected in previous observations during the ``anomalous'' state in \gro\ \citep{1655:hunstead97iau}, \xtefifteen\ \citep{1550:wu02apj,1550:kubota04mnras} and \src\  \citep{1630:hjellming99apj}, supporting the identification of the state transition.  

We discuss extensively the properties of the wind in obs~1--4 and the changes leading to its disappearance in obs~5 in Sect.~\ref{subsec:wind} below. 
For a discussion of the emission lines that appear in obs~6 and that we interpret as Doppler-shifted lines from a jet we refer the reader to \citet{1630:diaz13nat}. In Sect.~\ref{subsec:wind-jet} we present a possible scenario which relates the accretion state with the presence of a wind or a jet.
We also discuss the possibility that the radio emission arises in the interaction of a jet with the ISM instead of in an optically thin jet \citep{1630:neilsen14apjl}.

\subsection{The disc wind}
\label{subsec:wind}

Narrow absorption features consistent with \fetfive\ and \fetsix\  were first discovered in \src\ by $\suzaku$ \citep{1630:kubota07pasj}. The lines were blueshifted with velocities of $\sim$\,1000~km/s, consistent with the presence of an outflowing disc wind. Column densities of 7--10$\times$10$^{22}$~cm$^{-2}$ and an ionisation parameter  of \logxi\,$\sim$\,4.60--4.78 were determined. A plasma density of $\sim$\,10$^{12}$~cm$^{-3}$ and a launching radius for the wind of 1--5$\times$10$^{10}$~cm were inferred. The $\suzaku$ observations showed evolution of the wind, a marginal decrease in column density and ionisation parameter, as the luminosity decreased. 

We found that the disc wind reported for the $\suzaku$ observations is also present in obs~1--4. Both the $\suzaku$ and \xmm\ observations 1--4 were performed during the HSS of accretion and are characterised by a disc with high temperature ($\sim$1.2--1.4~keV and 1.4--1.6~keV, respectively) and luminosity (1.5--2.8\,$\times$\,10$^{38}$ and 2--3\,$\times$\,10$^{38}$ erg/s, respectively). We found column densities of $\sim$9--14\,$\times$\,10$^{22}$~cm$^{-2}$ and an ionisation parameter  of \logxi\,$\sim$\,4.27--4.40 in the \xmm\ observations (see Table~\ref{tab:bestfit-warm-rfxconv}). Based on the \ew\ of the detected features, we infer that the ionisation state of the wind is similar in both $\suzaku$ and \xmm\ observations. In particular $\suzaku$ obs~1 is similar to \xmm\ obs~4 and $\suzaku$ obs~2--6 to \xmm\ obs~2--3, while \xmm\ obs~1 has a slightly larger column density and lower ionisation than any other observation. The reason for the differences found in the ionisation parameter is probably the use of a different ionising continuum for the warm absorber. Small differences in the SED or bolometric flux, likely caused by calibration differences between $\suzaku$ and \xmm, will yield different ionisation parameters even if the \ew\ of the absorption lines is similar.  Therefore, we limit our discussion to the relative changes between observations of the same instrument, which can be reliably compared since any systematic effect caused by the instrument calibration or our choice of continuum, abundances or spectral analysis should be the same for all observations.

We obtained line blueshifts of $\sim$\,2100--3600~\kms\ for obs~1--4. Such blueshifts seem too large, compared to the shift of $\sim$1000~\kms\ \citep{1630:kubota07pasj} reported for \src\ and of $\sim$300--1600~km~s$^{-1}$ in other BH LMXBs \citep[e.g.][]{1655:kallman09apj,1915:ueda09apj} \citep[see however][for a recently reported blueshift of 3100\,$\pm$\,400 km/s in a $Chandra$ HETGS observation of IGR~J17480--2446]{igrj17480:miller11apj}.
Unfortunately, we do not detect any discrete absorption features from the warm absorber in the RGS, which has a much better energy resolution than EPIC. The lack of spectral
features in the RGS wavelength range is consistent with the high degree of ionisation of the absorber and the high interstellar absorption in the direction of the source. Thus, a more accurate
measurement of the blueshifts is not possible for these observations. While the systematic uncertainty associated to the energy scale in the EPIC pn timing mode after applying the RDPHA correction is estimated to be $\approxlt$\,20~eV at 6~keV (Guainazzi 2014\footnote[10]{CCF release note {\it ``RDPHA calibration in the Fe line regime for EPIC-pn Timing Mode''} }), corresponding to $\sim$\,900~km~s$^{-1}$, the existence of residuals at the Au edge  (see Sect.~\ref{sec:observations}) point to residual calibration uncertainties, which could also affect the blueshifts of the absorption lines. 

\subsubsection{Wind launching mechanism}
\label{subsec:launch}

We can calculate the distance between the ionising source
and the slab of absorbing material, r. Since \xil\ = L/n$_e$r$^2$ and n$_e \sim$ N$^{warmabs}_H/d$ (where
$d$ represents the thickness of the slab of ionised absorbing material),
we can calculate r as (L/ \xil\  N$^{warmabs}$)($d/r$). Considering
plausible values of $d/r$ range between 0.1 and 1, we obtain values
of $r$ between 9\,$\times$\,10$^{9}$~cm and 1.3\,$\times$\,10$^{11}$~cm. 
Furthermore, assuming that the value of r probably does not change significantly
between observations, we estimate r\,$\sim$\,1--9\,$\times$\,10$^{10}$~cm,
in agreement with \citet{1630:kubota07pasj}.

The characteristics of the wind allow us to distinguish among
the possible launching mechanisms, namely thermal, radiative,
magnetic, or a combination of them. Thermal driving is effective
at large distances from the central compact object, BH
or NS, where the thermal velocity exceeds the local escape velocity.
Thermally driven models are consequently most effective
in producing slow winds at large radii. 

\citet{woods96apj} determined that a wind would be launched by thermal pressure at radii
larger than 0.25~R$_{IC}$ (where R$_{IC}$ denotes the Compton radius or distance at which the escape 
velocity equals the isothermal sound speed at the Compton temperature T$_{IC}$). For T$_{IC}\sim$\,1.3$\times$10$^7$~K, 
as expected in LMXBs, 0.25~R$_{IC}$ corresponds to a radius of 1.9$\times$10$^{10}M$ cm, where $M$ is the mass of
the compact object in units of solar masses. However, they also found that even above such radius
the wind could be gravity-inhibited if the luminosity were below twice a critical luminosity defined
as L$_{cr}$ = 2.88$\times$10$^{-2}$\,T$_{IC8}^{-1/2}$\,L$_{Edd}$ (see their eq. 4.4), where L$_{Edd}$ is the Eddington luminosity and T$_{IC8}$ is
the Compton temperature in units of 10$^8$ K. Therefore, for T$_{IC}$ = 1.3$\times$10$^7$ K the wind could be gravitationally inhibited for luminosities 
below 0.16 L$_{Edd}$ (see their Fig. 17). \citet{proga02apj} found that the radiation force due to electrons could be important for very luminous systems: when included in their simulations, it lowered the effective gravity and subsequently the escape velocity and allowed a hot robust disc wind to be already produced at 0.01\,R$_{IC}$ (see their Fig.~5), well inside the Compton radius and previous estimates by \citet{woods96apj}.

Indeed, for NS LMXBs, for which the critical luminosity can be precisely calculated, \citet{review:diaz12} showed that
winds were only launched for radii above 0.25~R$_{IC}$ for values of the luminosity above the critical luminosity, while the photoionised plasma
stayed bound as an atmosphere for sources below that luminosity.  
The distance at which the wind is launched in \src, 1--9\,$\times$\,10$^{10}$~cm, points to a
thermal-radiative pressure origin, as discussed by \citet{1630:kubota07pasj}. 
For obs~1--4, assuming T$_{IC}$~$\sim$~1.3\,$\times$\,10$^7$ K, we infer values of R$_{IC} \sim$ 23(76)\,$\times$10$^{10}$~cm and Eddington fractions of 0.7-0.8 (0.2-0.3) for a 3(10)$\Msun$ BH. Thus, the launching radii in obs~1--4 correspond to 0.04--0.39(0.01-0.12) R$_{IC}$ for a 3(10)$\Msun$ BH. \citet{1630:tomsick98apj} estimated a mass of 3$\Msun$ for the BH in \src, based on the assumption that the maximum flux observed during outbursts \citep{1630:parmar95apj} corresponded to the Eddington luminosity at a distance of 10~kpc. However, \citet{1630:seifina14apj} have recently estimated a mass of 10$\Msun$, based on scaling techniques and taking as reference sources \gro, \seventeen\ and \xtefifteen. We find that the wind in \src\ can be launched via a thermal mechanism even if the mass of the BH is larger than 3$\Msun$. For BH masses as large as 10$\Msun$, a composite of thermal and radiation pressure would be probably required to launch the wind. 

\subsubsection{Variability of the wind as a function of accretion state}
\label{sec:wind-hss}

\begin{figure*}[!ht]
\includegraphics[angle=0,width=0.5\textwidth]{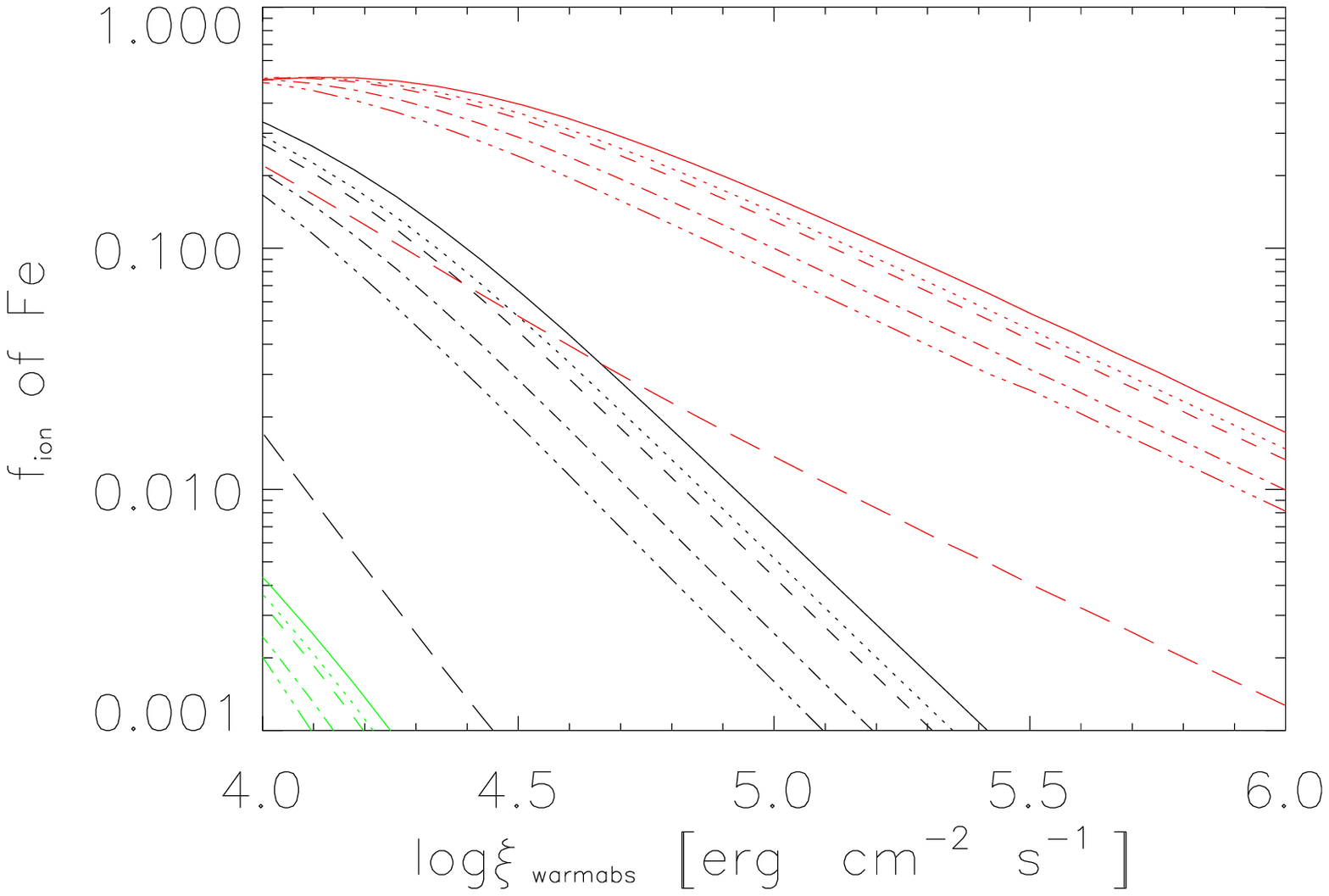}
\includegraphics[angle=0,width=0.5\textwidth]{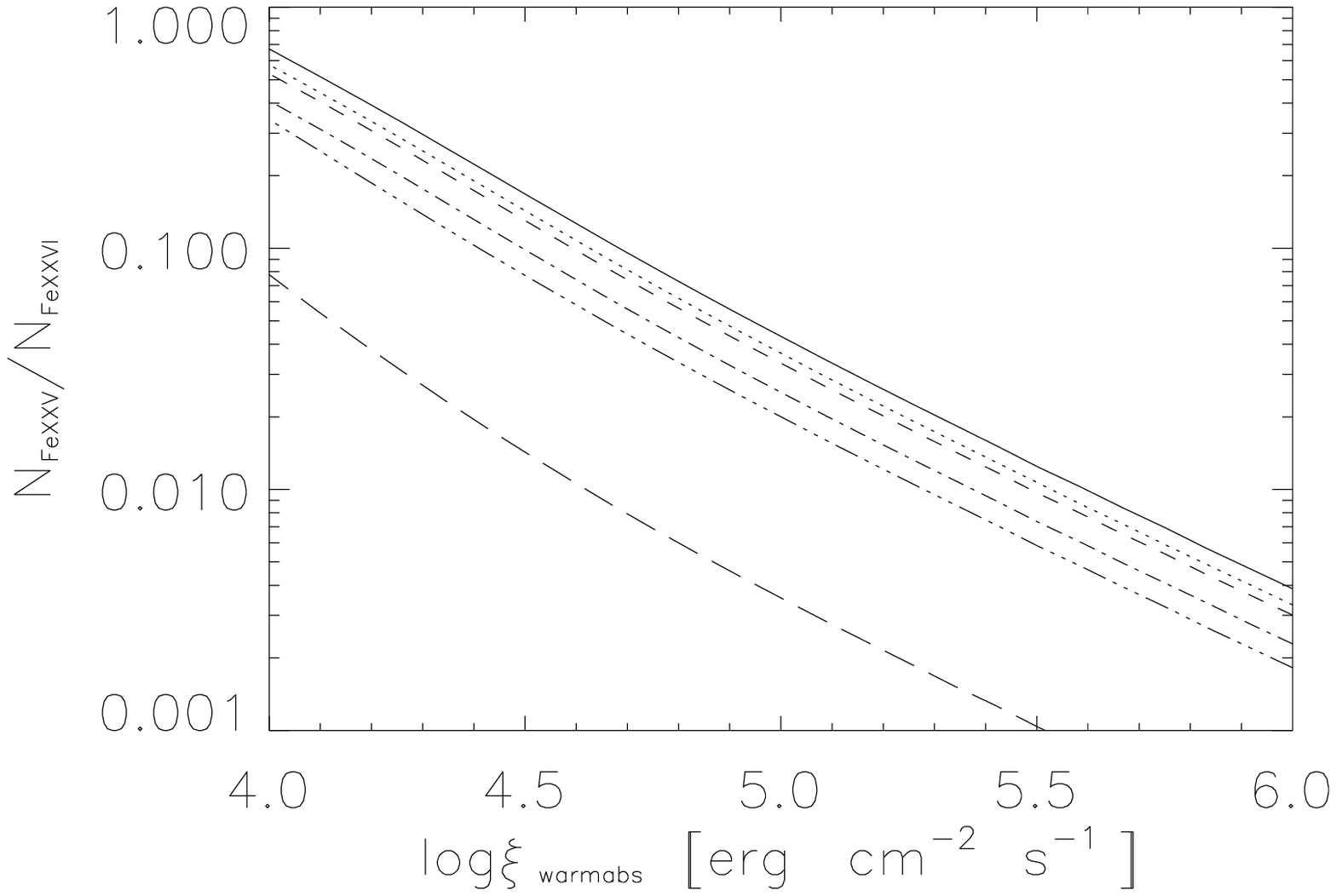}
\caption{Relative ionisation population of iron (left panel) and ratio of He-like to H-like iron (right panel) as a function of the ionisation parameter. In the left panel, the different colours show the populations of \fetfour\ (green), \fetfive\ (black) and \fetsix\ (red), respectively. In both panels the solid, dotted, dashed, dotted-dashed and dot-long dashed curves represent increasing disc temperatures of 1.40, 1.44, 1.46, 1.56 and 1.64~keV. The long-dashed lines show the same parameters for obs~6, for which the SED is composed of the disc and a non-thermal component.} 
\label{fig:ionisation}
\end{figure*}

The \xmm\ observations reported here provide the opportunity to study the evolution of a wind during a state transition and its ionisation state as a function of spectral hardness and luminosity. In particular, we are interested in 
determining whether disc winds may disappear as a consequence of ionisation \citep{1915:ueda09apj,review:diaz12}. We used the illuminating spectrum (or SED) derived from \xmm\ obs~1--4 to calculate the self-consistent ion populations of iron using XSTAR. The SED of obs~5 is harder compared to obs~4 (see Fig.~6). However, as explained in Sect.~\ref{sec:model2}, the decrease of the inner disc radius from obs~4T to obs~4B and the strong increase in temperature may be a consequence of cross-calibration differences between timing and burst mode (but note that such differences are model dependent, highlighting the uncertainty on whether the discrepancy is related to calibration or due to model deficiencies). Therefore, to generate the SED corresponding to obs~5 we considered a temperature of 1.64~keV (obtained by weighting the temperature of obs~5 with the ratio of temperatures between obs~4T and 4B, see Table~2) and a radius of $\sim$22~km, as measured for obs~4T.  
Fig.~\ref{fig:ionisation} shows the relative \fetfive\ and \fetsix\ ion fractions and the \fetfive\ to \fetsix\ ratio expected as a function of ionisation parameter for the five different SEDs. Clearly, as the disc temperature increases from obs~1 to 5 (i.e. the SED becomes harder and the Compton temperature higher), the \fetfive\ to \fetsix\ ratio {\it and} the ion fraction decrease for a given ionisation parameter \citep[see also][Fig.~8 for a similar study]{1630:kubota07pasj}. This implies that even if the luminosity stayed constant, the plasma would be more ionised and have a smaller column density as the disc becomes hotter during the HSS. 

Next, we attempted to establish if the ``non-detection'' of the wind in obs~5 is simply due to a higher ionisation of the plasma. Since $\xi = L /n_{\rm e} \, r^{2}$, if the density, $n_{\rm e}$, and the launching radius, $r$, of the wind stay roughly constant, we expect the ionisation parameter to increase to at least \logxi\,$\sim$\,4.5 for obs~5 as a consequence of the luminosity increase of $\sim$24\% from obs~4 to obs~5 (see Table~2).  
Next, we used the SED of obs~5 (see above) to generate a plasma with a column density of 6.2\,$\times$\,10$^{22}$~cm$^{-2}$ and a higher ionisation parameter of \logxi\,=\,4.5 (as expected from the evolution of hardness and luminosity from obs~4T to obs~5). 

\begin{figure}[!ht]
\includegraphics[angle=0,width=0.24\textwidth]{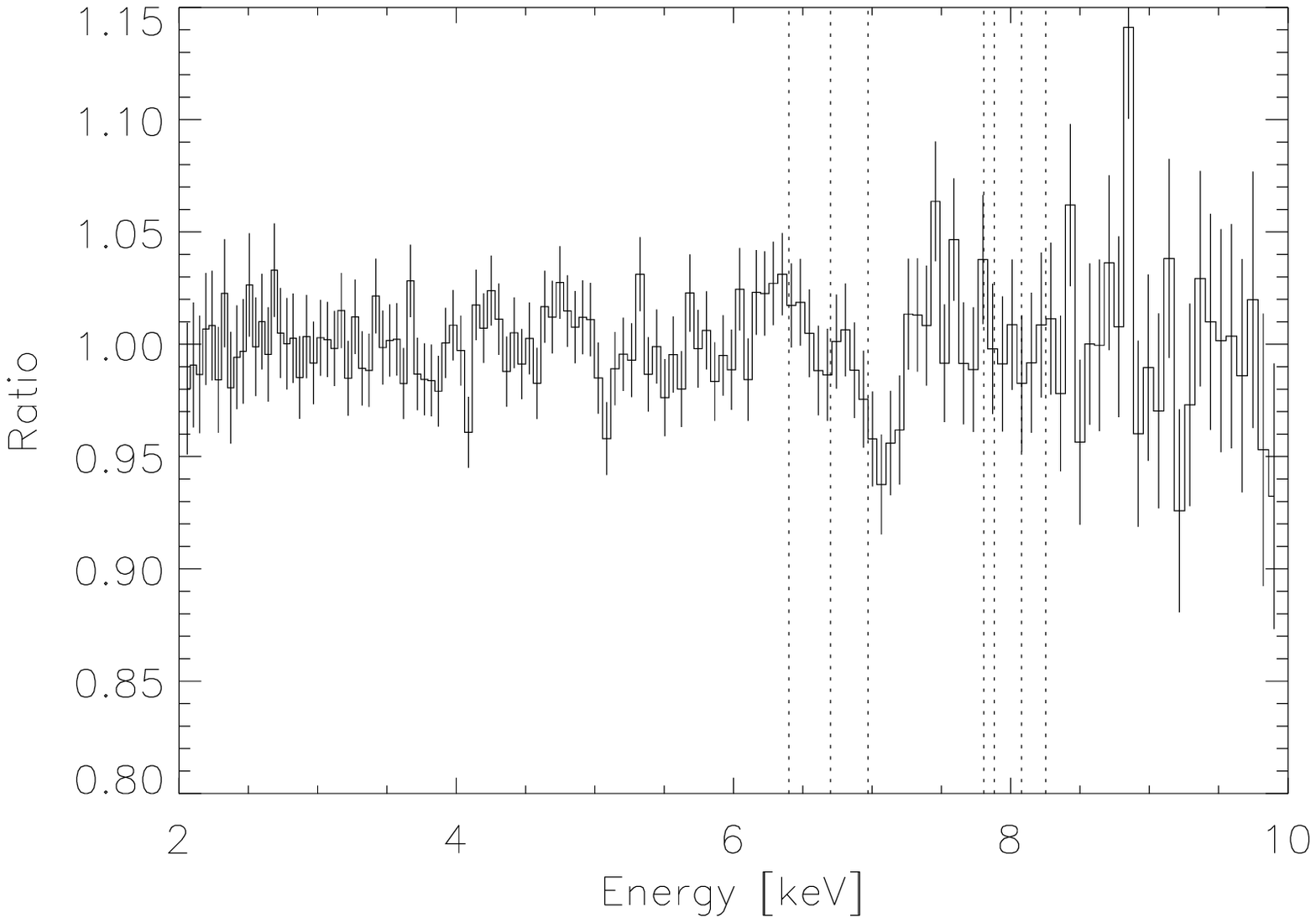}
\includegraphics[angle=0,width=0.24\textwidth]{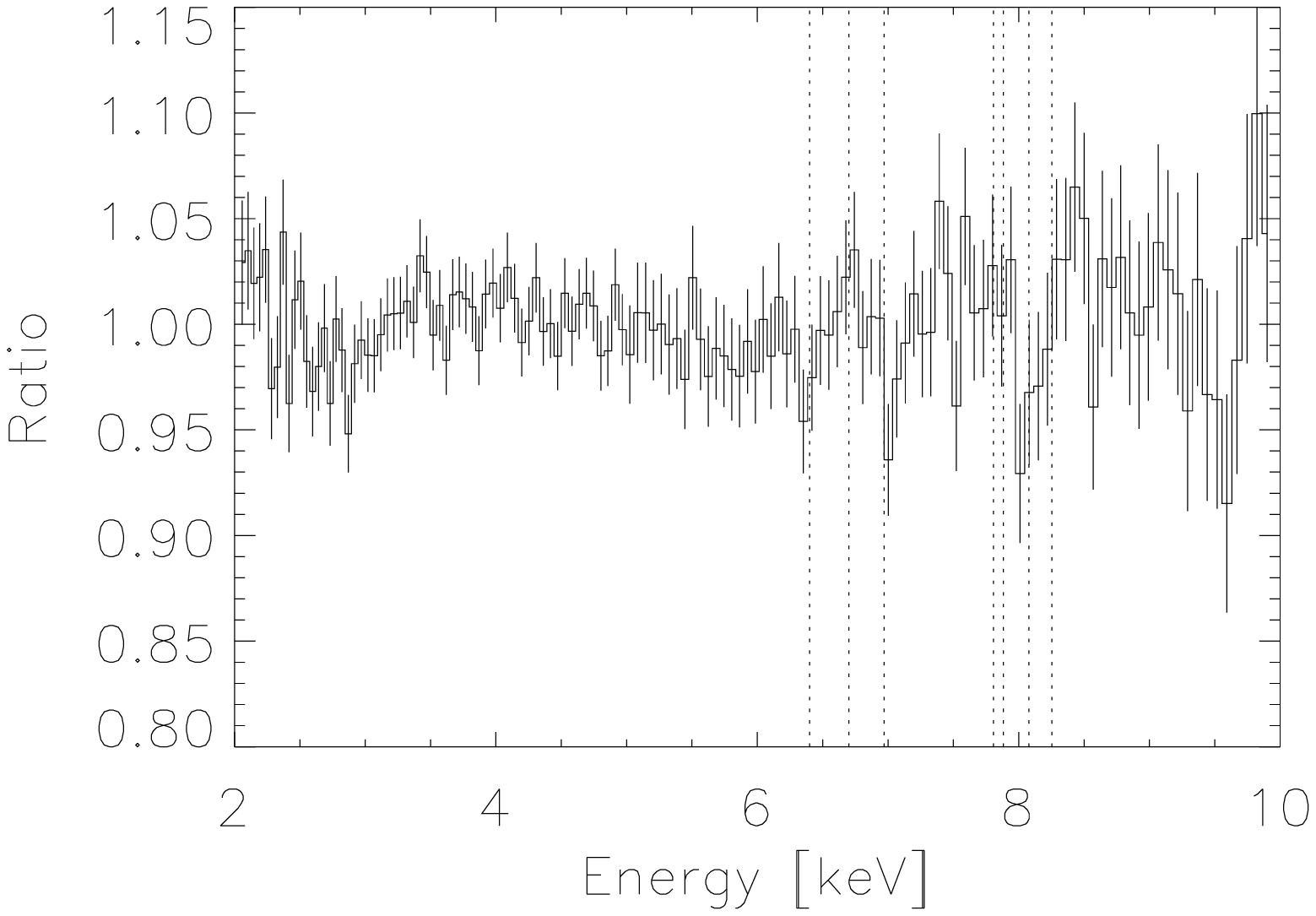}
\caption{Ratio of the data to the continuum model at the Fe K region for simulated (left) and real (right) spectra of obs~5.} 
\label{fig:ionisation2}
\end{figure}

Fig.~\ref{fig:ionisation2} shows the residuals of fitting the simulated (left) and real (right) obs~5 spectra with a model consistent of a disc blackbody absorbed only by neutral absorption, i.e. without the ionised plasma. The real spectrum is consistent with the simulated one, implying that the non-significant detection of the plasma in obs~5 is expected given its column density and degree of ionisation. However, the fact that the simulated spectrum shows a weak \fetsix\ absorption feature implies that the plasma would still have been detected with a longer exposure time. 
In particular, the \fetsix\ absorption feature has an \ew\ of 16\,$\pm$\,8~eV in the simulated spectrum. Similarly, we can fit an absorption feature at 7.02\,$\pm$\,0.6~keV (consistent with \fetsix) in the real spectrum with an \ew\ of  8\,$\pm$\,5~eV. Although the latter feature does not improve the quality of the fit (and is therefore not included in the fit in Table~2), it is consistent with the warm absorber expected in obs~5.
In conclusion, the evolution of the wind within the HSS is well understood in terms of changes in the continuum and the ionising SED. A consequence of the evolution is that at the time the source is transiting to the ``anomalous'' state the wind is already ``transparent'', i.e. it has apparently disappeared. 
In reality, as the wind becomes more and more ionised, the amount of electrons available for comptonisation increases. Therefore, the transition to a very luminous spectral state with strong disc emission and comptonisation is the logical consequence of the full ionisation of the wind during the soft state due to the increasing temperature and luminosity of the disc. In other words, we could be witnessing the transformation of the disc atmosphere or wind into a ``hot corona'', which will scatter and comptonise disc photons, and as such will further photoionise any remaining plasma, and may contribute to the continuum via bremsstrahlung. In particular, we note that while \citet{1630:hori14apj} estimate a decrease by a factor of 4 in the \fetsix\ ion fraction from the HSS to the very high state based on $\suzaku$ observations of \src, we find that the presence of the ``corona" in obs~6 makes the \fetsix\ ion fraction drop by more than an order of magnitude between obs~4 and obs~6 (see Fig.~\ref{fig:ionisation}) and is therefore consistent with the absence of absorption lines in the ``anomalous'', very high, state.

A consequence of the above is that the evolution of the wind in the HSS of other BH LMXBs should follow the same pattern as for \src\ if the density of the plasma and the launching radius remain approximately constant. Thus, observations of the HSS of high inclination sources as the luminosity increases should show how the wind becomes increasingly ionised until absorption lines disappear and the source transits to a state with strong comptonisation. Indeed, the wind in the BH transient \seventeen\ seems to follow the same evolution as in \src. \citet{1743:miller06apj} observed this source in four different epochs with $\chandra$/HETGS during its 2003 outburst. One of the observations (2) corresponds to a SPL state while the other three (1, 3, 4) are classified as a HSS based on spectral (power law index and flux) and timing properties \citep{1743:mcclintock09apj}. We identify the SPL observation with an ``anomalous'' state similar to that seen in \src\ or \gro\ based on the deviation of the disc luminosity and temperature from the $L \propto T^4$ relation proper from the HSS (see e.g. Fig.~2 of \citet{1655:kubota01apjl} compared to Fig.~7 of \citet{1743:mcclintock09apj}.
Interestingly for the three HSS observations, as the disc temperature increases the \fetfive\ to \fetsix\ ratio decreases, indicating that the wind is becoming more ionised (see Fig.~\ref{fig:ionisation}). In contrast, the SPL observation  does not show any signature of a wind. This is consistent with the trend observed for \src\ in this work. 

This picture could be altered if e.g. there are significant changes in the density of the plasma, which could be caused by the feedback mechanism between illumination and atmospheric structure \citep{jimenez02apj}.

Finally, we note that while the LHS is also characterised by strong comptonisation, the picture above may not be applicable to such states (i.e. that the wind is not observed due to full ionisation). The reason is that in the LHS the disc is cooler than in the HSS and most likely truncated. Therefore, heating/ionisation of the wind cannot be associated to the increasing temperature/luminosity of the disc. Observations of high inclination sources during the hard-to-soft and soft-to-hard transitions should shed more light on this topic.

\subsubsection{Relation of mass outflow rate to accreting power}

The mass outflow rate carried by the wind can be evaluated as 
 \begin{eqnarray} 
\dot{M}_{out} = \Omega_{wind}r^2nm_pv_{wind} = \Omega_{wind}(L/\xi)m_pv_{wind}
\end{eqnarray}
where m$_p$ is the proton mass, $\Omega_{wind}$ the solid angle of the wind, and $v_{wind}$ the outflow velocity \citep[see e.g.][]{gx13:ueda04apj}. 

Thus, for obs~1--4, we obtain an outflow rate of 
 \begin{eqnarray} 
\dot{M}_{out}~\sim~1\times10^{19}~\left(\frac{\Omega/4\pi}{0.4}\right)~\left(\frac{v}{1000~km~s^{-1}}\right)~g~s^{-1}.
 \end{eqnarray}

Assuming an efficiency of 10\% to convert accretion power in luminosity, the above outflow rate corresponds to 2.4 (0.7) times the Eddington accretion rate for a 3 (10) $\Msun$ BH. 

We can estimate the outflow rate in other accretion states of \src\ under the assumption that the wind is fully ionised and that its location and density do not vary significantly compared to the HSS. Based on previous outbursts \citep[see e.g.][]{dunn10mnras} the luminosity can be as high as $\sim$1.2$\times$10$^{38}$ erg s$^{-1}$ for the initial LHS, $\sim$4$\times$10$^{38}$ erg s$^{-1}$ during the ``anomalous'' state and $\sim$1$\times$10$^{36}$ erg s$^{-1}$ at the final LHS. Thus, if the outflow velocity were the same during the HSS and the states at which the wind is fully ionised, we would expect outflow rates lower by at least two orders of magnitude during the final LHS and similar or slightly higher during the ``anomalous'' state compared to the HSS (the initial LHS would cover all the range of outflow rates already considered in other states). 

A larger outflow velocity for the more ionised plasma was found in observations of \grs\ and \gro\ \citep{1915:ueda09apj,1655:kallman09apj} and explained as different parts of the wind being traced by different ions. \citet{1915:ueda09apj} further concluded that this was consistent with the ionisation being stronger, and consequently the more ionised ions being faster, at outer parts of the wind in thermally driven wind models (see their Sect.~4.1, but note that the relation invoked by \citet{1915:ueda09apj} is only correct in the region around and beyond the sonic point; \citet{1655:netzer06apj}). Moreover, 
an increase of luminosity and, consequently, of the radiation pressure, could allow to launch the wind from progressively smaller radii as the disc becomes hotter. 
However, 
simulations of ``overionised'' winds in AGN show that the wind will fail, i.e. will fall back before it is expelled from the system, under high ionisation conditions \citep{proga04apj,schurch06mnras} \citep[see also Fig.~6 of][for an example of different correlations between outflow velocity and ionisation of the wind for a BH LMXB at different parts of the wind]{1655:luketic10apj}. Therefore, it is plausible that in the states in which the ``wind'' is fully ionised, the plasma actually stays bound as a hot corona or a failed wind and as such does not contribute to mass loss in the system. 

\subsubsection{The broad iron line}
\label{sec:broadline}

We found a broad, $\sigma\sim$\,0.8--1~keV, emission line in obs~1--4. When fitted with a Gaussian, the energy centroid varied between 6.4 and 6.97~keV, consistent with emission of neutral or ionised iron. The \ew\ of the line decreased progressively from 175~eV in obs~1 to 56~eV in obs~4 (see Table~\ref{tab:bestfit-warm}).

The presence of broad Fe lines in LMXBs has been extensively studied. While several systematic studies have been performed for NS LMXBs \citep[see][]{white86mnras,hirano87pasj,asai00apjs,ng10aa}, such studies are challenging for BH LMXBs due to their transient nature. In both NS and BH systems, broad lines could arise in the inner accretion disc by fluorescence following illumination by an external source of X-rays, and be broadened by relativistic effects near the compact object \citep[e.g.][]{reynolds03ps, fabian05, matt06an}. Alternatively, they could originate in the inner
part of the so-called accretion disc corona, formed by
evaporation of the outer layers of the disc illuminated by the
emission of the central object \citep[e.g.,][]{white82apj, kallman89apj} and be broadened predominantly by Compton scattering \citep[e.g.][]{pozdnyakov79aa, sunyaev80aa}. A third possibility is that they originate in a partially ionised wind as a 
result of illumination by the central source continuum photons and are
broadened by electron downscattering in the wind environment \citep{laming04apjl,laurent07apj}. 

In the first interpretation, if relativistic effects represent the major contribution to the line breadth or  relativistic and Compton scattering effects can be disentangled, this can be used to measure black hole spin \citep{dovciak04apj}. 
In the second and third interpretations, the breadth of the line will mostly vary as a function of the state of the plasma in which it is produced, since relativistic effects should not be significant. Therefore, in those scenarios we expect relatively broad lines in an ionised plasma where Compton scattering is significant. In contrast, in the first scenario, we expect the line to broaden as the disc radius approaches the BH Innermost Circular Orbit (ISCO) as a consequence of relativistic effects. However, even in this scenario, Compton scattering would broaden and shift the line, making relativistic effects more difficult to observe and interpret \citep[see e.g.][]{brandt94mnras}. 

Recently, \xmm\ observations of the NS LMXB \gx\ showed a correlation of the variations of the broad iron line and the state (column density and degree of ionisation) of the warm absorber, indicating that absorption and emission could occur in the same plasma \citep{gx13:diaz12aa}.
Similarly, the observations analysed in this work seem to show a contemporaneous evolution of the state of the wind and the flux and \ew\ of the broad iron emission line, indicating that they could form in the same environment. In particular, we find that the flux and \ew\ of the broad Fe line decrease as the spectrum hardens and the wind becomes more ionised and thinner (see Fig.~9), although the large errors on the column density of the warm absorber prevent a more definite conclusion. 

In the presence of a photoionised plasma such as a disc wind or atmosphere we expect three spectral components to arise: absorption, intrinsic emission and reflection \citep{netzer93apj,krolik95apj}. The absorption component is only visible if the plasma is in the line of sight of the observer. In contrast, the intrinsic emission and reflected components could be observed from any inclination, depending on the strength of such components with respect to the continuum. The reflected component is qualitatively similar to that produced by standard disc reflection models \citep{krolik95apj,sim10mnras}, e.g. a moderately strong Fe line is present in both models. Since the existence of a photoionised plasma in obs~1--4 is demonstrated by the detection of narrow absorption lines, it is therefore sensible to consider the emission and reflection of such plasma. Moreover, the contemporaneous evolution of the absorption and the reflection fraction (see Fig.~\ref{fig:rfl_warm}) point to an origin in the same plasma. In contrast, it is difficult to justify such changes if the broad iron line is produced at the inner disc, as expected in the classical reflection scenarios, and the wind at $\approxgt$10$^{4}$ gravitational radii, as shown in Sect.~\ref{subsec:launch}, since they could be subject to a different illumination from the compact object. For example, the plasma at 10$^{4}$ r$_g$ may not suffer self-shielding from a puffed-up central corona due to its height above the disc, in contrast to a reflector very close to the innermost stable radius \citep[]{1915:ueda10apj}. 
Furthermore, we do not find significant changes in the disc radius (inferred from the disc normalisation) during obs~1--4, as expected if the source is in its HSS. Therefore, we cannot explain the changes observed in the broad Fe line (see Fig.~\ref{fig:pndel}) as caused by changes in the inner disc radius. 

We also note that the radius inferred from relativistically blurring the reflection component is decreasing from 19\,$^{+6}_{-3}$ $r_g$ in obs~1 to $\sim$11--14 $r_g$ in obs~2--3 and $<$\,5 $r_g$ in obs~4 (see Table~4). For comparison, the radius inferred from the continuum is $\sim$6(2) $r_g$ for a 3(10) solar masses black hole.
While it is difficult to compare the absolute values of the radius obtained via the relativistic blurring or the disc continuum due to calibration uncertainties, it is clear that there is an evolution of the radius when we measure it considering reflection as opposed to the (expected) constant radius measured from the continuum. Therefore, we caution that black hole spin measurements based on the reflection component may be affected by a high uncertainty. 

Finally, since the existence of photoionised plasma is probably ubiquitous to all LMXBs \citep{ionabs:diaz06aa,ponti12mnras}, measurements of the line profile could also be affected by material reflected in the photoionised plasma even for low inclination sources, for which absorption is not observed.  Interestingly, \citet{gx339:dunn08mnras} studied the low-inclination BH GX~339--4 over more than a decade of RXTE observations and found that the \ew\ of the broad Fe line was largest in the low luminosity HSS (see their Fig.~7), in agreement with what we find in this work. Recently, \citet{gx339:plant14mnras} studied the reflection fraction in the same source over three outbursts and found that the reflection fraction was also largest during the low luminosity HSS and that the ionisation parameter of the reflection component was higher during the HSS compared to the LHS, in agreement with our results. They ruled out the same mechanism causing the changes in reflection during the LHS and HSS. While the former would be driven by changes in the inner disc radius, the latter would be caused by changes in the geometry of the corona. Similarly, we conclude that the changes during the HSS cannot be caused by changes in the inner disc radius and consequently measurements of BH spin based on the breadth of the iron line are challenging. 

A caveat to the above interpretation arises from the limitations of the reflection model we are using. Ideally, a self-consistent model including absorption, intrinsic emission and reflection in the wind should be used. For example, we cannot compare the values of $\xi$ found for the absorption and the reflecting plasma due the different ionisation continuum used (the self-consistent continuum model for the former and a power law for the latter). 
Also, most reflection studies in BH LMXBs use a power-law illumination continuum (but see e.g. \citet{gx339:kolehmainen11mnras} for a different approach including disc illumination). Therefore, it is now urgent to take into account the disc illumination and its atmosphere/wind in studies of LMXBs throughout an outburst episode in order to confirm or rule out the proposed scenario.

\subsection{The disappearance of the wind and the switching on of the jet}
\label{subsec:wind-jet}

The \xmm\ observations reported here show the disappearance of a wind due to increasing ionisation as the temperature and luminosity of the disc rise, probably with a constant inner radius at the ISCO, first during the HSS and then as the source transits to an ``anomalous'' state.  This shows that ionisation can render the wind ``invisible'' in a very short period of time, i.e. no significant geometrical or accretion flow changes have to be invoked to explain the absence of the wind, at least in observations showing a sufficiently high luminosity and hard spectrum.

The last observation of this campaign (obs~6), performed on September 28, 17 days after obs~5, showed for the first time a strong emission line at $\sim$7.27~keV and weaker lines at 4.04 and 8.14~keV \citep{1630:diaz13nat}. In addition, quasi-simultaneous radio observations showed significant optically thin emission that was absent during obs~5. We interpreted the emission lines and the radio spectrum as arising in a relativistic jet that was switching on. However, \citet{1630:neilsen14apjl} recently reported on the detection of significant optically thin radio emission from February--June 2012, during a HSS of the outburst, at flux densities higher than we observed during \xmm\ obs~6. This is at odds with the expected radio jet quenching seen in the soft state of black hole transients. Moreover, the optically thin emission they detected was first seen $\sim$50~days after a hard-to-soft state transition. For these reasons, they interpreted the radio emission as the interaction of a jet (presumably ejected during the transition from the hard to the soft state some weeks before) with the ISM.  

Should this be the case, then if the same origin were to be invoked for the radio emission reported in obs~6 by \citet{1630:diaz13nat}, it would imply that the existence of strongly shifted X-ray lines was unrelated to the quasi-simultaneous radio emission in \xmm\ obs~6.  However, during \xmm\ obs~6 the source was in an ``anomalous'' accretion state \citep[characterized by a very high luminosity, a hot disc and significant hard X-ray emission][]{1630:abe05pasj}, rather than the HSS (and possibly LHS) of the $\chandra$ observations reported by \citet{1630:neilsen14apjl}.  Radio flaring has been detected in several previous instances of the ``anomalous'' state in \src\ \citep{1630:hjellming99apj}, \xtefifteen\ \citep{1550:wu02apj} and \gro\ \citep{1655:hunstead97iau}, making it much more plausible that the radio emission is indeed associated with this accretion state, rather than arising from an unrelated jet/ISM interaction.  Indeed, in \xtefifteen, the downstream interactions with the ISM were directly observed, and occurred several years after the original X-ray outburst \citep{1550:corbel02science}. Moreover, the previous radio observation of \src\ occurring during an ``anomalous'' state shows a high degree of linear polarization and an optically thin spectrum, consistent with an origin in optically thin jets with highly ordered magnetic fields \citep{1630:hjellming99apj}.

Finally, even if the X-ray and radio emission are unrelated, it is difficult to interpret the Doppler-shifted X-ray lines in the context of a jet-ISM interaction or of a disc outflow. However, we note that with the systematic errors estimated for the EPIC pn effective area (first available in 2014) the significance of the most prominent emission line at 7.3~keV is reduced to 5.2~$\sigma$ (chance probability of 1.7\,$\times$\,10$^{-8}$ as measured by an F-test). This, the fact that the lines were not seen in a $\suzaku$ observation taken only 4 days after obs~6 and during the same spectral state \citep{1630:hori14apj}, and the discrepancy between the spectral continuum of obs~4T and 4B pointing to CTI calibration differences between timing and burst modes make it now urgent to determine whether the lines are also detected in grating observations or if they could be an artifact from the low resolution CCD cameras when operated at such high count rates.

Interestingly, there don't seem to be structural changes in the disc from obs~1 to 6. Moreover, the hard X-ray component present in obs~6 could simply be the same ionised plasma of obs~1--4 which is not detectable anymore through absorption lines due to the full ionisation of its ions but reveals itself as a Comptonised component or via bremsstrahlung radiation. This gives us a framework to study the conditions of appearance of a jet, to date unknown. If the radio emission is confirmed to originate in a jet, structural changes in the disc cannot then be a relevant factor for its launching. Instead, the presence of the fully ionised wind or corona could provide the necessary conditions for the incipient optically thin jet to be accelerated and collimated.

\section{Conclusions}
We performed six \xmm\ observations during the HSS and the ``anomalous'' state of the 2012-13 outburst of \src. Absorption lines characteristic of a disc wind were found in the first four observations.

We find that:

- the disc wind in \src\ originates at large radii, $\sim$1--9\,$\times$\,10$^{10}$~cm, and is consistent with being thermally and possibly radiation driven, consistent with work by \citet{1630:kubota07pasj}.

- the disc wind is present in all the observations taken during the HSS but disappears as \src\ makes a transition towards the ``anomalous'' state. We attribute the disappearance of the wind to strong ionisation of the wind, which is already signaled by the increasing ionisation parameter and decreasing column density as the disc becomes hotter and more luminous during the HSS. This trend is also observed for the BH transient \seventeen. We speculate that the full ionisation of the wind contributes to the increased Comptonisation characteristic of the ``anomalous'' (very high) states.

- contemporaneous changes of the reflection fraction and the warm absorber point to an origin in the same disc wind. In particular, the broad iron line can be significantly better modeled by models that consider reflection from a self-illuminated disc, rather than from an external, non-thermal source.

- if the origin of the emission during obs~6 is confirmed to arise in an optically thin jet, this would be the first time that the switching off of the wind and the switching on of the jet is observed. The condition for the appearance of the jet could be then supplied by the fully ionised wind. 


\begin{acknowledgements}
This work was based on observations obtained with
XMM-Newton, a European Space Agency (ESA) science mission with instruments and
contributions directly funded by ESA member states and the USA (NASA).We thank the
XMM-Newton team for the fast scheduling of these observations. This work was supported by the
Spanish Ministerio de Econom\'ia y Competitividad and European Social Funds through
a Ram\'on y Cajal Fellowship (S.M.) and the Spanish Ministerio de Ciencia e Innovaci\'on
(S.M.; grant AYA2010-21782-C03-01) and the Australian Research Council's
ÔDiscovery ProjectsÕ funding scheme (J.C.A.M.-J.; project number DP120102393). 
\end{acknowledgements}


\bibliographystyle{aa}
\bibliography{1630}

\bigskip\
{\it Note added in proof: }After submission of this paper we had the opportunity to discuss the discrepancy between the results of \citet{1630:hori14apj} and this paper regarding full ionisation of the wind during the VHS with the authors of the former paper. We conclude that the discrepancy between the papers is most likely due to the use of different SEDs below 50 keV; \citet{1630:hori14apj} used an SED with two Comptonisation components above 10 keV, with photon indices of 2.9 and 2.1 below and above 50 keV, respectively, whereas we used a single component with photon index 2.0. While the softer SED used by \citet{1630:hori14apj} would significantly ionise the wind during the VHS, it can not explain its complete disappearance, in contrast to the results of this paper. However, we note that fixing the index of the power-law component to 2.9 for our \xmm\ obs~6 significantly worsens the quality of the fit ($\chi^2$ of 1.67 (129 d.o.f.) compared to 1.43 (129) when using an index of 2.0), and underpredicts the 15--50 keV Swift/BAT flux by a factor of 2. This highlights the importance for wind studies of a full characterisation of the SED over a broad energy band.

\appendix

\section{Estimation of the systematic errors on the effective area of the EPIC-pn camera}
\label{sec:app1}

In this Appendix we present the analysis underpinning the estimate of the systematic error on the calibration of the effective area employed in this paper.
Our methodology is based on
analysing sizeable samples of sources whose X-ray spectrum in the EPIC sensitive bandpass can be reasonably approximated by simple phenomenological models on
astrophysical grounds. The results of our studies should be therefore interpreted as upper limits on the true average systematic uncertainties, because any
systematic deviation between the assumed astrophysical model and the intrinsic source emission could not be disentangled from calibration effects.

As baseline sample we used a set of 112 observations of Active Galactic Nuclei (AGN) performed in the framework of the XMM-Newton calibration program. They
are primarily radio-loud objects, whose X-ray spectrum is typically dominated by non-thermal emission from a jet, with a small contribution by thermal emission
and/or reflection from the accretion disk. EPIC-pn data were reduced using the same SAS version and calibration files employed for the reduction of the 4U1630-47
data discussed in this paper. The instrumental mode mostly used in the AGN sample was Small Window. However, the effective area calibration is not
dependent on the instrumental mode. We can therefore apply the results of the AGN study to the data discussed in this paper. Reduction procedures for the
extraction of source and background spectra and responses are discussed in Stuhlinger et al. (2010) \footnote[11]{See {\it Status of XMM-Newton instruments cross-calibration with SASv7.1} (XMM-SOC-CAL-TN-0052), available at: {\tt http://xmm2.esac.esa.int/docs/documents/}{\tt CAL-TN-0052.ps.gz}}.

We rebinned the AGN spectra using a combination of intrinsic and extrinsic criteria. As far as the former ones are concerned,
we impose that each background-subtracted spectral channel
contains at least 25 counts, and that the instrumental resolution is oversampled by a factor not larger than 3. These conditions ensure the applicability of the
chi-squared goodness-of-fit test. Additionally, we impose that the same binning scheme as Obs~4T discussed in this paper is employed whenever
it is more restrictive than
those applied on the base of the intrinsic spectra. Rebinned spectra were modelled using a logarithmic power law modified by photoelectric absorption. We added an
unresolved, or a broad (width constrained not to be higher than 0.5 keV) Gaussian
emission line to the best-fit continuum model, whenever the inclusion of this component yielded an improvement in the fit quality at
the 90\% level for one interesting parameter. Sources were discarded from subsequent analysis, whenever the reduced $\chi^2$
against the final best-fit model was larger than 1.25. We applied
this criterion to avoid including sources in the final analysis, whose X-ray spectrum was not well described by the assumed simple phenomenological parameterization.

\begin{figure}
\begin{center}
\includegraphics[angle=90,width=0.40\textheight]{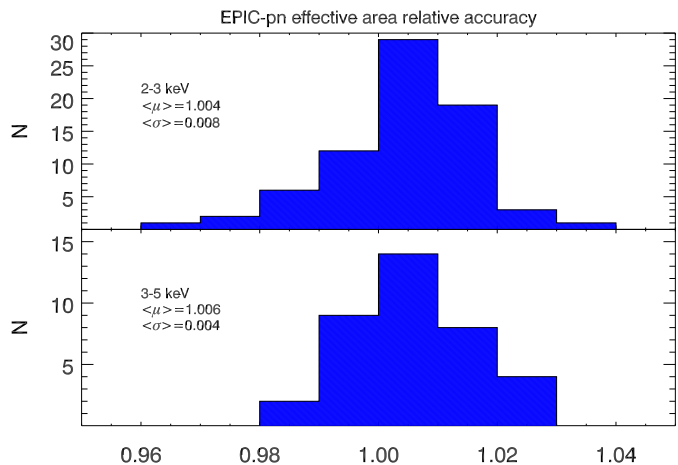}
\end{center}
\caption{
Distribution of spectra residuals in the cross-calibration AGN sample used to estimate the systematic error on the calibration of the EPIC-pn effective area
in the 2--3~keV ({\it upper panel}), and 3--5~keV ({\it lower panel}) energy band.
}
\label{fig1mg}
\end{figure}

The distribution of the residuals in the 2--3 and 3--5~keV band is shown in Fig.~\ref{fig1mg}. The averages are different from 1 by 0.4\% and 0.6\%, respectively. The
intrinsic standard deviations are 0.8\% and 0.4\%, respectively. They were estimated by subtracting in quadrature the average of the statistical error from
the observed standard deviation of the distributions. Therefore we use as final systematic errors 1.2 and 1.0\% in the 2--3 and 3--5 keV energy band, respectively.

The statistical quality of the AGN sample spectra is insufficient to provide constraints above 5~keV. The observed width of the distribution of the residuals is
dominated by the statistical scatter. Spectra of bright galaxy clusters such as Coma or Perseus do not provide better constraints.
In order to have an estimate of the effective area calibration systematics above 5~keV we applied
a similar method to a much smaller sample of X-Ray Binaries observed in EPIC-pn Timing Mode, whose high-energy spectrum is known to be unaffected by ionised
absorption, and can be well fitted by combinations of simple continua. The sample is constituted by the following sources (observations): \xtee\ (Obs.\#0157960101),
Aql~X$-$1 (Obs.\#0303220201), 4U\,1746$-$371 (Obs.\#0405510201), \twelve\ (Obs.\#0405510301). Spectra were extracted using the same procedure as described in this
paper for the \src\ data. Background spectra were extracted from a 4 column box corresponding to {\tt RAWX}~=~2-6.
We fit the spectra in the 5--12~keV energy range, except for \twelve\ that was fit above 7.5 keV to avoid an absorption feature at $\simeq$7~keV.
We employed either a
simple power law, or a combination of blackbody and multi-colour disk blackbody ({\tt diskbb} in XSPEC), and choose for each source the model yielding the
best fit quality. The scatter is dominated by the statistical quality of the data down to $\pm 1\%$. Hence we estimate that 1\% is an upper limit (probably
rather conservative) on the systematic uncertainties of the calibration of the effective area above 5~keV, in good agreement with our direct estimate at lower energies.

\end{document}